\documentclass{emulateapj}

\usepackage{graphics} 
\usepackage{epsfig}
\usepackage{natbib}
\usepackage{amsmath}
\usepackage{rotating} 
\usepackage{color}
\shorttitle{The Molecular Dissociation Continuum in Protoplanetary Disks}
\shortauthors{France et al.}
\begin{document}
\title{The 1600~\AA\ Emission Bump in Protoplanetary Disks: \\ A Spectral Signature of H$_{2}$O Dissociation\altaffilmark{*}}
\author{
Kevin France\altaffilmark{1}, Evelyne Roueff\altaffilmark{2}, Herv{\'e} Abgrall\altaffilmark{2}
}
\altaffiltext{*}{Based on observations made with the NASA/ESA $Hubble$~$Space$~$Telescope$, obtained from the data archive at the Space Telescope Science Institute. STScI is operated by the Association of Universities for Research in Astronomy, Inc. under NASA contract NAS 5-26555.}
\altaffiltext{1}{Laboratory for Atmospheric and Space Physics, University of Colorado, 600 UCB, Boulder, CO 80309;  kevin.france@colorado.edu}
\altaffiltext{2}{LERMA, Observatoire de Paris, PSL Research University, CNRS, Sorbonne Universités, UPMC Univ. Paris 06, F-92190, Meudon, France}
\begin{abstract}

The FUV continuum spectrum of many accreting pre-main sequence stars, Classical T Tauri Stars (CTTSs), does not continue smoothly from the well-studied Balmer continuum emission in the NUV, suggesting that additional processes contribute to the short-wavelength emission in these objects.  The most notable spectral feature in the FUV continuum of some CTTSs is a broad emission approximately centered at 1600~\AA, which has been referred to as the ``1600~\AA\ Bump''.  The origin of this feature remains unclear.  In an effort to better understand the molecular properties of planet-forming disks and the UV spectral properties of accreting protostars, we have assembled archival FUV spectra of 37 disk-hosting systems observed by the {\it Hubble Space Telescope}-Cosmic Origins Spectrograph.   Clear 1600~\AA\ Bump emission is observed above the smooth, underlying 1100~--~1800~\AA\ continuum spectrum in 19/37 Classical T Tauri disks in the $HST$-COS sample, with the detection rate in transition disks (8/8) being much higher than in primordial or non-transition sources (11/29).   We describe a spectral deconvolution analysis to separate the Bump (spanning 1490~--~1690~\AA) from the underlying FUV continuum, finding an average Bump luminosity, $L$(Bump)~$\approx$~7~$\times$~10$^{29}$ erg s$^{-1}$.  Parameterizing the Bump with a combination of Gaussian and polynomial components, we find that the 1600~\AA\ Bump is characterized by a peak wavelength $\lambda_{o}$~=~1598.6~$\pm$~3.3~\AA, with FWHM~=~35.8~$\pm$~19.1~\AA.

Contrary to previous studies, we find that this feature is inconsistent with models of H$_{2}$ excited by electron-impact.   We show that this Bump makes up between 5~--~50\% of the total FUV continuum emission in the 1490~--~1690~\AA\ band and emits roughly 10~--~80\% of the total fluorescent H$_{2}$ luminosity for stars with well-defined Bump features.  Energetically, this suggests that the carrier of the 1600~\AA\ Bump emission is powered by Ly$\alpha$ photons.  We argue that the most likely mechanism is Ly$\alpha$-driven dissociation of H$_{2}$O in the inner disk, $r$~$\lesssim$~2 AU.  We demonstrate that non-thermally populated H$_{2}$O fragments can qualitatively account for the observed emission (discrete and continuum), and find that the average Ly$\alpha$-driven H$_{2}$O dissociation rate is 1.7~$\times$~10$^{42}$ water molecules s$^{-1}$.    
\end{abstract}
\clearpage
\section{Introduction}

Measurements of the composition and physical state of protoplanetary gas disks form the basis for estimating the initial conditions of planet formation.  Molecular gas emission and absorption originating inside of 10 AU provide our best means of estimating the conditions at the radii where gas giant and rocky planet cores are forming and accreting their nascent atmospheres.  Over the past decade, surveys of molecular emission from the inner few AU have provided new constraints on the radial distribution, temperature, and composition of planet-forming disks.  Surveys of mid-IR emission from CO~\citep{salyk08, salyk09, brown13, banzatti15}, H$_{2}$O and organic molecules~\citep{pontoppidan10b,carr11, salyk11b}, UV emission and absorption of H$_{2}$ and CO~\citep{herczeg04, france11b, schindhelm12a, france14b}, and spectrally/spatially-resolved near-IR observations~\citep{pontoppidan11,carmona11,brittain15} have placed constraints on the relative abundance ratios of H$_{2}$, CO, and H$_{2}$O, the evolution of the inner gas disk radius, and the excitation conditions of the molecular gas.

An important caveat when considering molecular spectra of inner disks is that none of these tracers probe the disk midplane where planet-formation is active.   Similarly, very few of these surveys provide spatially-resolved maps of this region (although see, e.g., Eisner et al. 2008) and ALMA is not sensitive to warm gas emission inside of~$\sim$~10 AU~for disks at typical star-forming region distances of $d$~$\geq$~100 pc~\citep{hltau15group,andrews16}.   The inner disk tracers noted in the preceding paragraph all probe various heights in the inner disk atmosphere, mostly at vertical columns of 100~--~1000 lower than the optically thick and observationally inaccessible midplane.  Given these challenges, our best opportunity to converge on a three-dimensional view of the planet-forming regions around young stars is to assemble panchromatic tracers of the relevant disk molecules~\citep{aguilar16}.  Taken as an ensemble, the weaknesses of individual tracers can be mitigated and we can develop a suite of diagnostics that support a more complete picture of this phase of planet-formation. 

{\it James Webb Space Telescope} observations of water and other organic molecules are a highly anticipated contribution to this goal (e.g., Banzatti et al. 2017)~\nocite{banzatti17} and  there is work to be done placing panchromatic disk observations into a unified modeling framework~\citep{panchrom_model_willacy}.  In an effort to expand the observational database of UV molecular tracers, we present a new analysis of archival {\it Hubble Space Telescope} UV spectra of protoplanetary disks around M$_{*}$~$\lesssim$~2M$_{\odot}$ central stars (Classical T Tauri Stars; CTTSs) to propose a new origin for the broad, quasi-continuous, spectral feature near 1600~\AA\ (which we refer to as the ``1600~\AA\ Bump'', or simply the Bump hereafter).    This work represents the largest survey of high-resolution UV CTTS spectra obtained to date, and presents new measurements of key UV diagnostics (H$_{2}$, Ly$\alpha$, \ion{C}{4}, FUV continuum) in addition to the new analysis and model-based interpretation of the 1600~\AA\ Bump feature.

The FUV continuum in CTTSs has been shown to be in excess of what is predicted by scaling the Balmer continuum to shorter wavelengths~\citep{herczeg04,france14a} and several authors have noted the presence of the broad, likely molecular, quasi-continuum emission observed between 1450~$\lesssim$~$\lambda$~$\lesssim$~1650~\AA\ in many CTTSs. This emission was first noted by \citet{herczeg04} and \citet{bergin04} in $HST$-STIS observations of bright CTTSs. They argued that it originated from X-ray generated photo-electron impact on H$_{2}$, though it was noted that TW Hya did not show the characteristic discrete emission lines from the electron-impact process that are observed in Herbig-Haro objects~\citep{raymond97} and the aurorae of gas giant planets~\citep{liu96,gustin04,france10a}. Under the assumption that this feature was attributable to collisionally excited H$_{2}$, \citet{ingleby09} analyzed 33 low-resolution CTTS spectra to constrain the molecular surface density of the surrounding circumstellar material.

In previous work~\citep{france11a}, we have shown that while a combination of accretion continuum and electron-impact H$_{2}$ emission can reproduce some features of the spectra, there are significant discrepancies between the predicted and observed spectral features from this process.  Most notably, the dearth of bound-bound transitions (discrete emission lines) from this process and an observed shift in the peak wavelength of the observed H$_{2}$ collisional dissociation peak make it hard to assign the entire Bump spectrum to electron-impact excited H$_{2}$.  Deep, spectrally resolved observations have shown that CO fluorescence (excited by Ly$\alpha$ and \ion{C}{4} photons) can account for some of these discrepancies~\citep{france11b}, and in this work we propose a new mechanism to account for the excess emission in the quasi-continuous 1600~\AA\ Bump feature: emission from highly non-thermal H$_{2}$ molecules that are the fragments of water photodissociation by Ly$\alpha$ photons.

This paper is laid out as follows~--~Section 2 presents a brief overview of the $HST$ data sets used in this work.  In Section 3, we describe the new analysis of the UV spectra, including consideration of updated dust reddening parameters, new measurements of the 1600~\AA\ Bump and fluorescent H$_{2}$ emission, and extension of previous relations to estimate the intrinsic Ly$\alpha$ emission from these sources.   Section 3 concludes with an empirical characterization of the Bump properties and Section 4 places these properties into context with other parameters of the star + disk systems.   In Section 5, we make several arguments against an electron-impact excited H$_{2}$ origin for the Bump and present new calculations of Ly$\alpha$ excitation on the dissociation products of H$_{2}$O, demonstrating that this interpretation provides a consistent fit to both the quasi-continuous and discrete emissions present in UV spectral  observations of protoplanetary disks.   Section 5 also places the Bump in context of mid-IR water emission features observed by $Spitzer$ and Section 6 summarizes this work.


\section{Ultraviolet Spectroscopy} 

We use the medium spectral resolution G160M mode of $HST$-COS~\citep{green12} to measure the FUV continuum and separate broad spectral features from narrow emission lines for a sample of 37 disk-hosting stars.   Most have these observations have been analyzed as part of previous disk~\citep{france12b} or accretion diagnostic~\citep{ardila13} studies.  We refer the reader to these works for an overview of the observations.  We have augmented the COS Guaranteed Time (PIDs 11533 and 12036; PI - J. Green) and DAO samples (PID 11616; PI - G. Herczeg) with observations of DQ Tau and UZ Tau E (PID 12161; PI - D. Ardila), young stars in the Orion star-forming region (PID 13363; PI - N. Calvet), the binary PMS star AK Sco (PID 13372; PI - A. Gomez de Castro; Gomez de Castro et al. 2016), the young brown dwarfs 2M1207 and SCH0439 (PIDs 11531 and 11616; PI - J. Green and G. Herczeg), and follow-up observations of AA Tau, RECX-15, and RW Aur (PID 12876; PI - K. France).  Finally, we have included two observations of TW Hya, the well-studied $HST$-STIS spectrum~\citep{herczeg02,herczeg04} and archival observations from $HST$-COS (PID 12315; PI - H. M. Guenther).   This brings the sample to a total of 41 unique, high-resolution FUV observations capable of isolating and characterizing the 1600~\AA\ spectral feature.~\nocite{france10b,castro16,france14b}

\begin{figure}
\figurenum{1}
\begin{center}
\epsfig{figure=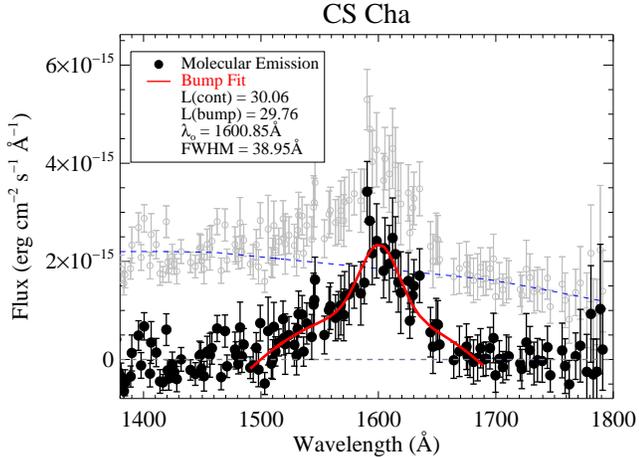,width=2.5in,angle=90}
\vspace{-0.1in}
\caption{A prototypical 1600~\AA\ Bump spectrum in the transition disk CS Cha.  The reddening-corrected extracted continuum spectrum is shown in gray, the second order polynomial fit to the underlying FUV continuum spectrum is shown as the dashed blue line.  The continuum-subtracted Bump spectrum is displayed as the solid black circles and a Gaussian plus second-order polynomial fit to the Bump is shown in red.  The central Gaussian component of the fit is used to characterize the central wavelength and the FWHM of the Bump in the targets.  Clear Bump emission is observed in 19/37 
 ($\sim$~50\%) of protoplanetary disk targets observed by $HST$-COS, and we show the extracted Bump spectra for all targets in Appendix A.        
\label{cosovly}}
\end{center}
\end{figure}

While the majority of the observations used in this study have broader FUV wavelength coverage, we only present new analyses of the G160M data (and STIS E140M for TW Hya). The COS G160M mode spans the wavelength range 1390~--~1780~\AA\ for most stars, and the medium resolution mode provides point-source resolution of $\Delta$$v$~$\approx$~17 km s$^{-1}$ with 7 pixels per resolution element~\citep{osterman11}.   The STIS observations were acquired using the E140M echelle mode ($\Delta$$v$~$\approx$~8 km s$^{-1}$) through the 0.2\arcsec~$\times$~0.2\arcsec\ aperture.  The relatively high spectral resolution and signal-to-noise ratio of these data are critical for being able to separate the forest of H$_{2}$, CO, and atomic emission and absorption lines~\citep{herczeg02,herczeg05,france11b} from the quasi-continuous Bump feature.   Table 1 lists relevant system parameters and assumed distances for the objects studied in this paper.



\section{Analysis}

\subsection{Spectral Measurements from the FUV Disk Sample} 

There are five key spectral features that we extract from this $HST$ FUV spectral disk survey to study the origin of the 1600~\AA\ Bump feature:  The Bump, the underlying FUV continuum spectrum, the fluorescent H$_{2}$ luminosity, the Ly$\alpha$ luminosity, and the \ion{C}{4} luminosity.  
 We use all of these observations to constrain the chemical and spatial origin of the Bump emission.

\subsubsection{FUV Continuum and 1600~\AA\ Bump}

We have created new FUV continuum spectra for the 41 observations (37 targets) analyzed in this work, following the continuum extraction methodology described in~\citep{france14a}.   We create a grid of 131 unique spectral points between 1381 and 1781 \AA, selected by hand to avoid discrete molecular and atomic emission and absorption features, where 0.75~\AA\ (approximately 10 spectral resolution elements) spectral continuum windows can be cleanly measured.  We measure the mean and standard deviation of the observed spectra, and these points define the binned flux spectrum and error array (see Figure 1).  These binned spectra are then corrected for interstellar reddening (see Section 3.2) and the FUV continuum and 1600~\AA\ Bump are separated as described below.     

We measure the FUV continuum emission by fitting the binned spectrum with a second order polynomial at wavelengths away from the Bump, in the regions $\Delta$$\lambda$~=~1395~--~1401~\AA, 1420~--~1465~\AA, 1690~--~1710~\AA, and 1730~--~1760~\AA.  Some spectra did not have this standard COS G160M wavelength coverage (e.g., the Orion stars and the STIS observations of TW Hya), had broad hot gas lines that had to be avoided in the continuum spectra (e.g., RW Aur), or were of early spectral type such that the photospheric contribution contaminated the long-wavelength FUV continuum fit (e.g., HD 134344B).  In these cases, small adjustments were made to the exact wavelength bounds, but visual inspection indicated that this did not have a significant impact on the Bump extraction region.    The FUV continuum flux, $F_{FUV Cont}$, is the integral of the polynomial over the 1490~--~1690~\AA\ wavelength region for direct comparison with the integrated 1600~\AA\ Bump fluxes. The FUV continuum luminosity is defined as $L$(FUV Cont)~=~4$\pi$$d^{2}$$F_{FUV Cont}$.

The most prominent feature in the binned spectra of some CTTSs is the 1600~\AA\ Bump, a spectral feature that spans $\sim$~30~--~150~\AA\ and has a peak flux $\sim$~0~--~4 times the smooth underlying continuum level at 1600~\AA.  We define the 1600~\AA\ Bump as the excess emission above the FUV continuum in the 1490~--~1690~\AA\ spectral region, where the Bump spectrum is the binned data minus the FUV continuum fit.  An example of the well-defined Bump spectrum of CS Cha is shown in Figure 1.  We found that a Gaussian line profile and a second-order polynomial were a good representation of the Bump spectrum in objects with well-defined Bumps, and we fitted this function to all of the observations in our sample (see the Appendix for plots of all 41 observations).  For broad Bumps however, these two components can be degenerate.   
We define the total Bump flux,  $F_{Bump}$, as the integral of the continuum-subtracted Bump spectrum from 1490 to 1690~\AA, and 
$L$(Bump)~=~4$\pi$$d^{2}$$F_{Bump}$.  We use the parameters of the Central Gaussian (CG) to quantify the full-width at half-maximum (FWHM) and central wavelength ($\lambda_{o}$(CG)) of the feature.  Given the asymmetric shape of the Bump spectrum, these values do not describe the FWHM and line center of the feature as a whole, but allow us to define the brightest peak of the emission seen in most of the Bump-detected sources and enable us to discriminate between different origins for this emission based on the peak of the continuum emission spectrum.   

This procedure captures the majority of the clear Bump sources, however this automated Bump-fitting routine is not perfect.   There are examples of stars without clear Bump emission that returned good fits (e.g., CSVO109, DK Tau, DR Tau, and TW Hya-STIS), and stars that demonstrate Bumps that are missed by this procedure (e.g., DN Tau, DQ Tau, and LkCa15).  In the Appendix, we display the Bump region spectra for all of the sources studied here, and Gaussian fits are included for stars with FWHM $>$~0 and good visual fits to the data.   For the plots in Figures 4, 5, 7, and 8, we plot  the 24 spectra with visually-defined Bumps  (stars with FWHM $>$~0, except CSVO109, DK Tau, DR Tau, and TW Hya-STIS, plus DN Tau, DQ Tau, and LkCa15) as solid black diamonds and non-detections (stars with FWHM $<$~0, plus CSVO109, DK Tau, DR Tau, and TW Hya-STIS) as blue upper limit symbols.   Stars with visually well-defined Bumps are noted in Table 2.

\subsubsection{H$_{2}$, Ly$\alpha$, and \ion{C}{4}}
In order to explore excitation mechanisms for the Bump emission, in particular photoexcitation/photodissociation versus excitation/dissociation by non-thermal electrons, we need to quantify the \ion{H}{1} Ly$\alpha$ flux as this is the dominant UV emission component in accreting protostars.  Ly$\alpha$ contributes, on average, $\sim$~10 times the intrinsic flux as the second largest contributor to the FUV luminosity, the continuum emission~\citep{france14a}.   However, direct measurements of the intrinsic Ly$\alpha$ luminosity are impossible owing to resonant scattering by circumstellar and interstellar hydrogen atoms~\citep{herczeg04, lamzin06, mcj14}.  The best indirect method for measuring the Ly$\alpha$ emission line profile and the total Ly$\alpha$ power is a fluorescent H$_{2}$-based profile reconstruction.  The flux distribution of the numerous H$_{2}$ progressions pumped by stellar/accretion-generated Ly$\alpha$ photons can provide constraints on the H$_{2}$ rovibrational temperature, the H$_{2}$ column density, and the neutral hydrogen outflow from the protostar.  These parameters can be solved simultaneously with a parameterized Ly$\alpha$ profile to calculate the intrinsic Ly$\alpha$ flux from the accreting protostar~\citep{herczeg04,schindhelm12b}.   

The Ly$\alpha$ reconstruction technique relies on constraints from the observed high-velocity wings of the line, but not all of the stars in our sample have the requisite FUV spectral coverage to carry out the full reconstruction.  Instead, we extrapolate the H$_{2}$-to-Ly$\alpha$ flux relationship developed by~\citet{schindhelm12b} to predict the Ly$\alpha$ flux from stars in our sample without reconstructed Ly$\alpha$ profiles in the literature.   We measure intrinsic Ly$\alpha$ fluxes for all of the targets with reconstructed profiles and create new measurements of the H$_{2}$ fluorescence luminosity\footnote{We estimate that the brightest 12 H$_{2}$ progressions contribute $\geq$~80\% of the total fluorescent H$_{2}$ emission from CTTSs.} following the prescription described in~\citet{france12b}.    We measure 3 to 4 individual emission lines from the 12 brightest H$_{2}$ progressions that are pumped out of transitions spanning the Ly$\alpha$ emission profile (1213.3~--~1219.1~\AA).  

H$_{2}$ thermally excited to several thousand degree K dominate the discrete H$_{2}$ fluorescence spectra of CTTSs~\citep{herczeg04,mcj16} and present significant populations in a range of rovibrational levels, up to roughly $v^{''}$~$\approx$~4, meaning that self-absorption can impact individual H$_{2}$ emission lines at shorter wavelengths~\citep{herczeg04,mcj16}.   The self-absorption concern, in concert with the fact that the strongest H$_{2}$ emission lines from most progressions arise at $\lambda$~$\gtrsim$~1340~\AA, limits us to the measurement of lines ending in vibrational levels $v^{''}$~$\gtrsim$~5.   The total flux in each progression $m$, $F_{m}$(H$_{2}$), is defined as the emission line flux divided by its branching ratio, and the uncertainty on the total progression flux is the taken as the standard deviation of all of the measured progression fluxes.  The total H$_{2}$ flux, $F_{tot}$(H$_{2}$), is the sum over all progressions, and the H$_{2}$ luminosity is then $L$(H$_{2}$)~=~4$\pi$$d^{2}$$F_{tot}$(H$_{2}$).  We refer the reader to Section 3.1 of~\citet{france12b} for a complete description of the H$_{2}$ emission lines studied here.   

Given the wide range of \ion{C}{4} doublet ($\lambda_{o}$~=~1548.20, 1550.77~\AA) line morphologies~\citep{ardila13}, we do not attempt to parameterize these lines and simply integrate the reddening-corrected spectra over a wavelength range 1547.6~--~1553.0~\AA\ and subtract the flux from a nearby continuum region to measure the \ion{C}{4} fluxes presented in Table 2.   Some uncertainty is introduced by contamination from the H$_{2}$ $B$~--~$X$ (1~--~8)$R$(3) $\lambda$1547.34~\AA\ and $B$~--~$X$ (3~--~7)$P$(17) $\lambda$1551.76~\AA\ lines (see, e.g., Figure 9 of France et al. 2014), but these lines typically make up less than 10\% of the total \ion{C}{4} flux, so we ignore this effect.   

\begin{figure}
\figurenum{2}
\begin{center}
\epsfig{figure=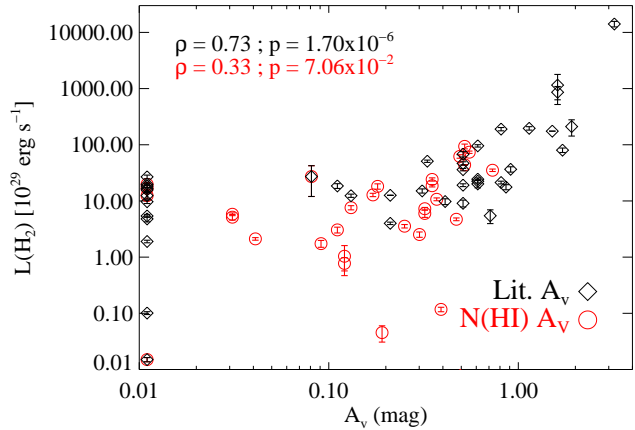,width=2.5in,angle=90}
\vspace{-0.1in}
\caption{ A plot of the total fluorescent H$_{2}$ luminosity (pumped by Ly$\alpha$ photons) versus the visual extinction for each target.  The black diamonds are the array of literature values presented in Table 1 and the red circles are the N(\ion{H}{1})-based reddenings from~\citet{mcj14}.  A strong correlation is observed between A$_{V}$ and luminosity (flux) for the literature reddenings, and this will imprint a significant correlation bias into the sample~--~``A$_{V}$ correlates with A$_{V}$''.  In this work, we adopt the ISM-based reddening curves from~\citet{mcj14} which do not introduce this bias.    Here and in subsequent plots, the Spearman rho ($\rho$) quantifies the degree of correlation and the p-value quantifies the likelihood that the two quantities are uncorrelated.   
\label{cosovly}}
\end{center}
\end{figure}

\subsection{Optical Reddening Values and the Extinction Correction}

The reddening correction is an important component to the luminosity calculations presented here.  This is particularly true because we are working in the FUV bandpass, where dust attenuation curves have significantly larger impact than at optical and infrared wavelengths.  For instance, at optical extinction values approaching A$_{V}$~=~2, the FUV reddening correction can be as high as 20~--~30, depending on the assumed dust grain distribution of the intervening interstellar and circumstellar media.   Traditionally, extinction estimates for PMS stars are derived by measuring optical or infrared color excesses, in photometric or spectrophotometric data, relative to stellar photospheric templates or stellar photosphere templates coupled with additional accretion continuum (``veiling''; e.g., Hartigan \& Kenyon 2003; Herczeg \& Hillenbrand 2014).\nocite{herczeg14,hartigan03}  Temporal variability in optical and near-IR observations of CTTSs~\citep{herbst94, cody13, venuti15} can happen on timescales of hours and can be accompanied by changes in the spectral slope, complicating simultaneous stellar model and extinction fitting for non-simultaneous observations.   

An alternative method for calculating the color excess in CTTSs has been proposed by~\citet{mcj14}~--~combining measurements of the atomic hydrogen along the sightline to the central star with column density-to-color excess relationships derived for the diffuse and translucent ISM~\citep{bohlin78,diplas94}.   This technique uses the central star as the background source and measures the N(\ion{H}{1}) from the heavily damped Ly$\alpha$ absorption line profile observed against the red side of the broad Ly$\alpha$ emission line (see also Lamzin 2006).  Since the absorbing material is fixed in velocity\footnote{Variable outflows from the star may enhance the \ion{H}{1} absorption on the blue side of the Ly$\alpha$ profile, but the outflows are likely not associated with significant dust columns and are 1~--~2 orders of magnitude smaller column densities than the stationary ISM component}, this method is insensitive to changes in the accretion rate (i.e., the background Ly$\alpha$ spectrum) and veiling does not contribute at these wavelengths.   Comparing the N(\ion{H}{1})-based A$_{V}$ with optical/IR-derived values from the literature,~\citet{mcj14} found A$_{V}$ values  on average 0.6 magnitudes lower, corresponding to factors of 1.5-to-10 lower extinction corrections, again depending on the adopted interstellar grain distribution. Two significant uncertainties with this approach are that 1) the grain distribution (and thus the UV opacity), approximated by the ratio of total to selective extinction ($R_{V}$), may be significantly different in star-forming regions than in the average diffuse ISM~\citep{calvet04}, and 2) the gas-to-dust ratio may be different than the interstellar value.   

Because we are investigating potential causal correlations for the Bump flux, we compared these two reddening estimation methods to evaluate potential biases.  A comparison of the correlation between fluorescent H$_{2}$ luminosity ($L$(H$_{2}$)) and A$_{V}$ is shown in Figure 2.   The two primary drivers for the H$_{2}$ luminosity are the amount of hot H$_{2}$ in the circumstellar disk surface and/or the Ly$\alpha$ flux incident on the disk, and neither of these should be related to the amount of dust along the line-of-sight.  Therefore, the reddening and the H$_{2}$ luminosity should be uncorrelated.  Figure 2 shows however that there is a strong correlation between $L$(H$_{2}$) and A$_{V}$ for the literature reddening values (open black diamonds), indicating that the extinction correction and not the intrinsic H$_{2}$ flux level will likely drive luminosity-luminosity relationships.  The Spearman correlation coefficient is $\rho$~=+0.73 with a very low probability of these variables being uncorrelated (p = 1.70~$\times$~10$^{-6}$).  A similar comparison using N(\ion{H}{1})-based A$_{V}$ shows a very weak correlation [$\rho$~=+0.33, p = 7.06~$\times$~10$^{-2}$].   In order to obtain unbiased flux measurements for the star and disk emission components studied here, we adopt the~\citet{mcj14} N(\ion{H}{1})-based A$_{V}$ values, assuming the standard ISM $R_{V}$ value of 3.1, for all stars with N(\ion{H}{1}) measurements.  The \ion{C}{4}, 1600~\AA\ Bump, and FUV continuum measurements were made with the N(\ion{H}{1})-based A$_{V}$ values.  The Ly$\alpha$ reconstructions were derived using literature reddening values~\citep{schindhelm12b,france12b}, and these literature H$_{2}$ and Ly$\alpha$ fluxes were scaled by the ratio of the literature and N(\ion{H}{1})-based extinction curves, evaluated at 1500~\AA\ and 1216~\AA, respectively.  
We note that the two accreting brown dwarfs do not show significant Ly$\alpha$ emission, and that DQ Tau, UZ Tau, and the Orion stars do not have Ly$\alpha$ observations available; we adopt the literature value for those sources.    

\begin{figure}
\figurenum{3}
\begin{center}
\epsfig{figure=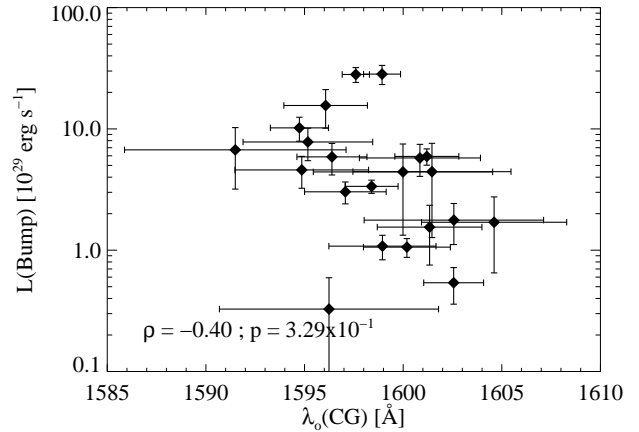,width=2.5in,angle=90}
\vspace{-0.1in}
\caption{ The Bump luminosity shows no significant correlation with the wavelength of Central Gaussian emission component for the 21 observations with well-defined Bumps.   This suggests that the population distribution of the emitting gas does not change significantly with total emitting power, consistent with an external photo-illumination origin for the Bump carrier.   The average Bump central wavelengths are $\lambda_{o}$(CG)~=~1598.6~$\pm$~3.3~\AA, with an average FWHM~=~35.8~$\pm$~19.1~\AA.
\label{cosovly}}
\end{center}
\end{figure}

\begin{figure}
\figurenum{4}
\begin{center}
\epsfig{figure=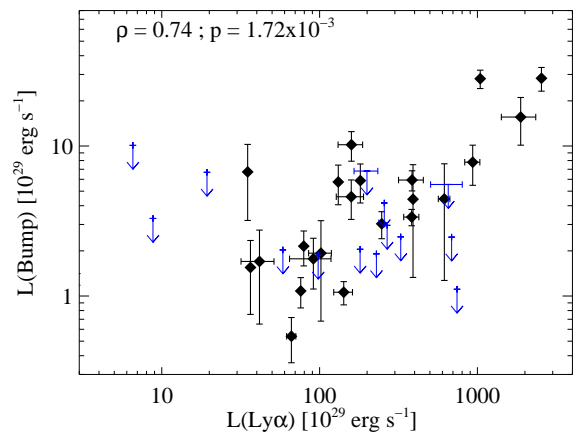,width=2.5in,angle=90}
\vspace{-0.1in}
\caption{ The 1600~\AA\ Bump luminosity compared to the reconstructed Ly$\alpha$ emission line luminosity for the stars in our survey.  The Ly$\alpha$ fluxes are taken from~\citet{schindhelm12b} or extrapolated from those fluxes as described in Section 3.1.2.   This correlation suggests that Ly$\alpha$ may be directly responsible for producing/exciting the Bump emission.
 For this plots as well as Figures 5, 7, and 8, sources with clearly defined Bumps (stars with FWHM $>$~0, except CSVO109, DK Tau, DR Tau, and TW Hya-STIS, plus DN Tau, DQ Tau, and LkCa15) are plotted as solid black diamonds and non-detections (stars with FWHM $<$~0, plus CSVO109, DK Tau, DR Tau, and TW Hya-STIS) are plotted as blue upper limit symbols.     
\label{cosovly}}
\end{center}
\end{figure}

\subsection{The 1600~\AA\ Bump Properties}

We find that 19/37 sources in this study display an unambiguously detected 1600~\AA\ Bump in their FUV spectra.  The criteria for an unambiguous detection include fits to the two-component Bump fit:  $F$(Bump)~$>$~0, FWHM~$>$~8~\AA, and 1580~$\leq$~$\lambda_{o}$(CG)~$\leq$~1620~\AA, and clear Bump upon visual inspection.  Given the size of the sample, we propose that this trend would hold for larger samples so that we can conclude that $\gtrsim$~50\% of all CTTSs show Bump emission.  The systems that display clear bumps span a range of stellar masses, from M2 (RECX-15) to G1 (SU Aur), a range of inner disk dust content from primordial disks (BP Tau, HN Tau) to prototypical transition disks (DM Tau, GM Aur), and a range of mass accretion rates ($>$~10$^{-8}$ M$_{\odot}$ yr$^{-1}$, BP Tau to $<$~10$^{-9}$ M$_{\odot}$ yr$^{-1}$, RECX-11).  The list of clear Bump sources is AA Tau, BP Tau, CS Cha, CSVO090, DE Tau, DF Tau, DM Tau,  GM Aur, HN Tau, RECX-11, RECX-15, RY Lup, SU Aur, SZ 102, TW Hya (COS), UX Tau, UZ Tau, V4046 Sgr, and V836 Tau.   In addition to the list of confident Bump detections, three sources appear to display Bumps, but did not meet the parameterized requirement for a well-defined Bump~--~DN Tau, DQ Tau, and LkCa15 (see Section 3.1.2).   

The average parameterized Bump properties are $\lambda_{o}$(CG)~= 1598.6 $\pm$~3.3~\AA, with FWHM~= 35.8$\pm$~19.1~\AA.  The average Bump Luminosity in the sample is $\langle$log$_{10}$$L$(Bump)$\rangle$~=~29.83, or 1.8~$\times$~10$^{-4}$~$L_{\odot}$.   For comparison with a strong accretion tracer, the 1600~\AA\ Bump has a similar emitted power to the \ion{C}{4} doublet, $\langle$$L$(Bump)/$L$(\ion{C}{4})$\rangle$~=~1.25 for the 24 observations with visually well-defined Bumps.   Figure 3 shows that the central wavelength of the Bump is relatively invariant to the total Bump flux, suggesting that the composition and/or rovibrational population of the Bump carrier does not change dramatically in different star + disk environments.  It also shows that the Bump is never centered near the 1575~\AA\ dissociation peak associated with electron-impact H$_{2}$, as will be discussed in the context of H$_{2}$ models below.   
We present the integrated 1600~\AA\ Bump luminosities and Gaussian parameters in Table 2, and present a discussion about the spectral characteristics of the Bump in Sections 4 and 5. 


\section{Discussion:  The 1600~\AA\ Bump in the Star + Disk System}

In these subsections, we look at the relationship between the measured Bump parameters with the properties of the star and disk systems.   In Section 4.1, we compare the Bump luminosity with the mass of the central star, the X-ray luminosity, and the mass-accretion rate.  Section 4.2 presents a comparison of the 1600~\AA\ Bump with tracers of the gas and dust distributions in the inner disks of CTTSs.

\subsection{Correlations with Stellar Parameters and Radiation Field}

Figure 4 shows the correlation between the Bump luminosity and the accretion-dominated Ly$\alpha$ luminosity generated near the stellar surface.   There is a strong correlation [$\rho$~=+0.74, p = 1.72~$\times$~10$^{-3}$], suggesting that Ly$\alpha$ photoexcitation may play a role in the production of the Bump emission and/or the carrier of the Bump.   The average fractional Bump luminosity is $\langle$$L$(Bump)/$L$(Ly$\alpha$)$\rangle$~=~0.013, however this Ly$\alpha$ luminosity is the intrinsic Ly$\alpha$ that does not account for absorption by neutral outflows.  When considering that outflows block $\sim$~30~--~60\% of the Ly$\alpha$ from reaching the disk surface, the Bump likely emits $\sim$~2~--~4\% of the Ly$\alpha$ luminosity.  The magnitude of the Bump/Ly$\alpha$ ratio, combined with the correlation with the Ly$\alpha$ luminosity, is a strong clue to the origin of the Bump emission.      If photoexcitation is responsible for producing the Bump, there are very few other emission sources in the system that can produce a few percent of the total FUV output\footnote{Ly$\alpha$ emission makes up, on average, $\approx$~88\% of the total stellar FUV output from CTTSs~\citep{france14a}.}.  The next brightest UV photon source, the FUV continuum, makes up about 8\% of the total stellar FUV output, however, at the observed flux levels, the FUV continuum-to-Bump would have to have an energy conversion efficiency of 10~--~20\% of the total panchromatic energy into the Bump, which seems unrealistic.  This large FUV continuum-to-Bump conversion is particularly unlikely because most of that energy would have to come from the shorter-wavelength continuum ($\lambda$~$<$~1500~\AA; the Bump has a specific intensity higher than the FUV continuum in many sources), and the FUV continuum decreases to the blue in most CTTS sources.   In short, it appears that the underlying FUV continuum has neither the spectral distribution nor the total power available to explain the 1600~\AA\ Bump.   

Figure 5 compares the Bump luminosity with both the stellar mass ($left$) and the X-ray luminosity ($middle$) of the star.   There is no clear correlation with either of these stellar quantities.  The Spearman correlation coefficient and null-correlation probability with the stellar mass are [$\rho$~=-0.04, p = ~6.93$\times$~10$^{-1}$] for the 24 spectra with visually-defined Bumps (Section 3.1.1).    Separate Bump populations in higher- and lower-mass CTTSs may be expected if the carrier of the Bump is destroyed by the elevated NUV and red-FUV ($\lambda$~$>$~1700~\AA) flux from the hotter stars.  However, no statistically significant difference is seen in our sample.  Figure 5 shows an arbitrarily defined division at 1.1 M$_{\odot}$, with a 52\% (15/29 sources) detection rate for M$_{*}$~$<$~1.1 M$_{\odot}$ and a 50\% (4/8 sources) detection rate with M$_{*}$~$>$~1.1 M$_{\odot}$.    A complementary sample of Herbig Ae stars would be useful to fill in the higher-mass portion of this diagram, but in practice the detectability of the Bump will likely be decreased due to contrast with the brighter photospheres of those hotter stars.    Similarly, there is no correlation between the X-ray luminosity and the 1600~\AA\ Bump luminosity [$\rho$~=-0.18, p = ~4.34$\times$~10$^{-1}$ for the 24 spectra with visually-defined Bumps (Section 3.1.1)] as shown in Figure 5, $right$.   Most Bump detections display $L$(Bump)/$L_{X}$ of order unity.  $L$(Bump)/$L_{X}$ for 20/24 detections are in the range 0.2~--~4 (note that 7 spectra show $L$(Bump)~$>$~$L_{X}$).  Of the sample of clearly detected Bump spectra, 4/24 detections (V836 Tau, RECX-11, RY Lup, and SU Aur) have low $L$(Bump)/$L_{X}$, 0.01~--~0.02. 

Figure 5 ($right$) also compares the Bump Lumonsity with mass accretion rate measurments from the literature.  We observe a weak but significant [$\rho$~=+0.44, p = ~5.04$\times$~10$^{-2}$] correlation between these two quantities when considering sources with well-defined Bumps.   We argue that this is an indirect correlation instead of production of the Bump in the immediate vicinity of the accretion flows.  In Section 5 we describe a scenario where the Bump is powered by stellar+accretion Ly$\alpha$ photons; the Ly$\alpha$ flux level is largely driven by the mass accretion rate onto the central star.   

The stellar properties analysis also supports the argument that the underlying FUV continuum emission is generated by mass-accretion luminosity (either directly or is powered by illumination from the accreting star)~\citep{france14a}.  Figure 6 shows the strong correlation between the FUV continuum luminosity, $L$(FUV Cont), and the \ion{C}{4} luminosity, $L$(\ion{C}{4}), [$\rho$~=+0.84, p = 3.79~$\times$~10$^{-10}$].  Following the discussion in \S3.2, previous correlations between \ion{C}{4} and the accretion rate may have been driven in part by an overestimated extinction correction, but this plot demonstrates that $L$(FUV Cont) and $L$(\ion{C}{4}) are strongly correlated over more than three decades of \ion{C}{4} luminosity.   The FUV continuum has been shown to not be a direct extension of the NUV Balmer continuum in most CTTS~\citep{herczeg04,france14a}, but instead is likely contributed by specific high-temperature pre-shock regions where accretion columns are impacting the protostellar surface, likely with a small geometric filling faction over the stellar surface~(e.g., Ingleby et al. 2013).  Models of the FUV accretion continuum generation would be valuable inputs to studies of disk photochemistry and the photoevaporative evolution of gas disks.\nocite{ingleby13}

\begin{figure*}
\figurenum{5}
\begin{center}
\epsfig{figure=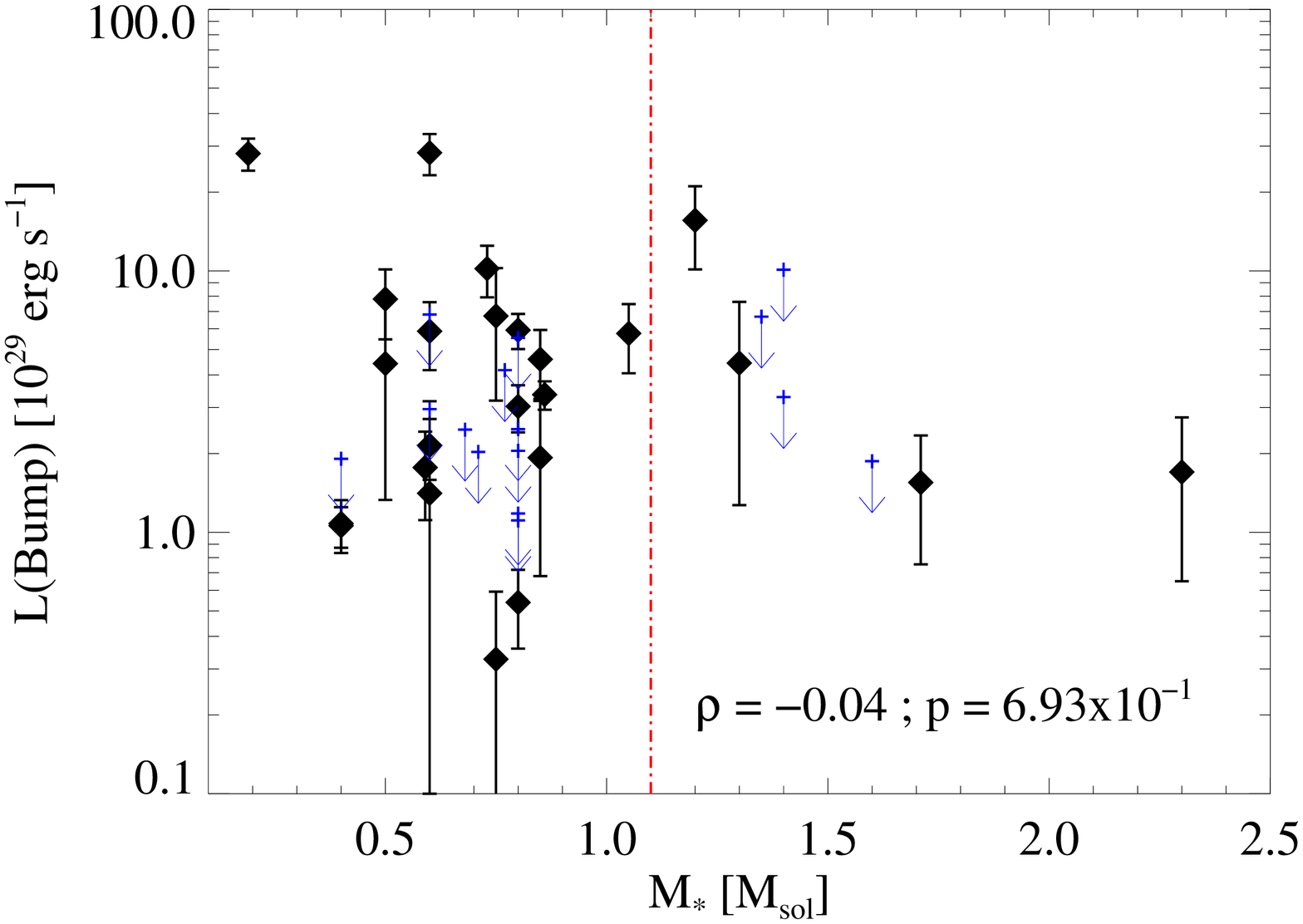,width=2.35in,angle=90}
\epsfig{figure=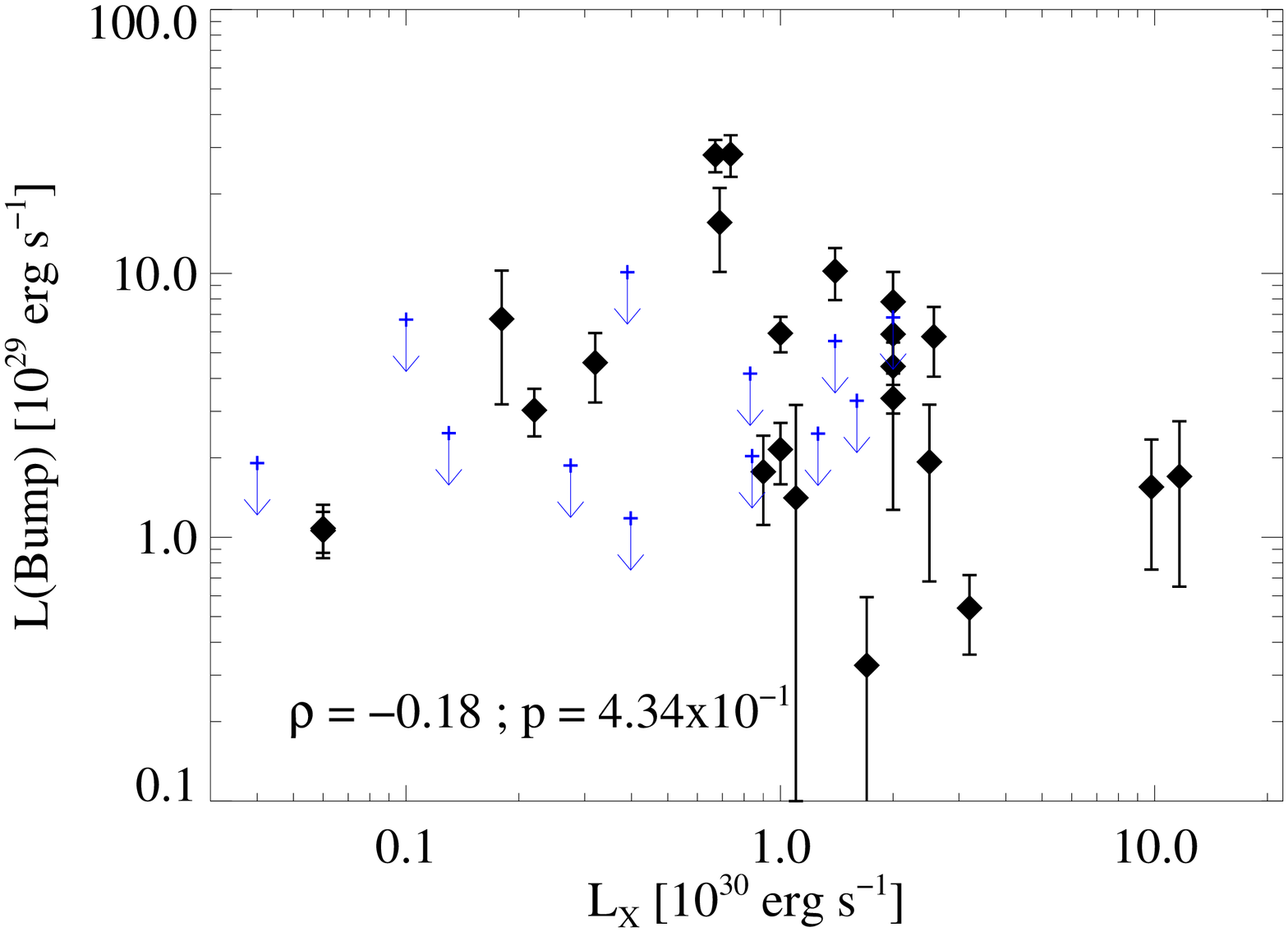,width=2.35in,angle=90}
\epsfig{figure=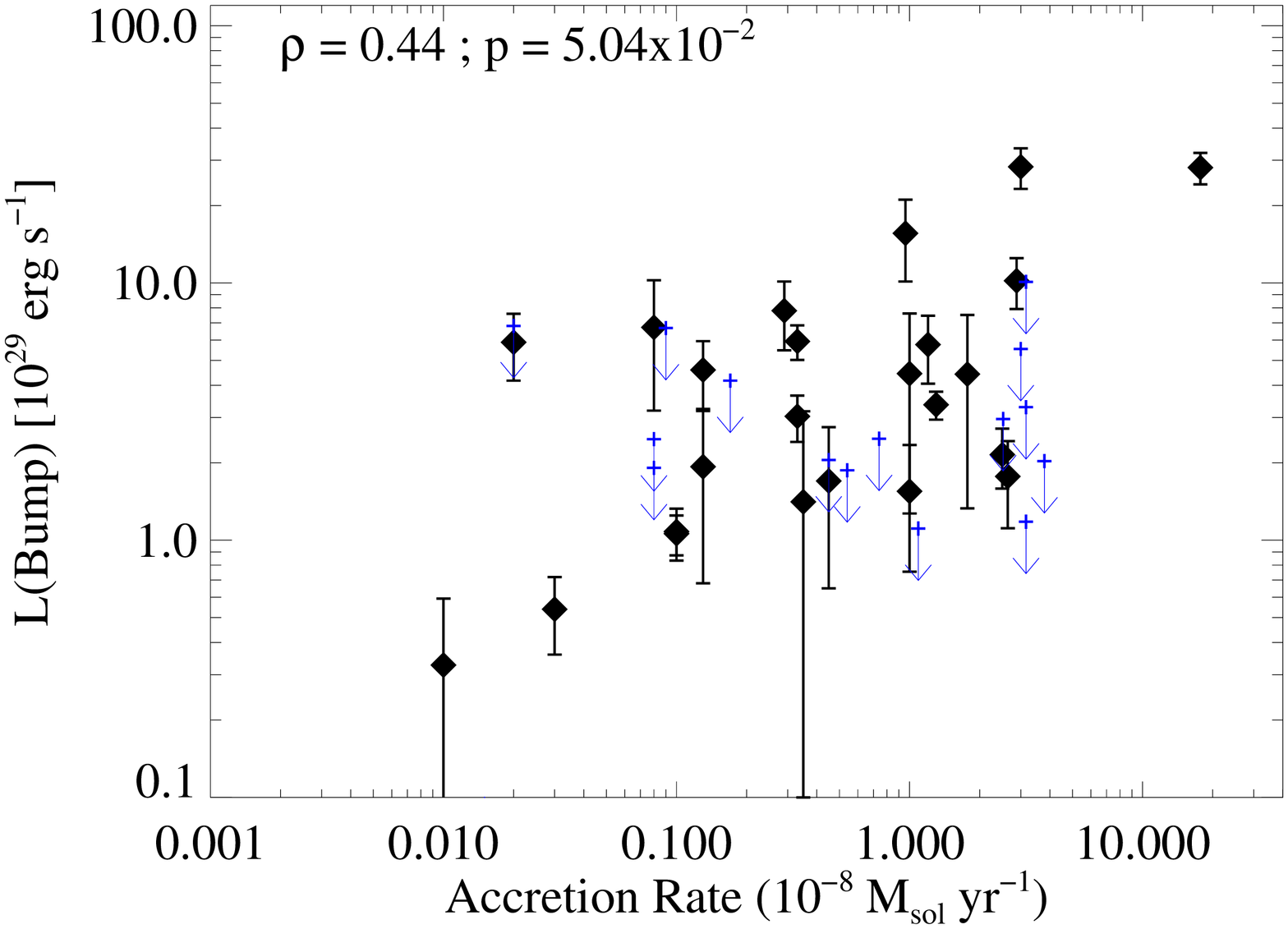,width=2.35in,angle=90}
\vspace{+0.0in}
\caption{ The Bump luminosity is not correlated with the stellar mass or the X-ray luminosity.  {\it At left}, we plot the bump luminosity against the stellar mass.  {\it At middle}, the Bump luminosity is compared with the X-ray luminosity of the source.  {\it At right}, the Bump luminosity is compared with the mass accretion rate of the source.  Well-resolved Bumps are shown as black diamonds and sources where the Gaussian fits to the molecular continuum spectrum found no detection are shown as blue crosses.  The bump luminosity is correlated with accretion rate, but no correlations are observed with mass of the central star or X-ray luminosity.   
\label{cosovly}}
\end{center}
\end{figure*}

\subsection{Correlations with Disk Parameters}

We investigate correlations between the Bump luminosity and tracers of inner disk evolution in Figure 7.  Figure 7 ($left$) displays $L$(Bump) as a function of the H$_{2}$ fluorescence inner and outer radii, R$_{H2}$~\citep{hoadley15}.    The H$_{2}$ radii are derived from radiative transfer modeling of the photo-illuminated disk surface; the radial flux distributions is retrieved by comparing the model output to the spectrally resolved H$_{2}$ emission line profiles.  The inner and outer R$_{H2}$ are defined as the disk radii where 90\% of the emission originates exterior to (inner radius) and 90\% of the emission originates interior to (outer radius), respectively.  By eye, there appears to be a weak anti-correlation between $L$(Bump) and the H$_{2}$ radius, in particular the outer radius, but no statistically strong correlation is observed [$\rho$~=-0.33, p = 4.31~$\times$~10$^{-1}$].   

Figure 7 ($right$) compares the Bump luminosity with the slope of the mid-IR spectral energy distribution, quantified by the difference in the 13~$\mu$m and 31~$\mu$m brightness, $n_{13-31}$~\citep{furlan09}.  Disks with $n_{13-31}$~$<$~0 are generally considered to be primordial, with little dissipation of their optically thick inner dust distributions while disks with  $n_{13-31}$~$>$~0 indicate that the inner disk dust has been (at least partially) cleared, likely by some combination of dynamical interaction with a forming planetary system~\citep{dong16} and photoevaporation by the high-energy irradiance of the accreting central star~\citep{alexander14}.  

There is a clear population bimodality in the detection rate of Bumps between so-called ``primordial'' and ``transition'' disks.   All eight of the transition disks (CS Cha, DM Tau, GM Aur, LkCa15, SU Aur, TW Hya (COS observation only), UX Tau A, and V4046 Sgr) display Bumps.   By contrast, only 11 of the remaining 29, non-transitional sources show a Bump.  This suggests that a degree of inner disk clearing creates a favorable environment for the production of the 1600~\AA\ Bump, possibly by facilitating the propagation of Ly$\alpha$ photons through the inner few astronomical units.   Disks in our sample span a range of primordial to transitional dust distribution ($-$1~$<$~$n_{13-31}$~$<$~3), and despite the obvious population bimodality, no correlation is found between $L$(Bump) and $n_{13-31}$ [$\rho$~=+0.08, p = 7.55~$\times$~10$^{-1}$].  Taken together, these results suggest that if the Bump carrier is a gas phase species, it is not coupled to the optically thick dust and does not significantly diminish as the hot H$_{2}$ disk atmosphere moves outward.  This would favor material in a residual gas disk or disk atmosphere that remains inside the transitional dust gap, as has been seen for fluorescent H$_{2}$~\citep{france12b} and collisional and fluorescent CO emission~\citep{salyk09,brown13,banzatti15,schindhelm12a}.   

\begin{figure}
\figurenum{6}
\begin{center}
\epsfig{figure=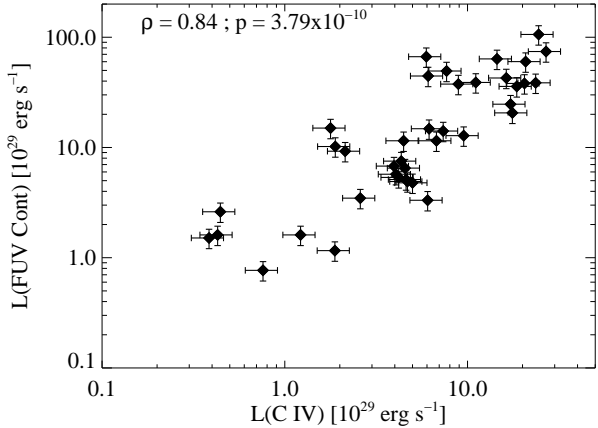,width=2.5in,angle=90}
\vspace{-0.1in}
\caption{ The FUV continuum (measured over 1490~--~1690~\AA) is strongly correlated with the \ion{C}{4} emission.  Building on existing studies that have established a relationship between \ion{C}{4} flux and mass-accretion rate, we conclude that the FUV continuum emission is a tracer of accretion luminosity across $\sim$~3 orders of magnitude in FUV continuum luminosity.    
\label{cosovly}}
\end{center}
\end{figure}

Related to the strong correlation between the Bump and Ly$\alpha$ is the comparison of the total luminosity from the Bump and $L$(H$_{2}$).  Figure 8 shows that targets with well-defined bumps span a range $L$(Bump)/$L$(H$_{2}$)~=~10~--~80\%, a number that is at first surprisingly large:  H$_{2}$ dissociation from a purely thermal population excited by Ly$\alpha$ photons is small but non-zero.  Taking a conservatively large dissociation fraction by assuming a hot H$_{2}$ distribution ($T_{rot}$~=~4000 K), the ratio of continuum (1500~--~1700~\AA) to bound-bound line flux (1200~--~1700~\AA) is approximately 2.4~\%.   Therefore, a thermal population of fluorescent H$_{2}$ falls far short of producing the observed, large $L$(Bump)/$L$(H$_{2}$) ratio.  
The $L$(Bump)/$L$(H$_{2}$) ratio is plotted versus the FUV continuum luminosity
in Figure 8, and no correlation is observed [$\rho$~=-0.14, p = 6.42~$\times$~10$^{-1}$].  

We pause here to remind the reader that it is only the large specific flux of Ly$\alpha$ that makes the hot H$_{2}$ in these systems observable.  For example, the brightest H$_{2}$ emission cascade in most CTTSs is the [$v^{'}$,$J^{'}$] = [1,4] progression pumped out of the $v$~=~2, $J$~=~5 level through the (1~--~2)P(5) absorption line ($\lambda_{abs}$~=1216.07~\AA).   The [2,5] rovibrational level of the ground electronic state contains $<$~0.2\% of the total H$_{2}$ column at $T(H_{2})$ = 2500 K.  Without the very strong Ly$\alpha$ pumping source~\citep{herczeg04,schindhelm12b}, these H$_{2}$ emission lines would not be observable.  Furthermore,~\citet{adam16} has argued that it is the strong Ly$\alpha$ irradiation itself that heats the H$_{2}$ to the high temperatures required for detectable H$_{2}$ fluorescence.  
Similarly, CO has a gas-phase abundance $\approx$~10$^{-4}$ that of H$_{2}$~\citep{france14b} and very weak absorbing overlap transitions (through the $A$~--~$X$ (14~--~0) band, where $A_{lu}$~$\approx$~10$^{4}$~s$^{-1}$) with Ly$\alpha$~\citep{france12b}.   However, the Ly$\alpha$ luminosity is large enough to enable the detection of fluorescent CO emission lines from this process for ~$\sim$~50\% of CTTSs~\citep{schindhelm12a}.  
The observation that the 1600~\AA\ Bump has a luminosity $\sim$~10~--~80~\% of the total H$_{2}$ fluorescent output in half of CTTSs permits a scenario where the carrier of the Bump may also be a relatively low-abundance species, provided that it has absorption transitions that overlap with Ly$\alpha$.        
In the following section, we compare the standard explanation for the Bump (collisional excitation by photoelectrons) with a possible Ly$\alpha$ photo-excitation scenario.

\begin{figure}
\figurenum{7}
\begin{center}
\epsfig{figure=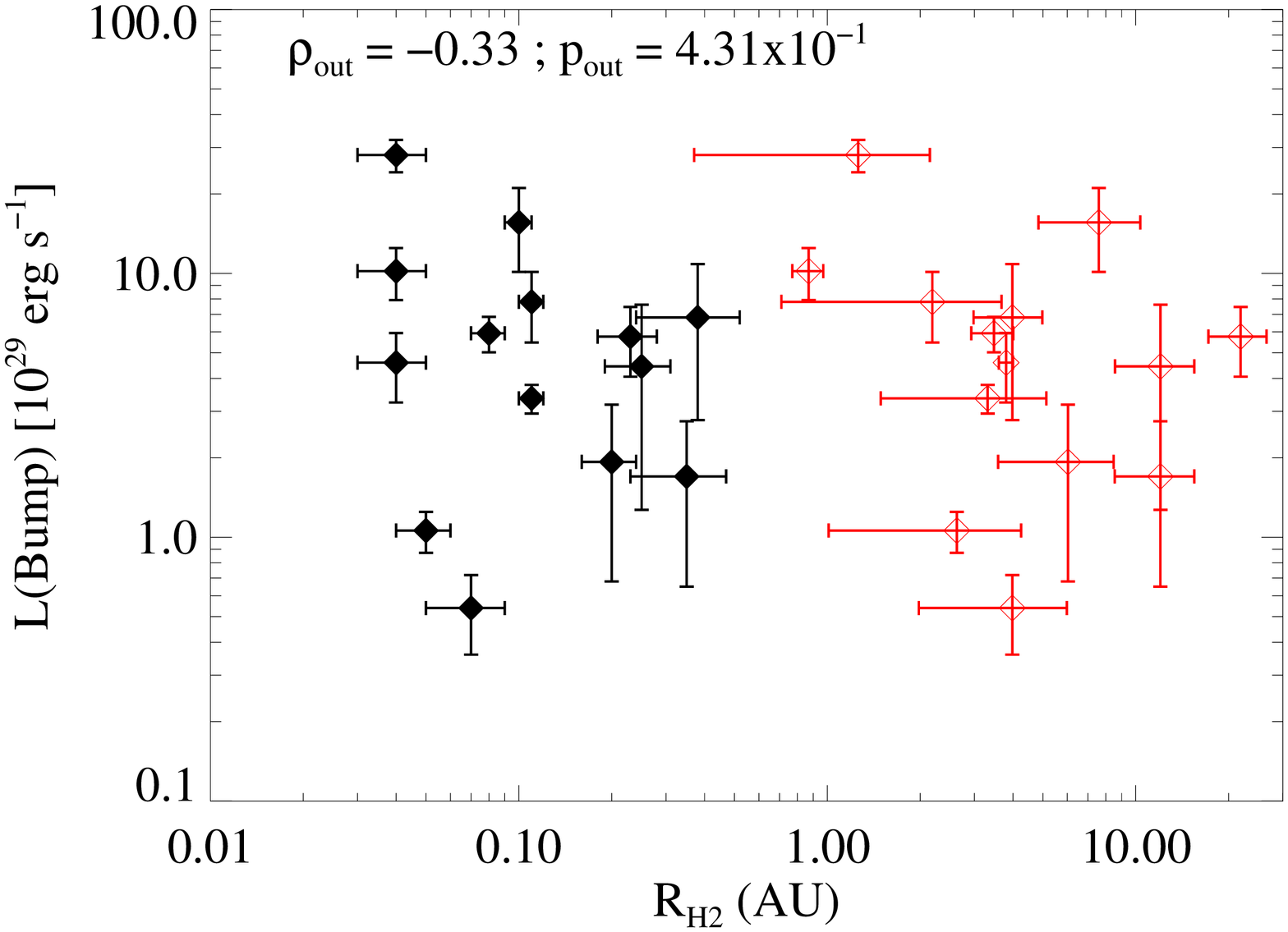,width=2.35in,angle=90}
\epsfig{figure=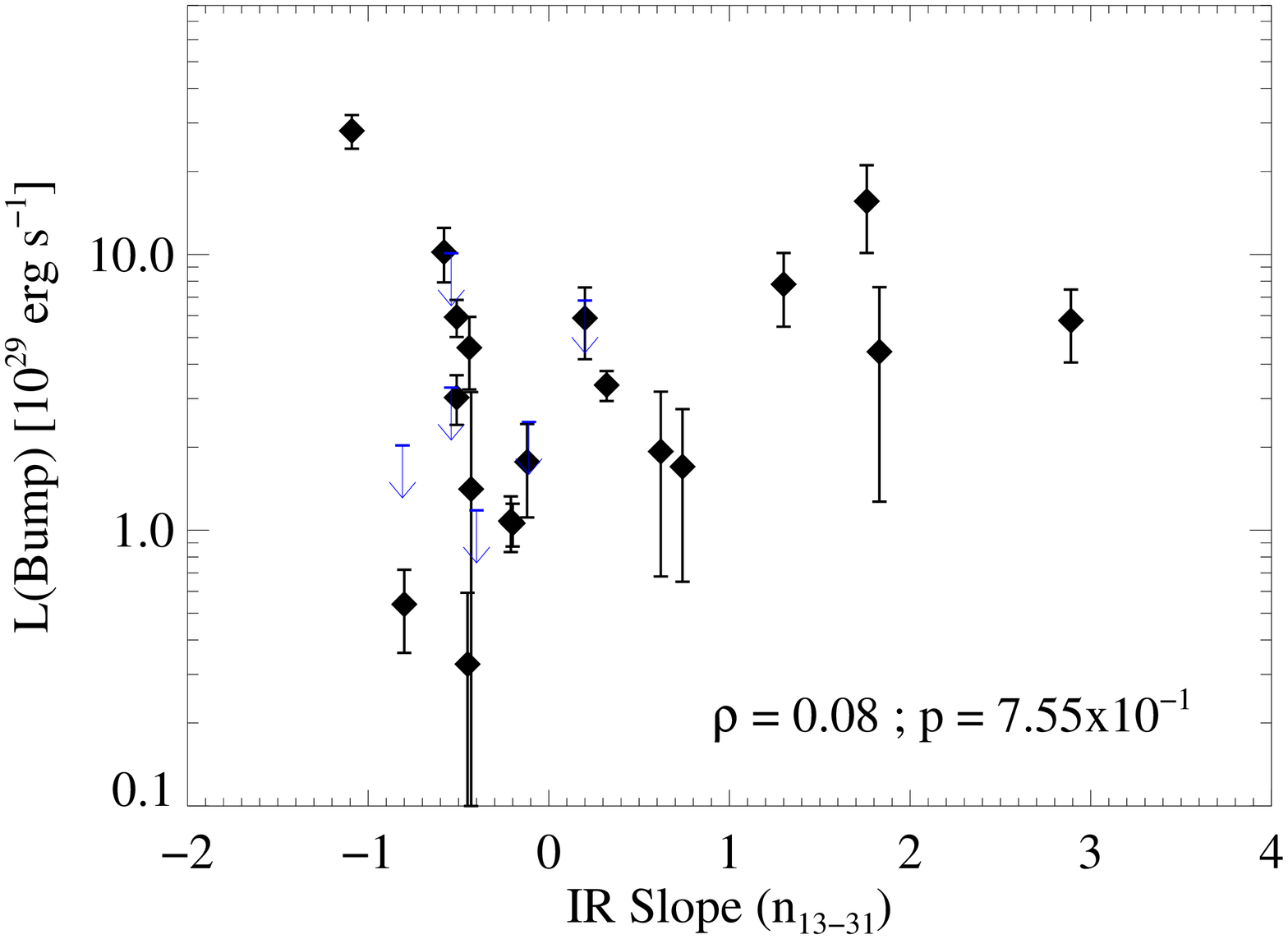,width=2.35in,angle=90}
\vspace{+0.0in}
\caption{ The Bump luminosity is not correlated with measures of gas or dust evolution within the disk.  {\it At top}, we plot the bump luminosity against the inner (filled black diamonds) and outer (open red diamonds) H$_{2}$ radius derived from observations of Ly$\alpha$-pumped fluorescent H$_{2}$~\citep{hoadley15}.  {\it At bottom}, the Bump luminosity is compared with $n_{13-31}$, the mid-IR SED slope that indicates the presence or absence of optically thick dust in the inner disk.  While there is no clear trend between Bump luminosity and the degree of dust clearing, there is a clear difference in the Bump detection rate between transition and primordial sources:  8/8 transition disks display strong Bumps whereas only 11/29 of non-transition sources display clearly detectable Bumps.    
\label{cosovly}}
\end{center}
\end{figure}

\begin{figure}
\figurenum{8}
\begin{center}
\epsfig{figure=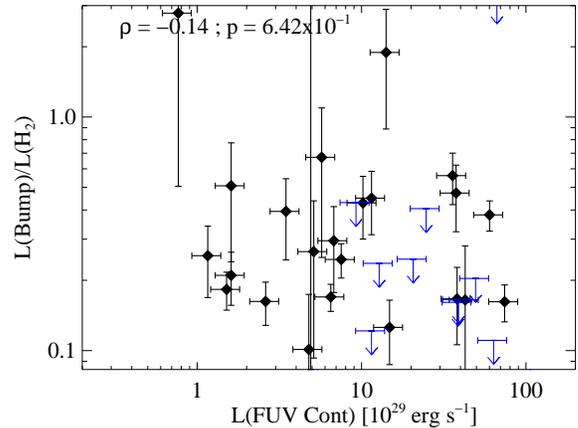,width=2.5in,angle=90}
\vspace{-0.1in}
\caption{A comparison of the ratio of the Bump flux to the fluorescent H$_{2}$ flux and the underlying FUV continuum.  We do not see any dependence with the Bump/H$_{2}$ ratio with continuum luminosity, and find that the total power in the Bump is between roughly 10~--~80\% of the total luminosity of the fluorescent H$_{2}$ in disks.   The strength of this emission also points to Ly$\alpha$-pumping origin for the Bump, but rules out direct dissociation from Ly$\alpha$ on a thermal ($T$(H$_{2}$)~$\sim$~2500 K) H$_{2}$ population because dissociation fractions are $<$~5\% for transitions with ground state energies $<$~2 eV.     Well-resolved Bumps are shown as black diamonds and sources where the Gaussian fits to the molecular continuum spectrum found no detection are shown as blue crosses.
\label{cosovly}}
\end{center}
\end{figure} 

\section{Physical Mechanism for Producing the 1600~\AA\ Bump}

\subsection{Comparison with Electron-Impact Models}

The assumed origin of the 1600~\AA\ Bump before COS was electron-impact excitation of H$_{2}$~\citep{bergin04,ingleby09}. \citet{bergin04} suggested that X-ray irradiation from the central source could produce hot photoelectrons capable of collisionally exciting H$_{2}$ near the region of planet formation. This assumption has been used to constrain the mass surface density of CTTS disks and draw inferences about the planetary architectures found therein \citep{ingleby09}. Conclusions about the electron-impact excited H$_{2}$ population have mainly relied on FUV spectra obtained with the low spectral resolution modes on $HST$ (STIS G140L and ACS PR130L).  CTTS observations with $HST$-COS have raised questions about this interpretation.  Using the deepest UV spectral observations of a CTTS obtained to date, \citet{france11a} analyzed the molecular continuum from the V4046 Sgr disk, finding that the peak in the dissociation continuum was shifted by~$\sim$~25 \AA\ to the red from the nominal 1575 \AA\ electron-impact dissociation peak observed in the well-studied Jovian aurorae~\citep{yung82,ajello84,abgrall97,gustin06}.  

The Bumps from two representative CTTSs (V4046 Sgr and AA Tau) are shown in Figure 9, compared with model electron-impact excited H$_{2}$ dissociation continua at 1000 K and 2500 K gas temperature (assuming a 100 eV secondary electron energy distribution; France et al. 2011b); one immediately notes the offset between the observed continuum peak and the H$_{2}$ dissociation continuum from electron-impact models (magenta and gray solid lines in Fig. 9).  Deep COS observations have indicated that fluorescent emission of warm CO (mainly the higher lying rotational levels of the $A$~--~$X$ (0~--~1)~$\lambda$1597~\AA\ band pumped by the \ion{C}{4} $\lambda$1548~\AA\ line; France et al. 2011b, Schindhelm et al. 2012a) likely contributes to this feature, although CO alone is insufficient to account for the observed shape and flux of the Bump.

\begin{figure}
\figurenum{9}
\begin{center}
\epsfig{figure=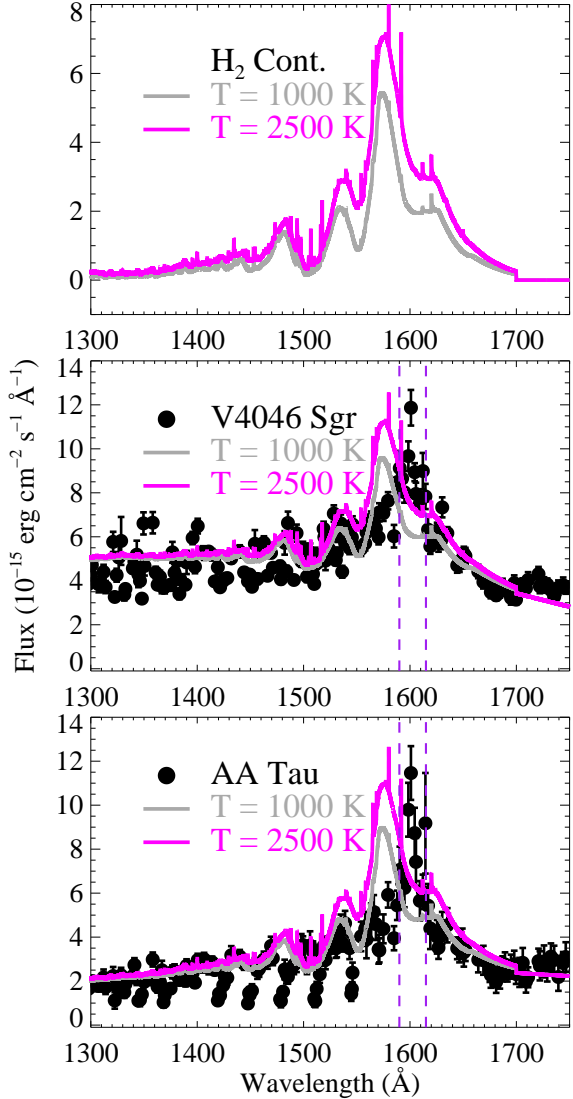,width=3.5in,angle=00}
\vspace{+0.12in}
\caption{A detailed comparison of the theoretical H$_{2}$ dissociation continuum and the high S/N binned continuum spectra of V4046 Sgr and AA Tau.  $(top)$ The theoretical dissociation continuum due to electron collision for two representative H$_{2}$ temperatures.  
{\it (middle) and (bottom)} We observe the average peak wavelength of the 1600~\AA\ Bump at 1598.6~\AA\ with a FWHM of the main peak of $\sim$~32.8~\AA.  This observed peak wavelength is offset from the peak of the theoretical electron-impact dissociation peak by 20~--~25~\AA\ for H$_{2}$ temperatures in the range $T$(H$_{2}$)~$\approx$~1000~--~4000 K.  This demonstrates that electron-impact H$_{2}$ emission alone is an insufficient mechanism to produce the excess 1600~\AA\ Bump emission in CTTSs.  
\label{cosovly}}
\end{center}
\end{figure}

\begin{figure*}
\figurenum{10}
\begin{center}
\epsfig{figure=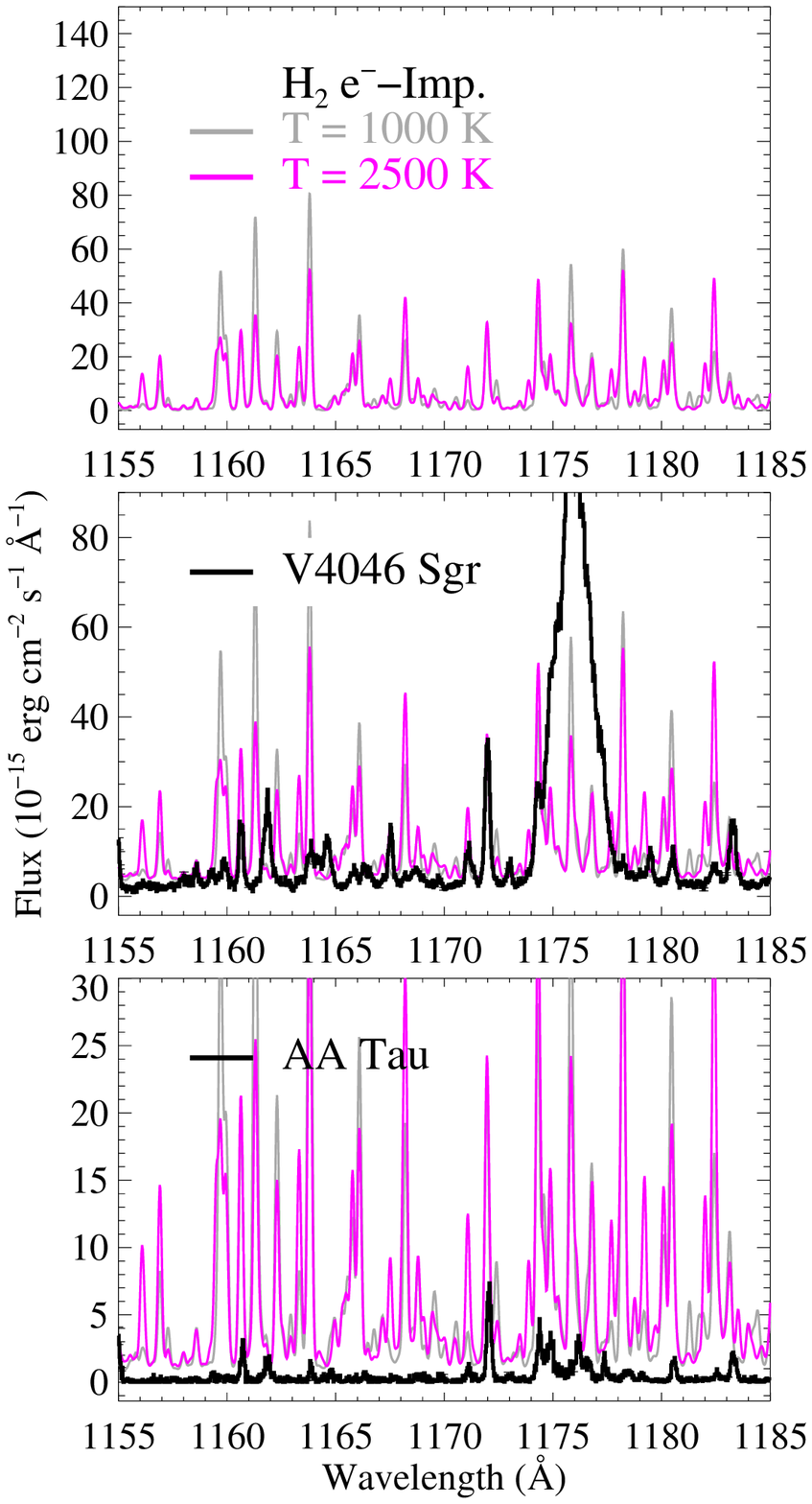,width=3.2in,angle=00}
\epsfig{figure=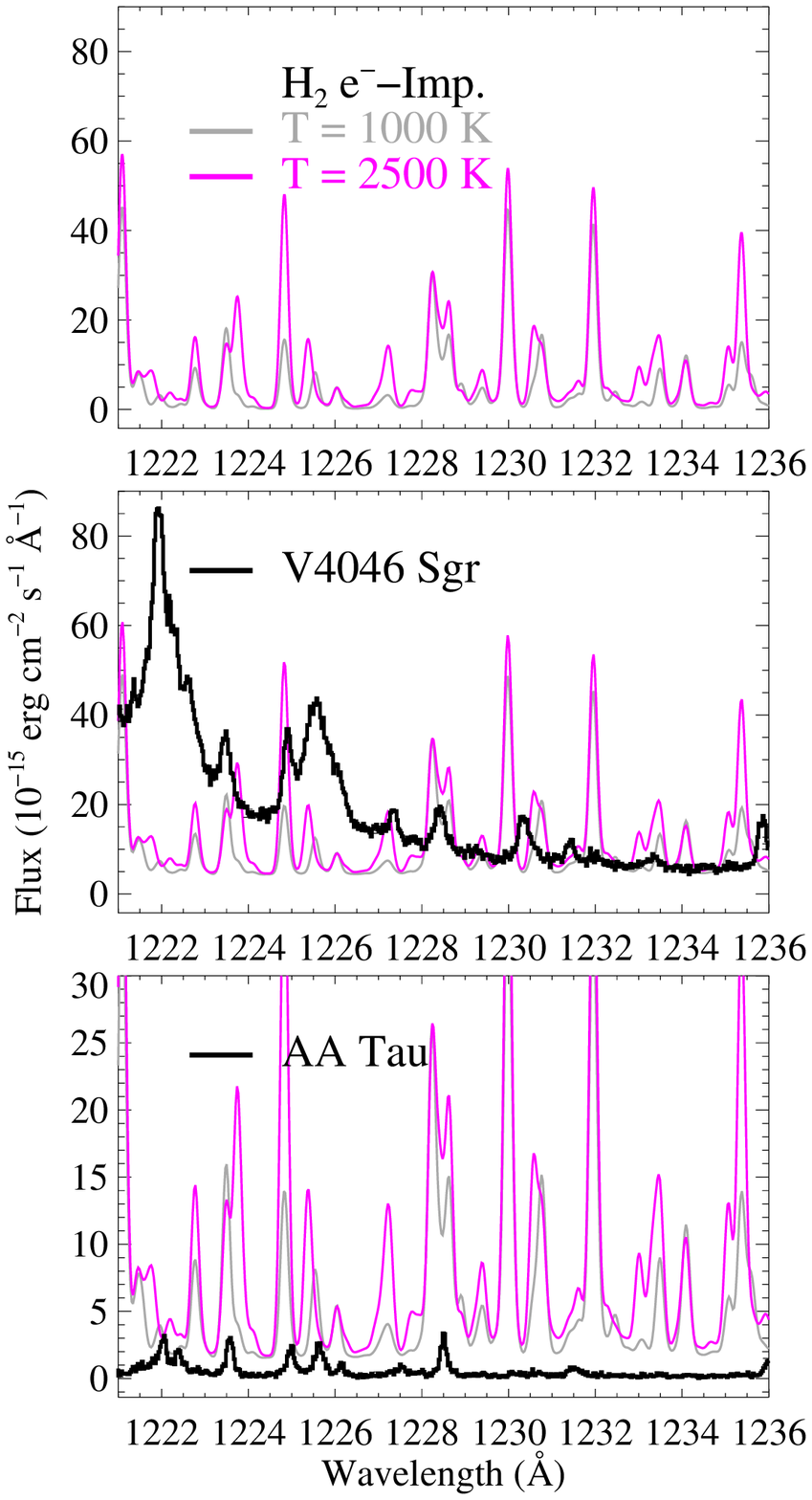,width=3.2in,angle=00}
\vspace{+0.2in}
\caption{Assuming the peak dissociation flux levels shown in Figure 9, one can calculate the intensity of the discrete line spectrum from collisionally excited H$_{2}$.  The bound-bound $M$~--~$X$ (where $M$ is an excited electronic state) transitions contain 75~--~90~\% of the total energy radiated during the H$_{2}$ emission cascade.  $(left)$ The peak of the discrete electron-impact H$_{2}$ line emission at $T$(H$_{2}$) = 1000 K is in the 1158 - 1185~\AA\ region. The predicted flux is not observed.   $(right)$ The peak of the discrete electron-impact H$_{2}$ line emission at $T$(H$_{2}$) = 2500 K is in the 
1221 - 1242~\AA\ spectral regions. Again, the predicted flux is not observed.  Combining this non-detection with the significant peak wavelength discrepancy shown in Figure 9 strongly suggests that H$_{2}$, electronically excited by collisions with electrons, does not contribute significantly to the FUV continuum spectrum of CTTSs.  
\label{cosovly}}
\end{center}
\end{figure*}


Our analysis of $\approx$~20 disks with well-defined Bumps has shown that this wavelength offset is a generic property of the continuum spectra (Section 3.3). Perhaps more problematic for the electron-impact interpretation is that none of the bound-bound transitions resulting from electron-impact excitation of H$_{2}$ have been unambiguously identified in the COS sample.  If the 1600~\AA\ Bump was attributable to electron-impact excitation, these discrete features should be easily observable:  more than 80\% of the emitted power from electron-impact excited H$_{2}$ should be contained in the bound-bound lines, giving rise to a rich H$_{2}$ emission spectrum at the resolution of $HST$-COS~\citep{gustin10}.   
Assuming collisional excitation by hot electrons, we can compute the discrete and quasi-continuous FUV H$_{2}$ emission spectrum and compare these synthetic spectra to the data.  Normalizing the model H$_{2}$ spectrum to the peak of the H$_{2}$ dissociation continuum sets the predicted flux level for the discrete transitions that we would expect to observe across the 912~--~1650~\AA\ bandpass.     In Figure 10, we demonstrate that none of these features are observed for our example objects.  In the remainder of this section, we propose a new mechanism that may explain the 1600~\AA\ Bump: Ly$\alpha$-driven 
 photodissociation of H$_{2}$O in protoplanetary disks.

\subsection{A Direct Probe of Water Dissociation in Protoplanetary Environments?}


As described in the introduction, H$_{2}$O has been shown to be abundant in CTTS disks~\citep{carr08,pontoppidan10b}.  As with H$_{2}$ and CO photoexcitation, H$_{2}$O has photoexcitation and photodissociation routes driven by Ly$\alpha$ photons.   The FUV absorption spectrum of H$_{2}$O consists of a broad continuum at long wavelengths (1400~--~1900 \AA) associated with the electronic transition 
$X(^{1}A_{1})$~$\rightarrow$~$A(^{1}B_{1})$.  A second broad absorption is centered around 1280 \AA, and the $X(^{1}A_{1})$~$\rightarrow$~$B(^{1}A_{1})$  transitions below 1240 \AA~\citep{yi07} overlap with Ly$\alpha$.
The possible channels for H$_{2}$O excitation/dissociation by Ly$\alpha$ photons are \\
\begin{equation}
\begin{split}
H_{2}O (X ^{1}\Sigma_{g}^{+}) + h \nu (Ly \alpha) \rightarrow H_{2}O^{*} \rightarrow  \\ H_{2}O (C^{1}B_{1}   \rightarrow X^{1}A_{1})~~~or~~~
B(^{1}A_{1}) \rightarrow X(^{1}A_{1})
\end{split}
\end{equation}
\begin{equation}
\begin{split}
H_{2}O (X ^{1}\Sigma_{g}^{+}) + h \nu (Ly \alpha) \rightarrow H_{2}O^{*} \\ \rightarrow  H + OH (A ^{2}\Sigma^{+}~~~or~~~X ^{2}\Pi )				
\end{split}
\end{equation}
\begin{equation}
\begin{split}
H_{2}O (X ^{1}\Sigma_{g}^{+}) + h \nu (Ly \alpha) \rightarrow H_{2}O^{*} \\ \rightarrow  H_{2}^{*}(X ^{1}\Sigma_{g}^{+}) + O(^{1}D)~~~or~~~2H(^{2}S) + O(^{3}P)	      
\end{split}
\end{equation}

 The available spin-allowed channels for H$_{2}$O photodissociation at 1216 \AA\ are dominated ($\sim$70 \%) by 
H + OH($X$ $^{2}\Pi$) and H + OH($A$ $^{2}\Sigma^{+}$) 
\citep{slanger82}. The experiments of Slanger and Black (1982) showed that branching ratios for the process described in Equation (3)  are 10\%  for ($^{1}D$) and 17\% for ($^{3}P$) channels.  Theoretical calculations\footnote{The fluorescence of OH fragments has been studied both experimentally and theoretically, but fluorescence from H$_{2}$ fragments has never been studied in laboratory experiments to our knowledge.} predict significant non-thermal rotational and vibrational excitation of H$_{2}$ from this process (van Harrevelt \& van Hemert 2008).  Therefore, $\sim$~10 \% of the Ly$\alpha$-irradiated H$_{2}$O population could end up as H$_{2}$ molecules with a highly non-thermal population distribution.  The ($^{1}D$) reaction in Equation 3, resulting in rovibrationally excited H$_{2}$, can be followed by
\begin{equation}
H_{2}^{*}(X ^{1}\Sigma_{g}^{+}) + h\nu(Ly\alpha) \rightarrow 2H(^{2}S) + Bump(1490~-~1690 \textnormal{\AA})	      
\end{equation}
This process will completely change the Lyman and Werner band excitation and emission during the Ly$\alpha$ pumping process Equation 3.  However, this mechanism has never been explored experimentally.  This H$_{2}$O $\rightarrow$ H$_{2}^{*}$ process may explain the observed shift in the molecular dissociation peak in CTTS environments.  Furthermore, this process is a natural production mechanism for the [\ion{O}{1}]~$\lambda$6300~\AA\ emission believed to originate near the molecular disk surface and the surrounding circumstellar environment~(e.g., Simon et al. 2016).\nocite{simon16}

\begin{figure}
\figurenum{11}
\begin{center}
\epsfig{figure=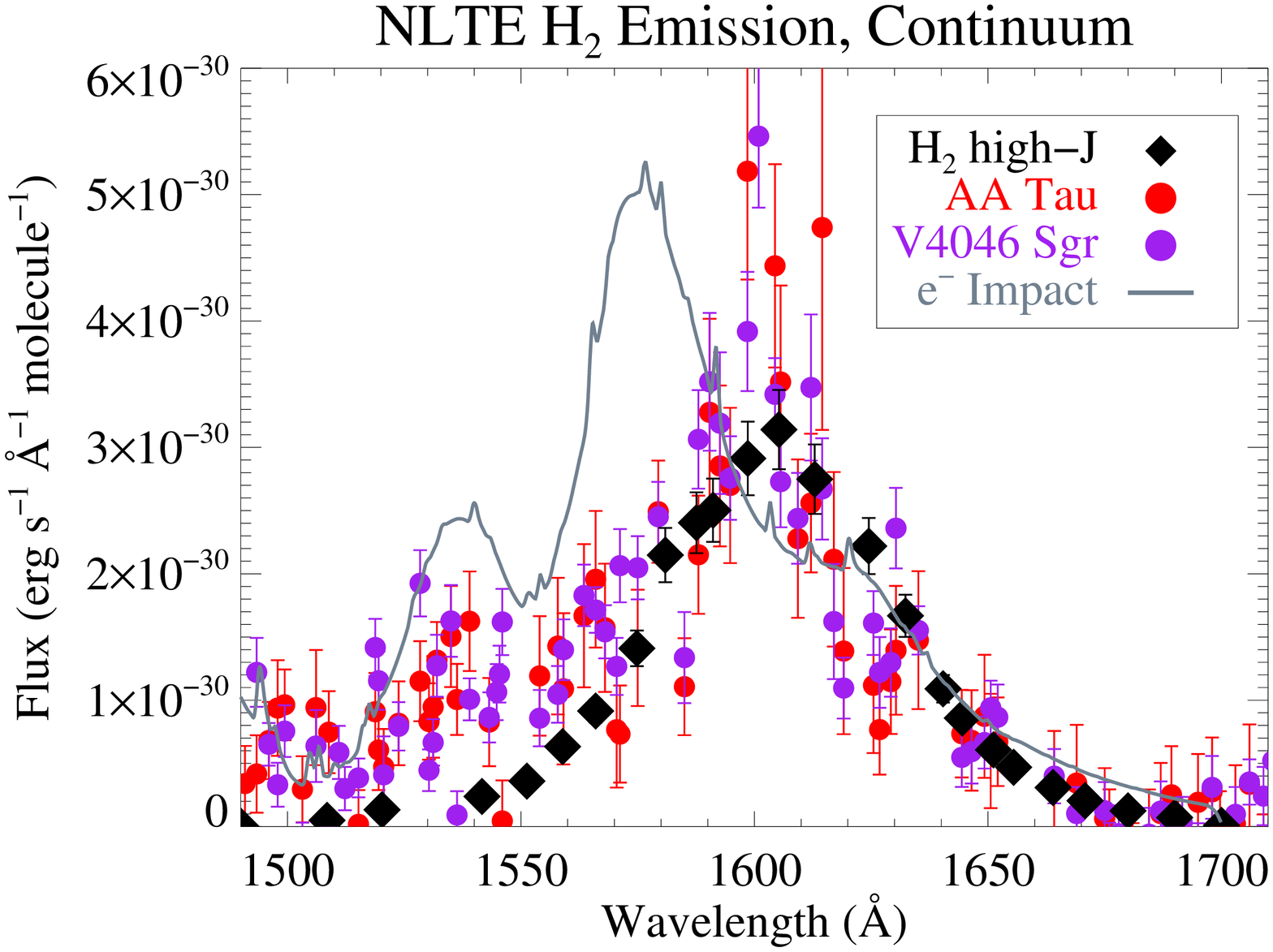,width=2.5in,angle=90}
\vspace{-0.1in}
\caption{ The theoretical H$_{2}$ Lyman band continuum coming from $v^{'}$~=~0~--~3 and $J^{'}$~=~19~--~35 is shown as black diamonds, demonstrating that H$_{2}$ continuum spectra originating from highly excited levels of the upper electronic state can approximately reproduce the observed 1600~\AA\ Bump.  The Bump spectra of AA Tau and V4046 Sgr are shown as red and purple circles, respectively.  For comparison, the electron-impact generated continuum spectrum (at $T$(H$_{2}$)~=~2500 K, binned to 1~\AA\ resolution) has been reproduced in gray.
\label{cosovly}}
\end{center}
\end{figure}

\subsubsection{Calculations of the H$_{2}$ Dissociation Continuum}

 Continuum fluorescence of  H$_{2}$ arises within Lyman and Werner band systems and involves transitions from the $B$ and $C$ electronically excited states to the continuum of the $X$ ground state above the dissociation limit. These transitions were first found in the laboratory  by \cite{stephens:70} and were subsequently interpreted by \cite{stephens:72} in a model where the rotational contribution was treated independently of vibrational and electronic motions. Such a simple approximation has been shown to be incomplete when trying to reproduce experimental spectra obtained from electronic impact in the laboratory~\citep{abgrall97}. \cite{abgrall:93,abgrall:00} have introduced the actual non-adiabatic couplings (radial and rotational) between the excited electronic states $B$ $^1\Sigma_u$, $B^{'}$ $^1\Sigma_u$, $C$ $^1\Pi_u$ and $D$ $^1\Pi_u$ as well as the Coriolis centrifugal term to solve the corresponding nuclear Schrodinger equations.  They 
 predict the theoretical FUV spectrum of H$_{2}$ by using the ground state experimental energy levels of \cite{dabrowski:84}, when available, complemented by ab-initio values for the remaining levels~\citep{wolniewicz95}. The continuum emission probability  
is given by:
\begin{equation}
A(v_u,e_l;J_u,J_l) = \frac{4}{3 \hbar^4 c^3 (2J_1+1)}\times  (E_{v_u,J_u}-E_{e_l,J_l})^3  \times  |{M_{S\alpha}^2}|
\end{equation}
where $E_{vJ}$ is the energy of level ($v$,$J$), $M_{S\alpha}$ is the electric dipole matrix element linking the excited $S$($v_u$,$J_u$) and ground $X$($e_l$,$J_l$) levels and $\alpha$ involves the dipole component responsible for P or R branches 
($\Delta J$ = $J_u$ - $J_l$ = -1, +1)\footnote{The Q branch, $\Delta J$ = 0, dissociation probabilities are very low.}. The continuum emission probability is normalized over energy and is expressed in s$^{-1}$ erg$^{-1}$.   

The wavefunction of the continuum ground state is labeled by the energy $e_l$, measured above the dissociation threshold H(1$s$) + H(1$s$) and rotational quantum number $J_l$, and is an oscillating function tending asymptotically to a Bessel function. Its normalization is performed over energy, which results in units of the emission probability in s$^{-1}$ per energy interval.  It should be noted that some quasi-bound states may occur for sufficiently large values of the rotational quantum number, giving rise to a well in the potential above the dissociation continuum. Such occurrences can produce resonances in the spectrum and theoretical calculations are required to predict the width and relative fluxes of such features.  We refer the reader to previous work (e.g., Abgrall et al. 1993, 2000) to recover the full theoretical treatments. 

We have revisited these calculations by extending the value of the rotational quantum number of the ground state up to its maximum value $J$~=~31, thanks to the recent highly accurate computations of energy level positions by~\citet{komasa11}. The energy transition wavenumbers and wavelengths are computed using these values for the ground state energy terms, and transition probabilities (matrix elements) are obtained with our wave functions computed from the ground state potential energy of~\citet{wolniewicz95}. The maximum rotational quantum number of the upper $B$ excited level is then $J$~=~32.  Comparison with high temperature plasma experiments~\citep{gabriel:09} involving high rotational quantum numbers has shown satisfactory agreement with the theoretical computations of discrete H$_{2}$, HD and D$_{2}$ transitions.  For the purpose of the present study, we have also carefully analyzed all possible quasi-bound levels in the ground state. The positions were previously reported by~\citet{schwenke88} and we have calculated the corresponding FUV emission probabilities from the upper electronic states with energy grid resolution corresponding to the width of the quasi-bound level. Then, we introduced the explicit dependence of the emission probability as a function of the wavelength instead of the energy, which simplifies the comparison with the observations. We consider all possible transitions emitted from a particular excited level and compute  
\begin{equation}
\begin{split}
A(v_u,J_u, \lambda) =  [A(v_u,e_l;J_u,(J_u-1))  + \\ A(v_u,e_l;J_u,J_u+1))] \times \frac{hc}{\lambda^2}, 
\end{split}
\end{equation}
 where 
$\lambda =  hc / (E(H 2s) + E (v_u,J_u)-E(H 1s) + e_l)$.  We note that $E(v_u,J_u)$ is negative  with the origin located at the energy of H(2$s$) level.

The fluorescence spectrum intensity depends on the population of the excited states:
\begin{equation}
I_{H2} = N_{v_u,J_u}A(v_u,J_u, \lambda)  \frac{hc}{\lambda},
\end{equation}
in units of erg s$^{-1}$ \AA$^{-1}$ molecule$^{-1}$.
The excited level population is derived from the excitation function of a given $X_l$  initial level through absorption
of the Ly$\alpha$ flux:
\begin{equation}
 N_{v_u,J_u}= N(H_2) \Sigma_{v_l,J_l}   I(\lambda) \sigma_{lu} (\lambda) \phi(\lambda) x_{v_l,J_l},
\end{equation}
where N(H$_{2}$) is the total column density, $I(\lambda)$ is the Ly$\alpha$ flux, $\phi(\lambda)$ is the absorption profile, and $x_{v_l,J_l}$ is the fractional population in the [$v_l,J_l$] state.   
The absorption cross section is expressed in cm$^2$  \AA\ as 
\begin{equation}
 \sigma  =  \frac{\lambda^4}{8 \pi c}Ê\frac{g_u}{g_l} \times A_{ul}.
 \end{equation}
In the following subsection, we compare the output of these new continuum calculations with the observed 1600~\AA\ Bump spectra of our sources.   

\subsubsection{Comparison of the Synthetic H$_{2}$O Dissociation Spectra with the FUV Data}

The two primary discrepancies in the electron-impact H$_{2}$ interpretation for the 1600~\AA\ Bump are the shifted continuum emission peak and the non-detection of the predicted discrete line emission.  In this subsection, we show rough quantitative agreement between H$_{2}$ emission from highly non-thermal H$_{2}$ ground state populations that result from H$_{2}$O photodissociation by Ly$\alpha$ photons~\citep{harrevelt08}.  The results presented here strongly suggest that Ly$\alpha$ pumping of highly non-thermal H$_{2}$ can approximately reproduce the observed continuum shift, that many of the discrete emission lines predicted by Ly$\alpha$ fluorescence from the H$_{2}$O fragment electronic ground state distribution [$v$,$J$] predicted by~\citet{harrevelt08} are detected in our CTTS spectra, and that the Bump emission fluxes are reasonable given typical abundances of H$_{2}$O in the inner disk.  

\begin{figure}
\figurenum{12}
\begin{center}
\epsfig{figure=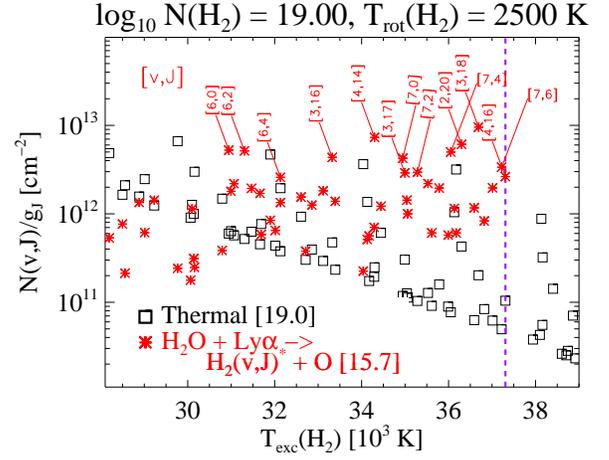,width=2.5in,angle=90}
\vspace{-0.1in}
\caption{ Comparison of a fiducial, thermal disk atmosphere H$_{2}$ population (N(H$_{2}$)~=~10$^{19}$ cm$^{-2}$, $T$(H$_{2}$)~=~2500 K; black squares) and H$_{2}$O fragment populations from~\citet{harrevelt08}, N(H$^{*}_{2}$)~=~10$^{15.7}$ cm$^{-2}$ (plotted as red asterix).  States with the largest non-thermal populations are labeled.   The purple dashed line near 37,300 K represents the highest energy level considered in the calculations of~\citet{harrevelt08}. 
\label{cosovly}}
\end{center}
\end{figure}

{\it Continuum Emission}~--~Figure 11 displays a comparison between our example CTTS Bump spectra (AA Tau and V4046 Sgr) and the H$_{2}$ dissociation continuum from highly excited rovibrational levels of the $B$$^{1}\Sigma^{+}_{u}$ state.  This particular example shows emission from $v^{'}$~=~0~--~3 and $J^{'}$~=~19~--~35, although there are several combinations of rovibrational levels of the Lyman band ($v^{'}$~=~0~--~7 and $J^{'}$~=~11~--~35) that reproduce this approximate shape and line strength.  In order to compare the model continua with the data, we summed all of the continuum profiles from [$v^{'}$,$J^{'}$] levels in the $B$$^{1}\Sigma^{+}_{u}$ state and then analyzed these model spectra in the same manner as the data, evaluating the continuum fluxes between sharp emission features.  In this way, we avoid confusion with both the bound-bound emission lines and resonances at quasi-bound continuum levels stabilized by the H$_{2}$ centrifugal potential barrier~\citep{abgrall97}.  Finally, we analyzed the model continuum spectrum with the same Gaussian profile fitting technique as presented in \S3.1.1, finding $\lambda_{mod}$~=~1605.3~\AA\ and $FWHM_{mod}$~=~53.4~\AA.  

It should be emphasized that while this example does not match the exact wavelength shift observed in the CTTS data (median $\lambda_{bump}$~=~1598.6~\AA), the flat distribution of intermediate $v$ and high $J$ levels is the expected population distribution for non-thermal H$_{2}$ water fragments~\citep{harrevelt08}, and the rotational level population should only increase as we go from cold (ground state, the only available calculations) to warm water more representative of the expected conditions in the warm molecular layer of protoplanetary disks.    This population distribution and resultant dissociation continuum should be testable in laboratory experiments that irradiate a gas of warm water molecules with a broad Ly$\alpha$ or scanning ultraviolet light source.  As an example, we identify three rovibrational levels of the $B$ electronic state that have dissociation continua spectrally coincident with the Bump:  [$v^{'}$, $J^{'}$] = [4,18], [5,10], and [6,9].  These progressions, pumped by (4~--~0)P(19) $\lambda$ 1217.41~\AA, (5~--~3)R(9) $\lambda$ 1219.11~\AA, and (6~--~3)P(10) $\lambda$ 1219.84~\AA, have relatively large dissociation fractions (42\%, 9\%, and 17\%, respectively) and produce continuum emission in a band consistent with the observations.    In reality, there are a wide range of available $v^{'}$~=~0~--~7 and $J^{'}$~$\geq$~10 levels that could be consistent with the observed continuum flux distribution, but identifying the exact subset of states that best reproduces the data will require continued molecular photoprocess modeling and laboratory experiments.   

{\it Discrete Line Emission}~--~Using our reconstructed population distribution above, we can search for the strongest discrete emission lines (i.e., those with the largest branching ratios) of the most promising continuum-producing states listed above.  Starting with a subsample of these levels (the [$v^{'}$, $J^{'}$] = [4,18], [5,10], and [6,9] progressions), we searched through the data to see if these lines are observed in approximately the line-strength ratios predicted by the branching ratios from these upper levels.  Indeed, we detect numerous discrete fluorescence lines from [4,18] in almost all of our CTTS sample with clear Bump emission. Similarly, we detect multiple discrete lines from both [5,10] and [6,9] in several sources, particularly those with broader red-wing Ly$\alpha$ emission as would be expected.

As a second, less biased, approach, we can use the calculated non-thermal H$_{2}$ ground state levels predicted to have the largest fractional population ($P(v,J)$) following H$_{2}$O dissociation by Ly$\alpha$ photons to determine if these levels are directly observable in our CTTS spectra.  
First, we identified the H$_{2}$ [$v$,$J$] levels with the highest populations following H$_{2}$O dissociation (Table 6 of van Harrevelt \& van Hemert 2008).   A comparison of the high excitation temperature ($T_{exc}$~$>$~25 kK) thermal populations at a fiducial H$_{2}$ column density of N(H$_{2}$)~=~10$^{19}$ cm$^{-2}$~\citep{schindhelm12b}, and the non-thermal H$_{2}$ populations for a column of 10$^{15.7}$ cm$^{-2}$ (see Section 5.2.3 below) are displayed in Figure 12.  A number of high-energy states display large abundances, as expected from the non-thermal excitation process.  Analyzing the H$_{2}$ states with significant overabundance relative to the primary thermal distribution, we identified strong absorption line ($A_{JJ'}$~$>$~2.5~$\times$~10$^{7}$ s$^{-1}$) coincidences with Ly$\alpha$~(within $\pm$~4~\AA\ of line-center) and excitation energies $T_{exc}$~$>$~25 kK.   21 candidate absorption lines were identified, and these are listed with the relevant transition information in Table 3.  

We then searched for the brightest potential fluorescent emission lines from these states.  For this calculation, we restricted the absorbing transitions to within $\pm$~3~\AA\ of Ly$\alpha$ line center (1212.67~--~1218.67~\AA;  to consider only the peak of the Ly$\alpha$ pumping flux for most CTTS), transition rates $A_{J'J''}$~$>$~2.5~$\times$~10$^{7}$~s$^{-1}$, and excitation energies $T_{exc}$~$>$~25 kK.  51 emission lines are identified (Table 4). 
Of these 51 predicted discrete emission lines, 12 (24\%) are unambiguously detected in more than one of our CTTS our spectra, and 8 (16\%) are marginally detected in at least one source.  23 (45\%) are ambiguous due to blends with other atomic and molecular emission lines (mainly emission from high-$T_{exc}$ lines on the tail of the thermal distribution at $T$(H$_{2}$)~$\approx$~2500 K), and only 8 are not detected (16\%).  
These lines are shown as a function of their branching ratios, emission wavelengths, and detection status in Figure 13.  Given the number of blends with emission from $\sim$~2500K H$_{2}$ emission lines, we conclude that thermally-populated H$_{2}$ fluorescence and dissocation emission likely contribute at a low level to the Bump and to some of the bound-bound transitions associated with the Bump.  However, neither Ly$\alpha$ photoexcitation nor electron-impact collisional excitation of a thermal population of H$_{2}$ appear to be capable of producing the observed 1600~\AA\ Bump emission and all of the assocated discrete emission features.

We present an example of two emission lines where H$_{2}$O dissociation fragments are likely identifications for the observed emission in Figure 14.   Figure 14, $left$, shows the spectrum of V4046 Sgr in the region around the Werner band 
line H$_{2}$ $C$~--~$X$ (1~--~4)~R(12) $\lambda$1186.23~\AA, and comparison with model fluorescence spectra using three thermal populations and the populations of~\citet{harrevelt08}.    The models were created using the fluorescence code presented by~\citet{mcj16} and the reconstructed Ly$\alpha$ line profiles from~\citet{schindhelm12b}.   The model spectra are scaled to the flux in the strong (1~--~1)~P(8) line.  The thermal models at 1500 and 2500 K are unable to reproduce the flux in the (1~--~4)~R(12) line.  It appears that the 3000 K spectrum can approximately describe the line complex near 1186~\AA, however this spectrum significantly overpredicts the observed data at other wavelengths, which is reflected in the results of previous fits to the H$_{2}$ spectrum, $T$(H$_{2}$)~$\leq$~2000K for V4046 Sgr~\citep{hoadley15, mcj16}.  Emission from non-thermal H$_{2}$ is able to augment the dominant thermal population and present a better overall fit to the observed spectrum.  A complementary example is given in Figure 14, $right$, where no thermal H$_{2}$ emission is predicted at the observed $B$~--~$X$ (9~--~4)~R(24) $\lambda$1373.95~\AA\ line.   The thermal models are scaled to the flux in the (1~--~4) P(11) line, and only the non-thermal H$_{2}$ population model predicts flux at the observed wavelengths.  These criteria were employed for all of the lines tallied in Figure 13.  Figure 15 shows larger swaths of spectra from V4046 Sgr and DF Tau, where non-thermally populated lines clearly or marginally detected are marked with a dashed green line, while non-detections or blends with stronger H$_{2}$, CO, or atomic lines are marked with dashed red lines.

\begin{figure}
\figurenum{13}
\begin{center}
\epsfig{figure=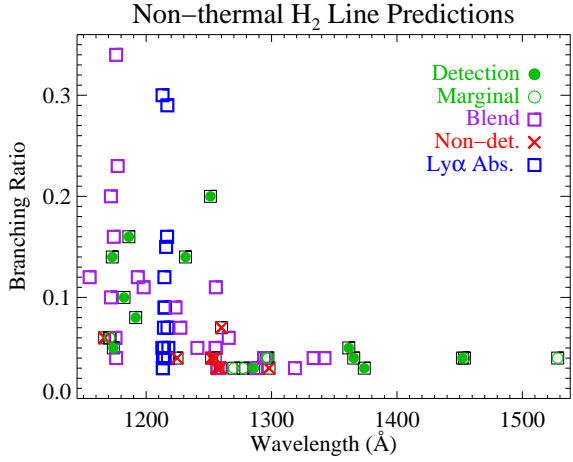,width=2.5in,angle=90}
\vspace{-0.1in}
\caption{ The branching ratios of the brightest predicted features from Ly$\alpha$ excited H$_{2}$O fragments (H$_{2}[v^{*},J^{*}]$) as a function of emission line wavelength.  Filled green circles indicate the line is detected and fluorescence from highly non-thermal molecules is the most likely line identification.  Open green circles indicate that the predicted line is only observed in one CTTS spectrum in our sample.  Purple squares indicate blends with other atomic/molecular features or a line where Ly$\alpha$ fluorescence of thermal H$_{2}$  ($T$(H$_{2}$)~$\sim$~2500 K) is the more likely line identification.  Lines labeled with blue squares are not detected because these are the coincident pumping transitions, and lines labeled with a red `X' are not detected.  We note that most of the high branching ratio discrete lines from the H$_{2}$O fragments are detected or blended with other H$_{2}$ lines, and true non-detections have branching ratios $\leq$~7\%.   
\label{cosovly}}
\end{center}
\end{figure}

\begin{figure}
\figurenum{14}
\begin{center}
\epsfig{figure=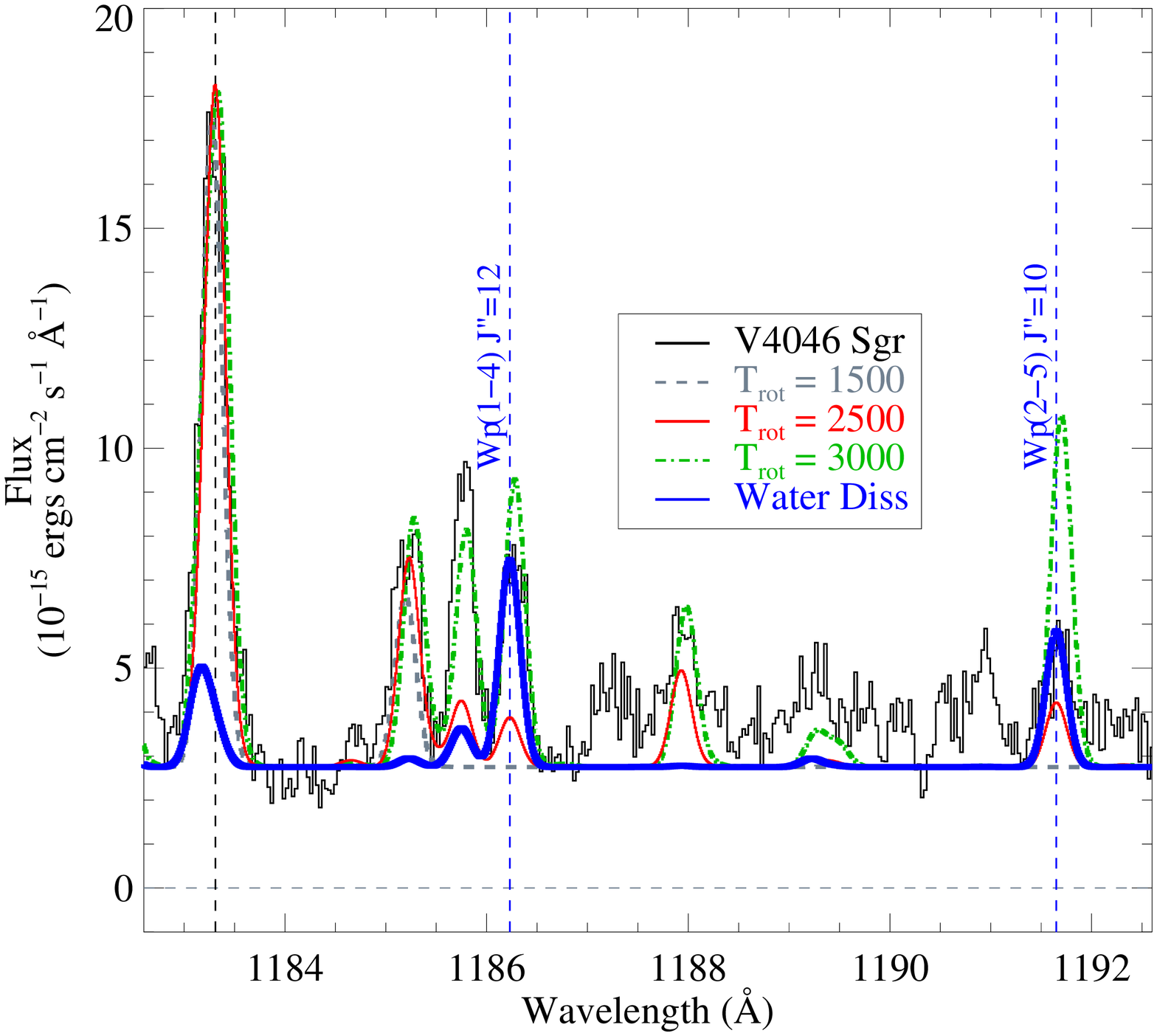,width=2.3in,angle=90}
\epsfig{figure=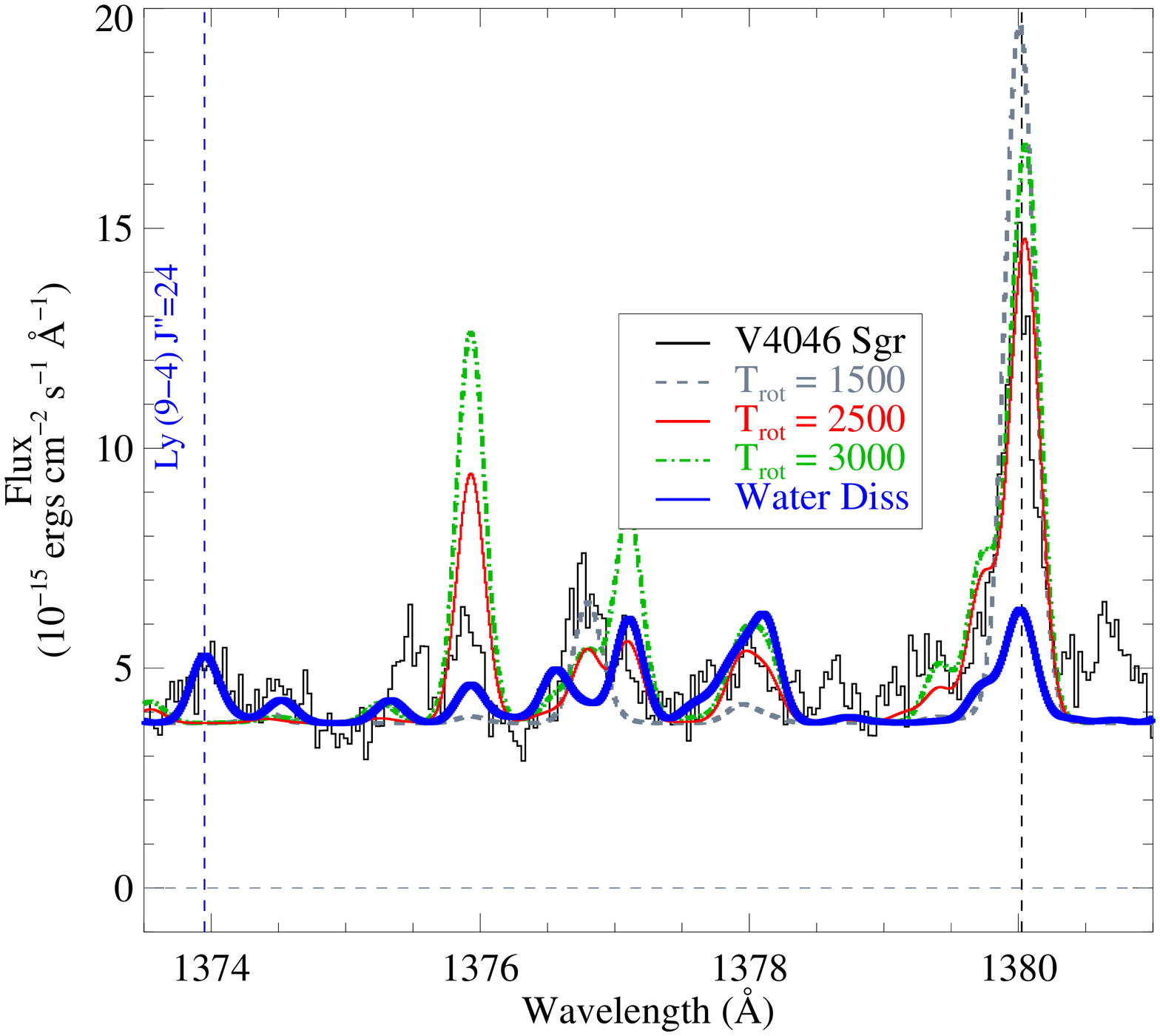,width=2.3in,angle=90}
\vspace{+0.2in}
\caption{ A comparison of $HST$-COS spectra of V4046 Sgr with modeled discrete emission lines coming from Ly$\alpha$-pumping of 1) thermal H$_{2}$ (gray, red, and green curves) and 2) non-thermal H$_{2}$ distributed according the H$_{2}$O dissociation fragment populations of~\citet{harrevelt08}, plotted as the blue curve.  {\it At top}, the thermal spectra are normalized to the $B$~--~$X$ (1~--~1) P(8) line flux ($\lambda$1183.31~\AA, marked by a black dashed vertical line), and the $C$~--~$X$ (1~--~4)~R(12) $\lambda$1186.23~\AA\ and $C$~--~$X$ (2~--~5)~R(10) $\lambda$1191.65~\AA\ lines resulting from a large non-thermal populations and spectral resonances with Ly$\alpha$ are marked with blue dashed vertical lines.  {\it At bottom}, the thermal spectra are normalized to the $B$~--~$X$ (1~--~4) P(11) line  ($\lambda$1380.10~\AA, marked by a black dashed vertical line).   The $B$~--~$X$ (9~--~4)~R(24) $\lambda$1373.95~\AA\ line resulting from a large non-thermal population and spectral resonance with Ly$\alpha$ is marked with blue dashed vertical line.  The predicted non-thermal states are consistent with observed lines that are not well-fit by thermal H$_{2}$ populations.    
}
\end{center}
\end{figure}
\subsubsection{Estimated H$_{2}$O Dissociation Rate and Inner Disk Origin}

In this subsection, we present an estimate of the average Ly$\alpha$-driven H$_{2}$O dissociation rate and attempt to constrain the spatial origin of the 1600~\AA\ Bump.   In Section 3.3, we measured an average Bump luminosity of log$_{10}$$L$(Bump) = 29.8, $L$(Bump)~=~6.7~$\times$~10$^{29}$ erg s$^{-1}$, or 5.4~$\times$~10$^{40}$ photons s$^{-1}$ at a fiducial wavelength of 1600~\AA.  We make an assumption that the Bump emission is optically thin and that the overlying dust does not contribute additional attenuation beyond what has been accounted for in the reddening correction.  We note that the latter assumption may not be correct, and the formation region for the Bump could be closer to the A$_{V}$~=~1 surface in the disk, which would make the above flux estimates and the following column density calculation lower limits to the true conditions in the water-dissociation region.  For example, taking a standard FUV extinction curve with ratio of total to selective reddening ($R_{V}$) of 4.0 to approximate some degree of grain growth, placing the emitting region at the A$_{V}$~=~1 surface would increase he calculated intrinsic photon arrival rate by a factor of 6.  

Assuming an optically thin medium, the number of arriving photons represents the total number of participating molecules and we can define the non-thermal H$_{2}$ dissociation rate as $D_{Bump}$~$\approx$~5~$\times$~10$^{40}$ molecules s$^{-1}$.   The total number of non-thermal H$_{2}$O fragments (H$^{*}_{2}$ molecules) can be defined as $C_{Bump}$, where $C_{Bump}$~=~$D_{Bump}$/$\langle$$P_{diss}$$\rangle$.  $\langle$$P_{diss}$$\rangle$ is the average dissociation fraction that results from transitions out of the non-thermal ground-state population (Table 3).  Calculating $\langle$$P_{diss}$$\rangle$ we see that lower $A_{lu}$ levels have higher dissociation fraction, such that $\langle$$P_{diss}$$\rangle$~$\sim$~40\% for all lines $A_{lu}$~$>$~10$^{7}$~s$^{-1}$ and $\langle$$P_{diss}$$\rangle$~$<$~10\% for all lines $A_{lu}$~$>$~10$^{8}$~s$^{-1}$.   Taking an average of $\langle$$P_{diss}$$\rangle$ weighted by the the relative abundances of their ground state populations and their $A_{lu}$-value, we find that the dissociation is dominated by absorbing transitions in with $A_{lu}$~$\approx$~(3~--~8)~$\times$~10$^{7}$~s$^{-1}$ and $\langle$$P_{diss}$$\rangle$~$\approx$~30\%.   Therefore, we estimate $C_{Bump}$~$\approx$~1.7~$\times$~10$^{41}$ H$_{2}$ molecules s$^{-1}$ produced by Ly$\alpha$ dissociation of water.   The total rate of dissociating water molecules responsible for the observed 1600~\AA\ Bump is then $C$(H$_{2}$O)$_{diss}$~=~$C_{Bump}$ / $\langle$$P_{H2}$$\rangle$, where $\langle$$P_{H2}$$\rangle$ is the daughter-product fraction into H$^{*}_{2}$ from the reaction H$_{2}$O + Ly$\alpha$ presented in Equation 3.   $\langle$$P_{H2}$$\rangle$ has been experimentally measured to be~10\%~\citep{slanger82,yi07}.   The total water dissociation rate by Ly$\alpha$ photons is $C$(H$_{2}$O)$_{diss}$~=~1.7~$\times$~10$^{42}$ molecules s$^{-1}$.  

\begin{figure*}
\figurenum{15a}
\begin{center}
\epsfig{figure=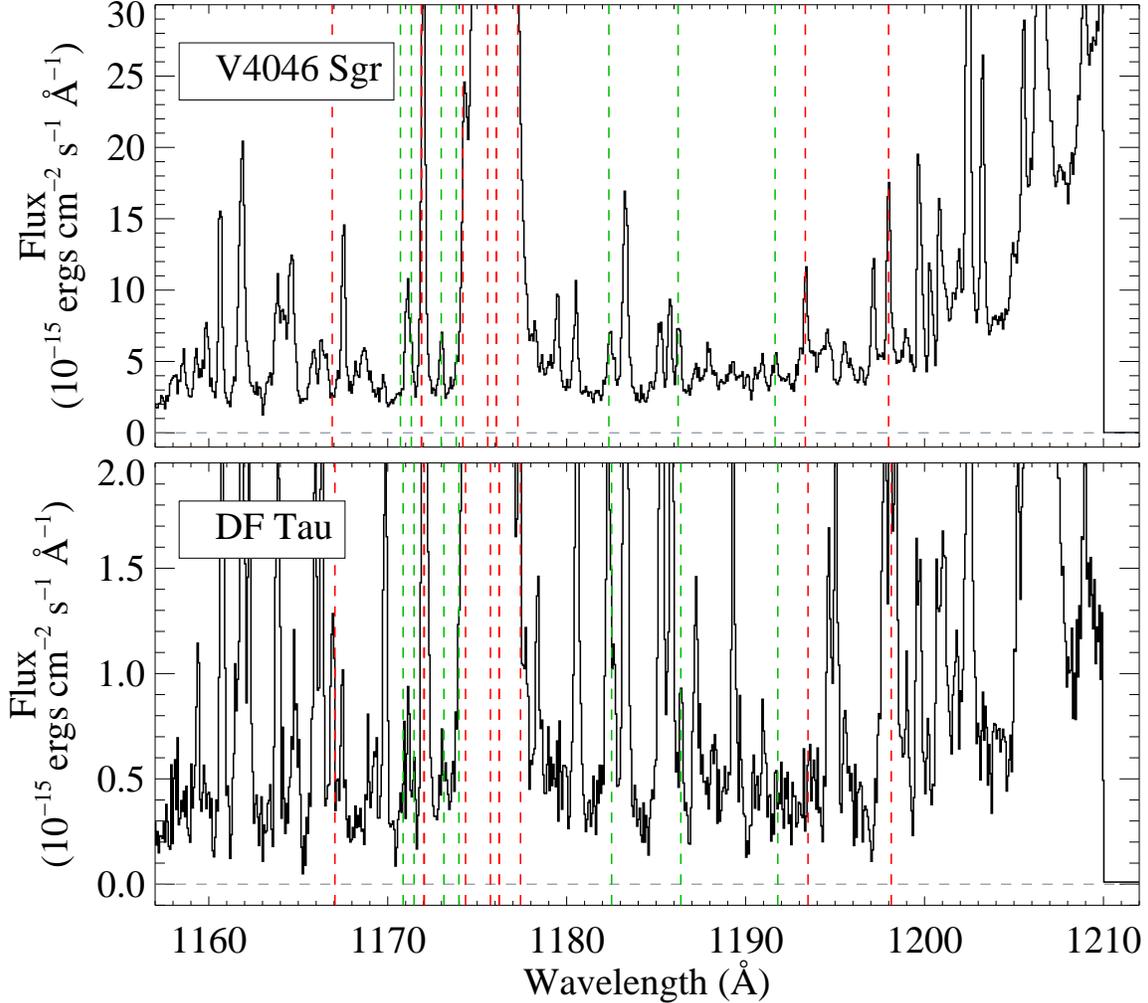,width=4.8in,angle=90}
\vspace{+0.4in}
\caption{ COS spectra of V4046 Sgr ($top$) and DF Tau ($bottom$), showing the wavelengths of the brightest predicted discrete emission lines from Ly$\alpha$-pumping of H$_{2}$O fragments.  The post-H$_{2}$O-dissociation non-thermal H$_{2}$ ground state populations from~\citet{harrevelt08} are assumed.   Dashed lines are overplotted at the predicted emission wavelengths; green lines indicate that Ly$\alpha$~fluorescence of non-thermal H$_{2}$ is the most likely line identification, while red lines indicate non-detections, blends with other atomic/molecular features, or a line where Ly$\alpha$ fluorescence of thermal H$_{2}$  ($T$(H$_{2}$)~$\sim$~2500 K) is the more likely line identification.   Table 4 lists the individual transitions marked here.   
\label{cosovly}}
\end{center}
\end{figure*}

\begin{figure*}
\figurenum{15b}
\begin{center}
\epsfig{figure=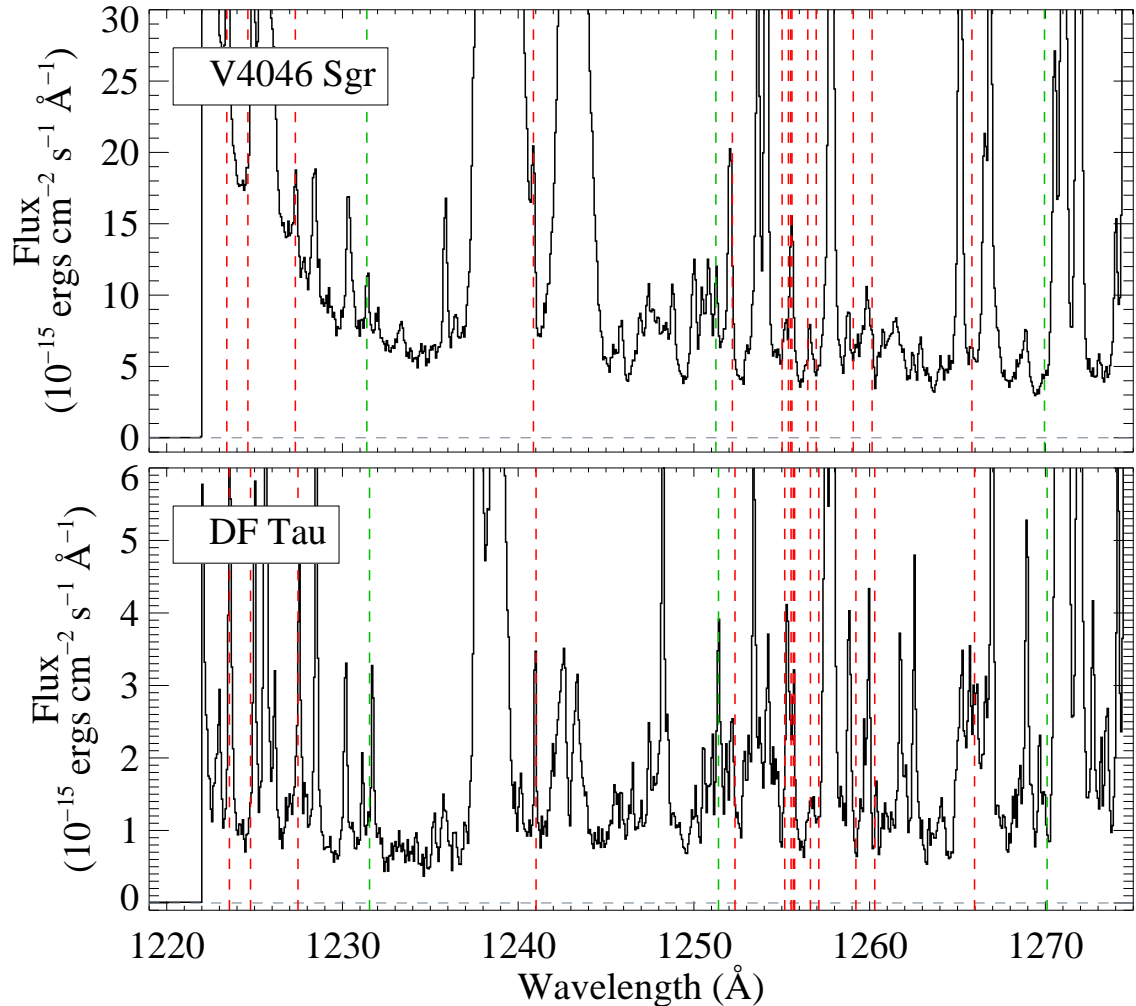,width=4.8in,angle=90}
\vspace{+0.4in}
\caption{ Same as Fig 15a, but for the 1220~--~1275~\AA\ window.    
\label{cosovly}}
\end{center}
\end{figure*}

\begin{figure*}
\figurenum{15c}
\begin{center}
\epsfig{figure=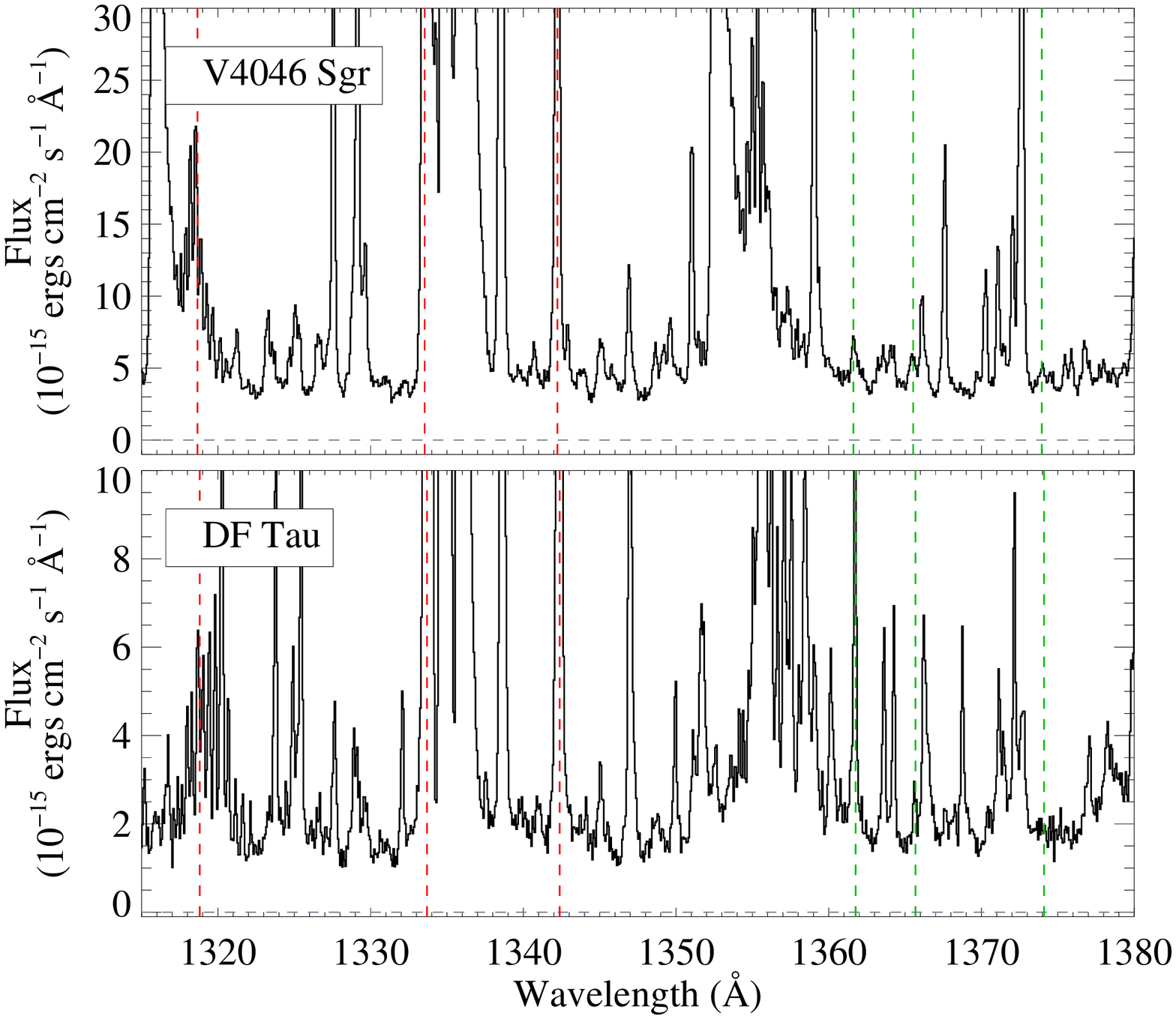,width=4.8in,angle=90}
\vspace{+0.4in}
\caption{ Same as Fig 15a, but for the 1315~--~1380~\AA\ window. 
\label{cosovly}}
\end{center}
\end{figure*}

The total dissociation rate from this process is large, with a rate of 2.8~$\times$~10$^{18}$ g s$^{-1}$ of water destroyed, or 0.045~M$_{\odot}$ Myr$^{-1}$.   This material is not lost to the system as the dissociation products go into heating the disk atmosphere~\citep{adam16}, catalyzing additional molecular formation in the warm molecular zone~\citep{tielens85,woitke16}, and likely supplying a fraction of the thermalized molecular hydrogen found in the upper disk atmospheres of all CTTSs (e.g., France et al. 2012b).   If we assume a steady state process of Ly$\alpha$-driven photodissociation balanced by water formation from neutral-neutral reactions of OH and H$_{2}$, as well as dissociative recombination of H$_{3}$O$^{+}$, we can use the total number of water molecules to estimate the column and volume densities of the emitting region.  \citet{antonellini15} model CTTS disks and find that the IR water emission (discussed in the following next subsection) is concentrated in the inner disk, with inner and outer radii of 0.1~--~0.15 and 0.35~--~0.5~AU, respectively, at a disk height of $z/r$~$\sim$~0.15~--~0.2.   Adopting average dissociation fractions, $\langle$$P_{diss}$$\rangle$, 20~--~30\%, we find that the total vertical column density of water undergoing Ly$\alpha$-driven dissociation is (1~--5)~$\times$~10$^{16}$ cm$^{-2}$, or a few percent of the total water column density inferred from $Spitzer$ observations of CTTS~\citep{carr11}.    Under these assumptions about the spatial origin, the volume density of the emitting gas is $n$(H$_{2}$O)~$\sim$~few~$\times$~10$^{4}$~cm$^{3}$, consistent with the upper vertical regions of high water concertation in the model of~\citet{antonellini15}.   

Is an inner disk origin correct for the 1600~\AA\ Bump?  This is a challenging question to answer as the three primary mechanisms for determining the origin of atomic and molecular emission from the inner regions of protoplanetary disks, spectroastrometry~\citep{pontoppidan11}, spectral line width modeling~\citep{salyk11a,hoadley15}, and interferometry~\citep{eisner09,eisner14} cannot be employed for the spatially and spectrally unresolved Bump emission.  To obtain a rough empirical constraint on the origin of the emission, we look for unique astrophysical environments that can provide information on disk emission as a function of radius.  Happily, the AA Tau system provided one such opportunity during the  long-duration dimming event that began in 2011~\citep{bouvier13}.     \citet{schneider15} present a panchromatic study of the spatial and compositional characteristics of the ``extra absorber'' responsible for the dimming, an azimuthally asymmetric enhancement in the disk height located several AU from the central star that has rotated into view\footnote{The extra absorber is not related to the inner disk warp in the AA Tau inner disk.}.   Comparing $HST$-COS observations of the Bump before the dimming and during the dimming (using the 2011 and 2013 AA Tau data sets analyzed here), \citet{schneider15} found that the Bump dimming was comparable to the dimming of the high-velocity wings of the Ly$\alpha$-pumped H$_{2}$ emission lines.  The H$_{2}$ emission lines are broadened by Keplerian rotation, so the velocity-dependent flux decrement was used to show that only gas at $r$~$\lesssim$~2 AU was being occulted by the extra absorber.  Given the similar levels of dimming, the 1600~\AA\ Bump likely also originates inside this radius.  This picture is supported by the theoretical work of~\citet{adam16}, who describe results of a thermal-chemical disk atmosphere model indicating that Ly$\alpha$-driven photodissociation of water produces sufficient heating to explain the Ly$\alpha$-pumped H$_{2}$ fluorescence observed in CTTSs.  They find that photochemical heating of inner molecular disks inside of $\sim$~0.5~AU is dominated by Ly$\alpha$-driven dissociation of H$_{2}$O and OH.  In this picture, the same process that produces the 1600~\AA\ Bump as a minor dissociation fragment is predicted to dominate the gas heating of the molecular transition zone, warming the local H$_{2}$ populations to temperatures that enable the Ly$\alpha$ fluorescence mechanism to operate (1500~--~2500~K).

\begin{figure}
\figurenum{16}
\begin{center}
\epsfig{figure=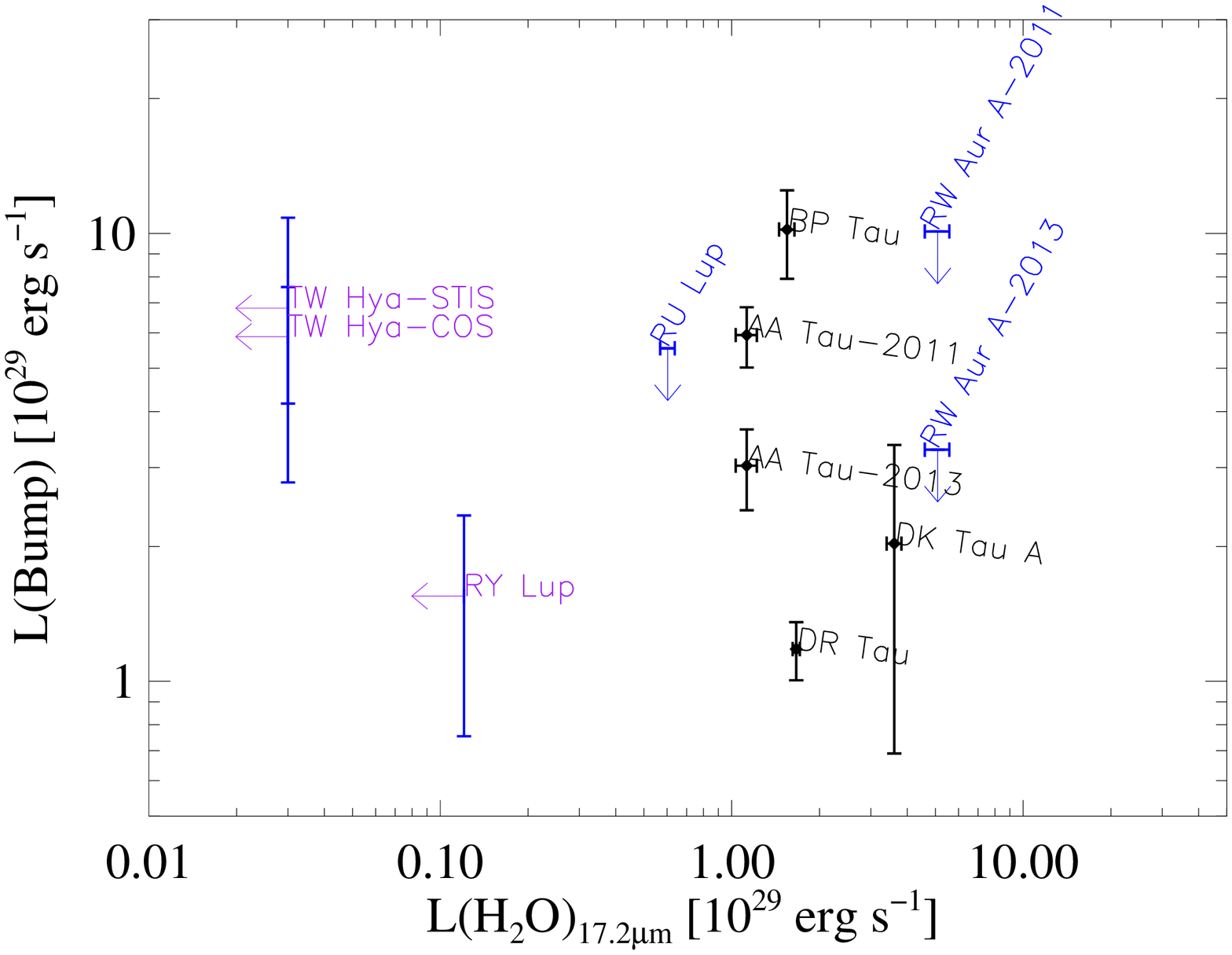,width=2.5in,angle=90}
\vspace{-0.1in}
\caption{A comparison between the 1600~\AA\ Bump luminosity and the emission luminosity in the 17.2~$\mu$m water lines observed in $Spitzer$-IRS observations of targets in common.     There are an insufficient number of detections in the common sample to determine a correlation, and none is observed with the data available.   
\label{cosovly}}
\end{center}
\end{figure}

\subsubsection{Comparison with Infrared Spectra}

Water emission is detected in between one-third and one-half of CTTSs (with central star spectral types G, K, and M) observed with $Spitzer$ mid-IR spectroscopy~\citep{pontoppidan10b}, comparable to the detection rate of the 1600~\AA\ Bump.   These observations show strong peaks between 12 and 34~$\mu$m, from which the vertical column density through the disk ($N$(H$_{2}$O)~$\sim$~10$^{18}$ cm$^{-2}$), the excitation temperature ($T$(H$_{2}$O)~$\sim$~500~--~1000 K), and the characteristic emitting radius ($R$(H$_{2}$O)~$\leq$~2 AU) can be derived~\citep{carr11}.  High-resolution water observations at near-IR wavelengths ($\lambda$~$\sim$~3~$\mu$m) have found similar results~\citep{salyk08,mandell12}.   Given the ubiquity of H$_{2}$ and CO in the inner disks of accreting CTTSs~\citep{france12b,brown13}, it seems clear that H$_{2}$O is equally as common with abundances between 0.1~--~1.0~times that of CO~\citep{carr08, salyk11b, mandell12}.   

Based on our proposed origin for the 1600~\AA\ Bump, one can imagine three potential relations between the Bump and the water emission in the mid-IR:  1) if the Bump is a byproduct of water dissociation, one could expect an anti-correlation between the Bump luminosity and the IR water luminosity, 2) a brighter bump might also indicate more water, and therefore one could expect a positive correlation, or 3) the IR water emission spectrum is produced in a region more embedded than the stellar Ly$\alpha$ can reach effectively, meaning that the two spectral tracers are spatially independent and therefore uncorrelated.  We attempted to distinguish between the above scenarios by comparing the observed water emission spectrum with the observed Bump emission for sources where both measurements were available.  We selected the $Spitzer$ mid-IR water emission complex near 17.2~$\mu$m, as it is the brightest mid-IR water band, and the $Spitzer$-IRS archive provides the largest uniform water database for comparison.  Extracting disk fluxes from~\citet{pontoppidan10b}, \citet{carr11}, and \citet{zhang13}, we identified 10 sources in common between the samples (AA Tau, BP Tau, DK Tau, DN Tau, DR Tau, HD 134344B, RU Lup, RW Aur, RY Lup, and TW Hya).  Unfortunately, the overlap in the two samples is not sufficient for any conclusions to be drawn.  Only AA Tau and BP Tau are strong detections in both mid-IR water emission and the FUV Bump (DR Tau and DK Tau have water detections but marginal Bump detections).  RU Lup and RW Aur have strong water emission and no Bump.  DN Tau, RY Lup, and TW Hya have intermediate Bump emission but no water detected, and HD134344B has neither.  These are displayed in Figure 16, with each target labeled in the figure and upper limits plotted where appropriate.

\subsubsection{Alternative Mechanisms for Producing Highly Non-thermal H$_{2}$ Populations in Inner Disks}

Previous studies have noted the presence of non-thermal H$_{2}$ fluorescence features, in particular emission pumped out of the [$v$,$J$]~=~[5,18] level by~\ion{C}{4} in the TW Hya disk~\citep{herczeg02} and others (Hoadley et al.~--~in prep.).  The \citet{harrevelt08} calculations do not predict a significant population in the  [$v$,$J$]~=~[5,18] level, however it should be emphasized that that work only presented model populations for H$_{2}$O with ground rotational states of 0$_{00}$ (para-water) and 1$_{01}$ (ortho-water), respectively,  a ``cold'' water population.  Emission from warm ($T$(H$_{2}$O)~$\approx$~575 K, see references in \S5.2.4), rotationally excited water has been observed in CTTS disks, therefore one expects that additional rovibrational H$_{2}$ states are appreciably populated by Ly$\alpha$ dissociation of warm water.  H$_{2}$O dissociation by Ly$\alpha$ photons may then be a natural explanation for the highly excited H$_{2}$ emission observed.   

It has been shown that the process of ``multiple pumping''\footnote{``Multiple pumping'' refers to the situation where  excitation by UV photons occurs at a rate faster than the molecules can decay via rovibrational emission lines or collisions.} by FUV continuum photons at $\lambda$~$<$~1120~\AA\ likely operates in protoplanetary disks~\citep{nomura05,france12a}.   Pumping by the FUV continuum is still a relatively slow way to climb the rotational ladder because the transitions must obey the $\Delta$$J$~=~$-$1, 0, +1 selection rules, and gas heating in these regions tends to thermalize the gas at $T$~$\gtrsim$~1000 K~\citep{nomura05,adam16}, resulting in the highly excited vibrational populations that are weighted towards low rotational states.  Models of UV pumping that incorporate X-ray irradiation and dust grain evolution (which regulates the penetration depth of FUV photons into the disk) show similar, high-$v$, low-$J$ excitation diagrams~\citep{nomura07}.     H$_{2}$ dissociation spectra in from these populations will be similar to the hot-star pumped H$_{2}$ fluorescence spectrum observed in photodissociation regions, where the continuum peak is clearly established at~$\lambda$~$\approx$~1575~--~1580~\AA~\citep{witt89,france05}, in contrast to the observed spectra in our CTTS survey.   

There may be other pathways to the formation of the highly non-thermal population distributions that give rise to the quasi-continuous and discrete emission lines observed in the CTTS sample, including the dissociation products of other molecular species and formation-pumping of H$_{2}$ molecules newly formed on dust grains.     Non-thermal H$_{2}$ distributions have been observed in $HST$ and {\it Far-Ultraviolet Spectroscopic Explorer} observations of comets~(e.g., Feldman et al. 2009 and references therein), suggesting a potential common excitation mechanism.  Initial modeling work argued that the majority of this population was produced by the dissociation of H$_{2}$O via Ly$\alpha$ photons~\citep{liu07}, analogous to the process we describe for protoplanetary disks.  More recent analyses by~\citet{feldman09} and \citet{feldman15} have demonstrated that both the H$_{2}$ and the rovibrationally excited CO fluorescence spectra are better described as the fragments of formaldehyde (H$_{2}$CO) dissociation.   We do not favor the formaldehyde hypothesis for the T Tauri star case because while the population resulting from H$_{2}$CO dissociation can produce highly excited vibrational levels~\citep{zhang05}, the rotational populations peak at $J$~=~5 or 7~\citep{chambreau06}, whereas $J$~$>$~15 is required to produce the observed 1600~\AA\ Bump.  Also, we (possibly naively) assume that the H$_{2}$CO abundance in the inner regions of protoplanetary disks will be much lower than H$_{2}$O.   

An additional possibility is that H$_{2}$ formation on dust grains in the inner disk is providing a source of non-thermal H$_{2}$~--~if this could be demonstrated it would constitute an exciting direct probe of molecular formation at planet-forming radii around young stars.    Several authors have computed grain-formation rates for H$_{2}$ under a variety of molecular cloud and PDR conditions~\citep{jura75,black87,bourlot95,draine96}, and the latter authors find the average rovibrational distribution of newly formed H$_{2}$ to be $\langle$[$v$,$J$]$\rangle$~$\approx$~[5,9].    This is still relatively low rotational excitation compared to what is required to produce the 1600~\AA\ Bump, however it should be noted that the calculations cited above were not carried out for the local conditions in a protoplanetary environment.  Additional calculations of the H$_{2}$ formation rates in disks and their nascent population distributions would be very useful.   

\section{Summary}

We have extracted and analyzed the quasi-continuous 1600~\AA\ Bump from a sample of 37 CTTSs observed with $HST$-COS, finding a detection rate of $\approx$~50\%.  The widths and fluxes of these features have been measured, and we determine that this emission is inconsistent with electron-impact excitation of H$_{2}$.   The large total flux in this emission argues that the 1600~\AA\ Bump is powered by Ly$\alpha$ photons, and we argue for a new mechanism:  H$_{2}$O dissociation by Ly$\alpha$ photons followed by photodissociation of the highly non-thermal water dissociation fragments.  This water dissociation hypothesis provides qualitative agreement with both the discrete and continuous emission features observed in the T Tauri star sample.   

We present new evidence demonstrating that water dissociation by the strong stellar+accretion Ly$\alpha$ radiation field is a viable explanation for the Bump emission observed in the $HST$-COS T Tauri star sample.  This conclusion is supported by several observational clues that match with the theory behind the water dissociation scenario, including the ability to account for the shifted, quasi-continuous emission spectrum of the Bump, being able to predict and verify observations of bound-bound H$_{2}$ emission features, and finding plausible agreement with observed fluxes of these components and their expected disk densities.   Therefore, the 1600~\AA\ Bump may be a new probe of water chemistry in Ly$\alpha$-irradiated protoplanetary  environments.  The total emitted power in the Bump is between 10\% and 90\% of the underlying FUV continuum emission in protoplanetary disk spectra.  In future work, we will build on this finding to constrain the spatial distribution of the emission and the physical conditions in the disk where the emission arises.   This can be combined with disk models and water observations at IR and sub-mm wavelengths to develop a more complete picture of the H$_{2}$O chemistry in protoplanetary environments.  Laboratory experiments exploring the dissociation products of warm water vapor illuminated by Ly$\alpha$ photons would also be very valuable for quantifying both the H$_{2}$O dissociation mechanism and the resultant non-thermal H$_{2}$ dissociation spectrum.

\acknowledgments
  This work received support from NASA grant NNX08AC146 to the University of Colorado at Boulder ($HST$ programs 11533 and 12036), made use of data from $HST$ GO programs 8041 and 11616, and was supported by HST GO programs 12876 and 13372.  The authors appreciate discussions with Dr. Rob van Harrevelt regarding H$_{2}$ dissociation fragments and thank him for making electronic versions of ortho and para water dissociation populations available.    KF acknowledges Paul Johnson and Xiaming Liu for discussions on H$_{2}$ structure.    KF thanks Christian Schneider, Greg Herczeg, Matt McJunkin, and Keri Hoadley for enjoyable discussion relating to FUV radiation fields and disk photochemistry.  Finally, KF appreciates the hospitality of White Sands Missile Range, where a portion of this work was carried out during integration and testing of NASA missions 36.271 UG and 36.323 UG.  

\appendix

\section{$HST$-COS CTTS 1600~\AA\ Bump Archive }

Figures A.1a~--~A.1c display the 1600~\AA\ Bump spectra for all 41 observations studied in this work.


\bibliography{ms}

\begin{thebibliography}{159}
\expandafter\ifx\csname natexlab\endcsname\relax\def\natexlab#1{#1}\fi

\bibitem[{{Abgrall} {et~al.}(2000){Abgrall}, {Roueff}, \& {Drira}}]{abgrall:00}
{Abgrall}, H., {Roueff}, E., \& {Drira}, I. 2000, \aaps, 141, 297

\bibitem[{{Abgrall} {et~al.}(1993){Abgrall}, {Roueff}, {Launay}, {Roncin}, \&
  {Subtil}}]{abgrall:93}
{Abgrall}, H., {Roueff}, E., {Launay}, F., {Roncin}, J.~Y., \& {Subtil}, J.~L.
  1993, Journal of Molecular Spectroscopy, 157, 512

\bibitem[{{Abgrall} {et~al.}(1997){Abgrall}, {Roueff}, {Liu}, \&
  {Shemansky}}]{abgrall97}
{Abgrall}, H., {Roueff}, E., {Liu}, X., \& {Shemansky}, D.~E. 1997, \apj, 481,
  557

\bibitem[{{{\'A}d{\'a}mkovics} {et~al.}(2016){{\'A}d{\'a}mkovics}, {Najita}, \&
  {Glassgold}}]{adam16}
{{\'A}d{\'a}mkovics}, M., {Najita}, J.~R., \& {Glassgold}, A.~E. 2016, \apj,
  817, 82

\bibitem[{{Ajello} {et~al.}(1984){Ajello}, {Shemansky}, {Kwok}, \&
  {Yung}}]{ajello84}
{Ajello}, J.~M., {Shemansky}, D., {Kwok}, T.~L., \& {Yung}, Y.~L. 1984, \pra,
  29, 636

\bibitem[{{Akeson} {et~al.}(2002){Akeson}, {Ciardi}, {van Belle}, \&
  {Creech-Eakman}}]{akeson02}
{Akeson}, R.~L., {Ciardi}, D.~R., {van Belle}, G.~T., \& {Creech-Eakman}, M.~J.
  2002, \apj, 566, 1124

\bibitem[{{Alencar} {et~al.}(2003){Alencar}, {Melo}, {Dullemond}, {Andersen},
  {Batalha}, {Vaz}, \& {Mathieu}}]{2003A&A...409.1037A}
{Alencar}, S.~H.~P., {Melo}, C.~H.~F., {Dullemond}, C.~P., {Andersen}, J.,
  {Batalha}, C., {Vaz}, L.~P.~R., \& {Mathieu}, R.~D. 2003, \aap, 409, 1037

\bibitem[{{Alexander} {et~al.}(2014){Alexander}, {Pascucci}, {Andrews},
  {Armitage}, \& {Cieza}}]{alexander14}
{Alexander}, R., {Pascucci}, I., {Andrews}, S., {Armitage}, P., \& {Cieza}, L.
  2014, Protostars and Planets VI, 475

\bibitem[{{ALMA Partnership} {et~al.}(2015){ALMA Partnership}, {Brogan},
  {P{\'e}rez}, {Hunter}, {Dent}, {Hales}, {Hills}, {Corder}, {Fomalont},
  {Vlahakis}, {Asaki}, {Barkats}, {Hirota}, {Hodge}, {Impellizzeri}, {Kneissl},
  {Liuzzo}, {Lucas}, {Marcelino}, {Matsushita}, {Nakanishi}, {Phillips},
  {Richards}, {Toledo}, {Aladro}, {Broguiere}, {Cortes}, {Cortes}, {Espada},
  {Galarza}, {Garcia-Appadoo}, {Guzman-Ramirez}, {Humphreys}, {Jung}, {Kameno},
  {Laing}, {Leon}, {Marconi}, {Mignano}, {Nikolic}, {Nyman}, {Radiszcz},
  {Remijan}, {Rod{\'o}n}, {Sawada}, {Takahashi}, {Tilanus}, {Vila Vilaro},
  {Watson}, {Wiklind}, {Akiyama}, {Chapillon}, {de Gregorio-Monsalvo}, {Di
  Francesco}, {Gueth}, {Kawamura}, {Lee}, {Nguyen Luong}, {Mangum}, {Pietu},
  {Sanhueza}, {Saigo}, {Takakuwa}, {Ubach}, {van Kempen}, {Wootten},
  {Castro-Carrizo}, {Francke}, {Gallardo}, {Garcia}, {Gonzalez}, {Hill},
  {Kaminski}, {Kurono}, {Liu}, {Lopez}, {Morales}, {Plarre}, {Schieven},
  {Testi}, {Videla}, {Villard}, {Andreani}, {Hibbard}, \&
  {Tatematsu}}]{hltau15group}
{ALMA Partnership}, {Brogan}, C.~L., {P{\'e}rez}, L.~M., {Hunter}, T.~R.,
  {Dent}, W.~R.~F., {Hales}, A.~S., {Hills}, R.~E., {Corder}, S., {Fomalont},
  E.~B., {Vlahakis}, C., {Asaki}, Y., {Barkats}, D., {Hirota}, A., {Hodge},
  J.~A., {Impellizzeri}, C.~M.~V., {Kneissl}, R., {Liuzzo}, E., {Lucas}, R.,
  {Marcelino}, N., {Matsushita}, S., {Nakanishi}, K., {Phillips}, N.,
  {Richards}, A.~M.~S., {Toledo}, I., {Aladro}, R., {Broguiere}, D., {Cortes},
  J.~R., {Cortes}, P.~C., {Espada}, D., {Galarza}, F., {Garcia-Appadoo}, D.,
  {Guzman-Ramirez}, L., {Humphreys}, E.~M., {Jung}, T., {Kameno}, S., {Laing},
  R.~A., {Leon}, S., {Marconi}, G., {Mignano}, A., {Nikolic}, B., {Nyman},
  L.-A., {Radiszcz}, M., {Remijan}, A., {Rod{\'o}n}, J.~A., {Sawada}, T.,
  {Takahashi}, S., {Tilanus}, R.~P.~J., {Vila Vilaro}, B., {Watson}, L.~C.,
  {Wiklind}, T., {Akiyama}, E., {Chapillon}, E., {de Gregorio-Monsalvo}, I.,
  {Di Francesco}, J., {Gueth}, F., {Kawamura}, A., {Lee}, C.-F., {Nguyen
  Luong}, Q., {Mangum}, J., {Pietu}, V., {Sanhueza}, P., {Saigo}, K.,
  {Takakuwa}, S., {Ubach}, C., {van Kempen}, T., {Wootten}, A.,
  {Castro-Carrizo}, A., {Francke}, H., {Gallardo}, J., {Garcia}, J.,
  {Gonzalez}, S., {Hill}, T., {Kaminski}, T., {Kurono}, Y., {Liu}, H.-Y.,
  {Lopez}, C., {Morales}, F., {Plarre}, K., {Schieven}, G., {Testi}, L.,
  {Videla}, L., {Villard}, E., {Andreani}, P., {Hibbard}, J.~E., \&
  {Tatematsu}, K. 2015, \apjl, 808, L3

\bibitem[{{Andrews} \& {Williams}(2007)}]{2007ApJ...659..705A}
{Andrews}, S.~M. \& {Williams}, J.~P. 2007, \apj, 659, 705

\bibitem[{{Andrews} {et~al.}(2011){Andrews}, {Wilner}, {Espaillat}, {Hughes},
  {Dullemond}, {McClure}, {Qi}, \& {Brown}}]{2011ApJ...732...42A}
{Andrews}, S.~M., {Wilner}, D.~J., {Espaillat}, C., {Hughes}, A.~M.,
  {Dullemond}, C.~P., {McClure}, M.~K., {Qi}, C., \& {Brown}, J.~M. 2011, \apj,
  732, 42

\bibitem[{{Andrews} {et~al.}(2016){Andrews}, {Wilner}, {Zhu}, {Birnstiel},
  {Carpenter}, {P{\'e}rez}, {Bai}, {{\"O}berg}, {Hughes}, {Isella}, \&
  {Ricci}}]{andrews16}
{Andrews}, S.~M., {Wilner}, D.~J., {Zhu}, Z., {Birnstiel}, T., {Carpenter},
  J.~M., {P{\'e}rez}, L.~M., {Bai}, X.-N., {{\"O}berg}, K.~I., {Hughes}, A.~M.,
  {Isella}, A., \& {Ricci}, L. 2016, \apjl, 820, L40

\bibitem[{{Antonellini} {et~al.}(2016){Antonellini}, {Kamp}, {Lahuis},
  {Woitke}, {Thi}, {Meijerink}, {Aresu}, {Spaans}, {G{\"u}del}, \&
  {Liebhart}}]{antonellini16}
{Antonellini}, S., {Kamp}, I., {Lahuis}, F., {Woitke}, P., {Thi}, W.-F.,
  {Meijerink}, R., {Aresu}, G., {Spaans}, M., {G{\"u}del}, M., \& {Liebhart},
  A. 2016, \aap, 585, A61

\bibitem[{{Antonellini} {et~al.}(2015){Antonellini}, {Kamp},
  {Riviere-Marichalar}, {Meijerink}, {Woitke}, {Thi}, {Spaans}, {Aresu}, \&
  {Lee}}]{antonellini15}
{Antonellini}, S., {Kamp}, I., {Riviere-Marichalar}, P., {Meijerink}, R.,
  {Woitke}, P., {Thi}, W.-F., {Spaans}, M., {Aresu}, G., \& {Lee}, G. 2015,
  \aap, 582, A105

\bibitem[{{Ardila} {et~al.}(2013){Ardila}, {Herczeg}, {Gregory}, {Ingleby},
  {France}, {Brown}, {Edwards}, {Johns-Krull}, {Linsky}, {Yang}, {Valenti},
  {Abgrall}, {Alexander}, {Bergin}, {Bethell}, {Brown}, {Calvet}, {Espaillat},
  {Hillenbrand}, {Hussain}, {Roueff}, {Schindhelm}, \& {Walter}}]{ardila13}
{Ardila}, D.~R., {Herczeg}, G.~J., {Gregory}, S.~G., {Ingleby}, L., {France},
  K., {Brown}, A., {Edwards}, S., {Johns-Krull}, C., {Linsky}, J.~L., {Yang},
  H., {Valenti}, J.~A., {Abgrall}, H., {Alexander}, R.~D., {Bergin}, E.,
  {Bethell}, T., {Brown}, J.~M., {Calvet}, N., {Espaillat}, C., {Hillenbrand},
  L.~A., {Hussain}, G., {Roueff}, E., {Schindhelm}, E.~R., \& {Walter}, F.~M.
  2013, ArXiv e-prints

\bibitem[{{Banzatti} \& {Pontoppidan}(2015)}]{banzatti15}
{Banzatti}, A. \& {Pontoppidan}, K.~M. 2015, \apj, 809, 167

\bibitem[{{Banzatti} {et~al.}(2017){Banzatti}, {Pontoppidan}, {Salyk},
  {Herczeg}, {van Dishoeck}, \& {Blake}}]{banzatti17}
{Banzatti}, A., {Pontoppidan}, K.~M., {Salyk}, C., {Herczeg}, G.~J., {van
  Dishoeck}, E.~F., \& {Blake}, G.~A. 2017, \apj, 834, 152

\bibitem[{{Bergin} {et~al.}(2004){Bergin}, {Calvet}, {Sitko}, {Abgrall},
  {D'Alessio}, {Herczeg}, {Roueff}, {Qi}, {Lynch}, {Russell}, {Brafford}, \&
  {Perry}}]{bergin04}
{Bergin}, E., {Calvet}, N., {Sitko}, M.~L., {Abgrall}, H., {D'Alessio}, P.,
  {Herczeg}, G.~J., {Roueff}, E., {Qi}, C., {Lynch}, D.~K., {Russell}, R.~W.,
  {Brafford}, S.~M., \& {Perry}, R.~B. 2004, \apjl, 614, L133

\bibitem[{{Bertout} {et~al.}(1988){Bertout}, {Basri}, \&
  {Bouvier}}]{1988ApJ...330..350B}
{Bertout}, C., {Basri}, G., \& {Bouvier}, J. 1988, \apj, 330, 350

\bibitem[{{Bertout} {et~al.}(1999){Bertout}, {Robichon}, \&
  {Arenou}}]{1999A&A...352..574B}
{Bertout}, C., {Robichon}, N., \& {Arenou}, F. 1999, \aap, 352, 574

\bibitem[{{Black} \& {van Dishoeck}(1987)}]{black87}
{Black}, J.~H. \& {van Dishoeck}, E.~F. 1987, \apj, 322, 412

\bibitem[{{Bohlin} {et~al.}(1978){Bohlin}, {Savage}, \& {Drake}}]{bohlin78}
{Bohlin}, R.~C., {Savage}, B.~D., \& {Drake}, J.~F. 1978, \apj, 224, 132

\bibitem[{{Bouvier} {et~al.}(1999){Bouvier}, {Chelli}, {Allain}, {Carrasco},
  {Costero}, {Cruz-Gonzalez}, {Dougados}, {Fern{\'a}ndez}, {Mart{\'{\i}}n},
  {M{\'e}nard}, {Mennessier}, {Mujica}, {Recillas}, {Salas}, {Schmidt}, \&
  {Wichmann}}]{1999A&A...349..619B}
{Bouvier}, J., {Chelli}, A., {Allain}, S., {Carrasco}, L., {Costero}, R.,
  {Cruz-Gonzalez}, I., {Dougados}, C., {Fern{\'a}ndez}, M., {Mart{\'{\i}}n},
  E.~L., {M{\'e}nard}, F., {Mennessier}, C., {Mujica}, R., {Recillas}, E.,
  {Salas}, L., {Schmidt}, G., \& {Wichmann}, R. 1999, \aap, 349, 619

\bibitem[{{Bouvier} {et~al.}(2013){Bouvier}, {Grankin}, {Ellerbroek}, {Bouy},
  \& {Barrado}}]{bouvier13}
{Bouvier}, J., {Grankin}, K., {Ellerbroek}, L.~E., {Bouy}, H., \& {Barrado}, D.
  2013, \aap, 557, A77

\bibitem[{{Brittain} {et~al.}(2015){Brittain}, {Najita}, \&
  {Carr}}]{brittain15}
{Brittain}, S.~D., {Najita}, J.~R., \& {Carr}, J.~S. 2015, \apss, 357, 54

\bibitem[{{Brown} {et~al.}(2013){Brown}, {Pontoppidan}, {van Dishoeck},
  {Herczeg}, {Blake}, \& {Smette}}]{brown13}
{Brown}, J.~M., {Pontoppidan}, K.~M., {van Dishoeck}, E.~F., {Herczeg}, G.~J.,
  {Blake}, G.~A., \& {Smette}, A. 2013, \apj, 770, 94

\bibitem[{{Calvet} {et~al.}(2005){Calvet}, {D'Alessio}, {Watson},
  {Franco-Hern{\'a}ndez}, {Furlan}, {Green}, {Sutter}, {Forrest}, {Hartmann},
  {Uchida}, {Keller}, {Sargent}, {Najita}, {Herter}, {Barry}, \&
  {Hall}}]{calvet05}
{Calvet}, N., {D'Alessio}, P., {Watson}, D.~M., {Franco-Hern{\'a}ndez}, R.,
  {Furlan}, E., {Green}, J., {Sutter}, P.~M., {Forrest}, W.~J., {Hartmann}, L.,
  {Uchida}, K.~I., {Keller}, L.~D., {Sargent}, B., {Najita}, J., {Herter},
  T.~L., {Barry}, D.~J., \& {Hall}, P. 2005, \apjl, 630, L185

\bibitem[{{Calvet} {et~al.}(2004){Calvet}, {Muzerolle}, {Brice{\~n}o},
  {Hern{\'a}ndez}, {Hartmann}, {Saucedo}, \& {Gordon}}]{calvet04}
{Calvet}, N., {Muzerolle}, J., {Brice{\~n}o}, C., {Hern{\'a}ndez}, J.,
  {Hartmann}, L., {Saucedo}, J.~L., \& {Gordon}, K.~D. 2004, \aj, 128, 1294

\bibitem[{{Carmona} {et~al.}(2011){Carmona}, {van der Plas}, {van den Ancker},
  {Audard}, {Waters}, {Fedele}, {Acke}, \& {Pantin}}]{carmona11}
{Carmona}, A., {van der Plas}, G., {van den Ancker}, M.~E., {Audard}, M.,
  {Waters}, L.~B.~F.~M., {Fedele}, D., {Acke}, B., \& {Pantin}, E. 2011, \aap,
  533, A39

\bibitem[{{Carr} \& {Najita}(2008)}]{carr08}
{Carr}, J.~S. \& {Najita}, J.~R. 2008, Science, 319, 1504

\bibitem[{{Carr} \& {Najita}(2011)}]{carr11}
---. 2011, \apj, 733, 102

\bibitem[{{Chambreau} {et~al.}(2006){Chambreau}, {Lahankar}, \&
  {Suits}}]{chambreau06}
{Chambreau}, S.~D., {Lahankar}, S.~A., \& {Suits}, A.~G. 2006, \jcp, 125,
  044302

\bibitem[{{Cody} {et~al.}(2013){Cody}, {Tayar}, {Hillenbrand}, {Matthews}, \&
  {Kallinger}}]{cody13}
{Cody}, A.~M., {Tayar}, J., {Hillenbrand}, L.~A., {Matthews}, J.~M., \&
  {Kallinger}, T. 2013, \aj, 145, 79

\bibitem[{{Coffey} {et~al.}(2004){Coffey}, {Bacciotti}, {Woitas}, {Ray}, \&
  {Eisl{\"o}ffel}}]{2004ApJ...604..758C}
{Coffey}, D., {Bacciotti}, F., {Woitas}, J., {Ray}, T.~P., \& {Eisl{\"o}ffel},
  J. 2004, \apj, 604, 758

\bibitem[{{Comer{\'o}n} \& {Fern{\'a}ndez}(2010)}]{2010A&A...511A..10C}
{Comer{\'o}n}, F. \& {Fern{\'a}ndez}, M. 2010, \aap, 511, A10

\bibitem[{{Comer{\'o}n} {et~al.}(2003){Comer{\'o}n}, {Fern{\'a}ndez},
  {Baraffe}, {Neuh{\"a}user}, \& {Kaas}}]{2003A&A...406.1001C}
{Comer{\'o}n}, F., {Fern{\'a}ndez}, M., {Baraffe}, I., {Neuh{\"a}user}, R., \&
  {Kaas}, A.~A. 2003, \aap, 406, 1001

\bibitem[{{Dabrowski}(1984)}]{dabrowski:84}
{Dabrowski}, I. 1984, Canadian Journal of Physics, 62, 1639

\bibitem[{{Dalgarno} {et~al.}(1970){Dalgarno}, {Herzberg}, \&
  {Stephens}}]{stephens:70}
{Dalgarno}, A., {Herzberg}, G., \& {Stephens}, T.~L. 1970, \apjl, 162, L49

\bibitem[{{Diplas} \& {Savage}(1994)}]{diplas94}
{Diplas}, A. \& {Savage}, B.~D. 1994, \apjs, 93, 211

\bibitem[{{Dong} \& {Dawson}(2016)}]{dong16}
{Dong}, R. \& {Dawson}, R. 2016, \apj, 825, 77

\bibitem[{{Draine} \& {Bertoldi}(1996)}]{draine96}
{Draine}, B.~T. \& {Bertoldi}, F. 1996, \apj, 468, 269

\bibitem[{{Eisner} {et~al.}(2009){Eisner}, {Graham}, {Akeson}, \&
  {Najita}}]{eisner09}
{Eisner}, J.~A., {Graham}, J.~R., {Akeson}, R.~L., \& {Najita}, J. 2009, \apj,
  692, 309

\bibitem[{{Eisner} {et~al.}(2014){Eisner}, {Hillenbrand}, \&
  {Stone}}]{eisner14}
{Eisner}, J.~A., {Hillenbrand}, L.~A., \& {Stone}, J.~M. 2014, \mnras, 443,
  1916

\bibitem[{{Espaillat} {et~al.}(2007){Espaillat}, {Calvet}, {D'Alessio},
  {Bergin}, {Hartmann}, {Watson}, {Furlan}, {Najita}, {Forrest}, {McClure},
  {Sargent}, {Bohac}, \& {Harrold}}]{2007ApJ...664L.111E}
{Espaillat}, C., {Calvet}, N., {D'Alessio}, P., {Bergin}, E., {Hartmann}, L.,
  {Watson}, D., {Furlan}, E., {Najita}, J., {Forrest}, W., {McClure}, M.,
  {Sargent}, B., {Bohac}, C., \& {Harrold}, S.~T. 2007, \apjl, 664, L111

\bibitem[{{Espaillat} {et~al.}(2010){Espaillat}, {D'Alessio}, {Hern{\'a}ndez},
  {Nagel}, {Luhman}, {Watson}, {Calvet}, {Muzerolle}, \&
  {McClure}}]{2010ApJ...717..441E}
{Espaillat}, C., {D'Alessio}, P., {Hern{\'a}ndez}, J., {Nagel}, E., {Luhman},
  K.~L., {Watson}, D.~M., {Calvet}, N., {Muzerolle}, J., \& {McClure}, M. 2010,
  \apj, 717, 441

\bibitem[{{Feldman}(2015)}]{feldman15}
{Feldman}, P.~D. 2015, \apj, 812, 115

\bibitem[{{Feldman} {et~al.}(2009){Feldman}, {Lupu}, {McCandliss}, \&
  {Weaver}}]{feldman09}
{Feldman}, P.~D., {Lupu}, R.~E., {McCandliss}, S.~R., \& {Weaver}, H.~A. 2009,
  \apj, 699, 1104

\bibitem[{{France} {et~al.}(2005){France}, {Andersson}, {McCandliss}, \&
  {Feldman}}]{france05}
{France}, K., {Andersson}, B.-G., {McCandliss}, S.~R., \& {Feldman}, P.~D.
  2005, \apj, 628, 750

\bibitem[{{France} {et~al.}(2012{\natexlab{a}}){France}, {Burgh}, {Herczeg},
  {Schindhelm}, {Yang}, {Abgrall}, {Roueff}, {Brown}, {Brown}, \&
  {Linsky}}]{france12a}
{France}, K., {Burgh}, E.~B., {Herczeg}, G.~J., {Schindhelm}, E., {Yang}, H.,
  {Abgrall}, H., {Roueff}, E., {Brown}, A., {Brown}, J.~M., \& {Linsky}, J.~L.
  2012{\natexlab{a}}, \apj, 744, 22

\bibitem[{{France} {et~al.}(2014{\natexlab{a}}){France}, {Herczeg}, {McJunkin},
  \& {Penton}}]{france14b}
{France}, K., {Herczeg}, G.~J., {McJunkin}, M., \& {Penton}, S.~V.
  2014{\natexlab{a}}, \apj, 794, 160

\bibitem[{{France} {et~al.}(2010{\natexlab{a}}){France}, {Linsky}, {Brown},
  {Froning}, \& {B{\'e}land}}]{france10b}
{France}, K., {Linsky}, J.~L., {Brown}, A., {Froning}, C.~S., \& {B{\'e}land},
  S. 2010{\natexlab{a}}, \apj, 715, 596

\bibitem[{{France} {et~al.}(2014{\natexlab{b}}){France}, {Schindhelm},
  {Bergin}, {Roueff}, \& {Abgrall}}]{france14a}
{France}, K., {Schindhelm}, E., {Bergin}, E.~A., {Roueff}, E., \& {Abgrall}, H.
  2014{\natexlab{b}}, \apj, 784, 127

\bibitem[{{France} {et~al.}(2011{\natexlab{a}}){France}, {Schindhelm}, {Burgh},
  {Herczeg}, {Harper}, {Brown}, {Green}, {Linsky}, {Yang}, {Abgrall}, {Ardila},
  {Bergin}, {Bethell}, {Brown}, {Calvet}, {Espaillat}, {Gregory},
  {Hillenbrand}, {Hussain}, {Ingleby}, {Johns-Krull}, {Roueff}, {Valenti}, \&
  {Walter}}]{france11b}
{France}, K., {Schindhelm}, E., {Burgh}, E.~B., {Herczeg}, G.~J., {Harper},
  G.~M., {Brown}, A., {Green}, J.~C., {Linsky}, J.~L., {Yang}, H., {Abgrall},
  H., {Ardila}, D.~R., {Bergin}, E., {Bethell}, T., {Brown}, J.~M., {Calvet},
  N., {Espaillat}, C., {Gregory}, S.~G., {Hillenbrand}, L.~A., {Hussain}, G.,
  {Ingleby}, L., {Johns-Krull}, C.~M., {Roueff}, E., {Valenti}, J.~A., \&
  {Walter}, F.~M. 2011{\natexlab{a}}, \apj, 734, 31

\bibitem[{{France} {et~al.}(2012{\natexlab{b}}){France}, {Schindhelm},
  {Herczeg}, {Brown}, {Abgrall}, {Alexander}, {Bergin}, {Brown}, {Linsky},
  {Roueff}, \& {Yang}}]{france12b}
{France}, K., {Schindhelm}, E., {Herczeg}, G.~J., {Brown}, A., {Abgrall}, H.,
  {Alexander}, R.~D., {Bergin}, E.~A., {Brown}, J.~M., {Linsky}, J.~L.,
  {Roueff}, E., \& {Yang}, H. 2012{\natexlab{b}}, \apj, 756, 171

\bibitem[{{France} {et~al.}(2010{\natexlab{b}}){France}, {Stocke}, {Yang},
  {Linsky}, {Wolven}, {Froning}, {Green}, \& {Osterman}}]{france10a}
{France}, K., {Stocke}, J.~T., {Yang}, H., {Linsky}, J.~L., {Wolven}, B.~C.,
  {Froning}, C.~S., {Green}, J.~C., \& {Osterman}, S.~N. 2010{\natexlab{b}},
  \apj, 712, 1277

\bibitem[{{France} {et~al.}(2011{\natexlab{b}}){France}, {Yang}, \&
  {Linsky}}]{france11a}
{France}, K., {Yang}, H., \& {Linsky}, J.~L. 2011{\natexlab{b}}, \apj, 729, 7

\bibitem[{{Furlan} {et~al.}(2009){Furlan}, {Watson}, {McClure}, {Manoj},
  {Espaillat}, {D'Alessio}, {Calvet}, {Kim}, {Sargent}, {Forrest}, \&
  {Hartmann}}]{furlan09}
{Furlan}, E., {Watson}, D.~M., {McClure}, M.~K., {Manoj}, P., {Espaillat}, C.,
  {D'Alessio}, P., {Calvet}, N., {Kim}, K.~H., {Sargent}, B.~A., {Forrest},
  W.~J., \& {Hartmann}, L. 2009, \apj, 703, 1964

\bibitem[{{Gabriel} {et~al.}(2009){Gabriel}, {van den Dungen}, {Roueff},
  {Abgrall}, \& {Engeln}}]{gabriel:09}
{Gabriel}, O., {van den Dungen}, J.~J.~A., {Roueff}, E., {Abgrall}, H., \&
  {Engeln}, R. 2009, Journal of Molecular Spectroscopy, 253, 64

\bibitem[{{Garcia Lopez} {et~al.}(2006){Garcia Lopez}, {Natta}, {Testi}, \&
  {Habart}}]{2006A&A...459..837G}
{Garcia Lopez}, R., {Natta}, A., {Testi}, L., \& {Habart}, E. 2006, \aap, 459,
  837

\bibitem[{{Getman} {et~al.}(2011){Getman}, {Broos}, {Salter}, {Garmire}, \&
  {Hogerheijde}}]{getman11}
{Getman}, K.~V., {Broos}, P.~S., {Salter}, D.~M., {Garmire}, G.~P., \&
  {Hogerheijde}, M.~R. 2011, \apj, 730, 6

\bibitem[{{Gizis} \& {Bharat}(2004)}]{gizis04}
{Gizis}, J.~E. \& {Bharat}, R. 2004, \apjl, 608, L113

\bibitem[{{G{\'o}mez de Castro}(2009)}]{2009ApJ...698L.108G}
{G{\'o}mez de Castro}, A.~I. 2009, \apjl, 698, L108

\bibitem[{{G{\'o}mez de Castro} {et~al.}(2013){G{\'o}mez de Castro},
  {L{\'o}pez-Santiago}, {Talavera}, {Sytov}, \& {Bisikalo}}]{castro13}
{G{\'o}mez de Castro}, A.~I., {L{\'o}pez-Santiago}, J., {Talavera}, A.,
  {Sytov}, A.~Y., \& {Bisikalo}, D. 2013, \apj, 766, 62

\bibitem[{{G{\'o}mez de Castro} {et~al.}(2016){G{\'o}mez de Castro}, {Loyd},
  {France}, {Sytov}, \& {Bisikalo}}]{castro16}
{G{\'o}mez de Castro}, A.~I., {Loyd}, R.~O.~P., {France}, K., {Sytov}, A., \&
  {Bisikalo}, D. 2016, \apjl, 818, L17

\bibitem[{{Grady} {et~al.}(2009){Grady}, {Schneider}, {Sitko}, {Williger},
  {Hamaguchi}, {Brittain}, {Ablordeppey}, {Apai}, {Beerman}, {Carpenter},
  {Collins}, {Fukagawa}, {Hammel}, {Henning}, {Hines}, {Kimes}, {Lynch},
  {M{\'e}nard}, {Pearson}, {Russell}, {Silverstone}, {Smith}, {Troutman},
  {Wilner}, {Woodgate}, \& {Clampin}}]{2009ApJ...699.1822G}
{Grady}, C.~A., {Schneider}, G., {Sitko}, M.~L., {Williger}, G.~M.,
  {Hamaguchi}, K., {Brittain}, S.~D., {Ablordeppey}, K., {Apai}, D., {Beerman},
  L., {Carpenter}, W.~J., {Collins}, K.~A., {Fukagawa}, M., {Hammel}, H.~B.,
  {Henning}, T., {Hines}, D., {Kimes}, R., {Lynch}, D.~K., {M{\'e}nard}, F.,
  {Pearson}, R., {Russell}, R.~W., {Silverstone}, M., {Smith}, P.~S.,
  {Troutman}, M., {Wilner}, D., {Woodgate}, B., \& {Clampin}, M. 2009, \apj,
  699, 1822

\bibitem[{{Grady} {et~al.}(2004){Grady}, {Woodgate}, {Torres}, {Henning},
  {Apai}, {Rodmann}, {Wang}, {Stecklum}, {Linz}, {Williger}, {Brown},
  {Wilkinson}, {Harper}, {Herczeg}, {Danks}, {Vieira}, {Malumuth}, {Collins},
  \& {Hill}}]{2004ApJ...608..809G}
{Grady}, C.~A., {Woodgate}, B., {Torres}, C.~A.~O., {Henning}, T., {Apai}, D.,
  {Rodmann}, J., {Wang}, H., {Stecklum}, B., {Linz}, H., {Williger}, G.~M.,
  {Brown}, A., {Wilkinson}, E., {Harper}, G.~M., {Herczeg}, G.~J., {Danks}, A.,
  {Vieira}, G.~L., {Malumuth}, E., {Collins}, N.~R., \& {Hill}, R.~S. 2004,
  \apj, 608, 809

\bibitem[{{Green} {et~al.}(2012){Green}, {Froning}, {Osterman}, {Ebbets},
  {Heap}, {Leitherer}, {Linsky}, {Savage}, {Sembach}, {Shull}, {Siegmund},
  {Snow}, {Spencer}, {Stern}, {Stocke}, {Welsh}, {B{\'e}land}, {Burgh},
  {Danforth}, {France}, {Keeney}, {McPhate}, {Penton}, {Andrews},
  {Brownsberger}, {Morse}, \& {Wilkinson}}]{green12}
{Green}, J.~C., {Froning}, C.~S., {Osterman}, S., {Ebbets}, D., {Heap}, S.~H.,
  {Leitherer}, C., {Linsky}, J.~L., {Savage}, B.~D., {Sembach}, K., {Shull},
  J.~M., {Siegmund}, O.~H.~W., {Snow}, T.~P., {Spencer}, J., {Stern}, S.~A.,
  {Stocke}, J., {Welsh}, B., {B{\'e}land}, S., {Burgh}, E.~B., {Danforth}, C.,
  {France}, K., {Keeney}, B., {McPhate}, J., {Penton}, S.~V., {Andrews}, J.,
  {Brownsberger}, K., {Morse}, J., \& {Wilkinson}, E. 2012, \apj, 744, 60

\bibitem[{{G{\"u}del} {et~al.}(2007){G{\"u}del}, {Briggs}, {Arzner}, {Audard},
  {Bouvier}, {Feigelson}, {Franciosini}, {Glauser}, {Grosso}, {Micela},
  {Monin}, {Montmerle}, {Padgett}, {Palla}, {Pillitteri}, {Rebull}, {Scelsi},
  {Silva}, {Skinner}, {Stelzer}, \& {Telleschi}}]{gudel07}
{G{\"u}del}, M., {Briggs}, K.~R., {Arzner}, K., {Audard}, M., {Bouvier}, J.,
  {Feigelson}, E.~D., {Franciosini}, E., {Glauser}, A., {Grosso}, N., {Micela},
  G., {Monin}, J.-L., {Montmerle}, T., {Padgett}, D.~L., {Palla}, F.,
  {Pillitteri}, I., {Rebull}, L., {Scelsi}, L., {Silva}, B., {Skinner}, S.~L.,
  {Stelzer}, B., \& {Telleschi}, A. 2007, \aap, 468, 353

\bibitem[{{G{\"u}del} {et~al.}(2010){G{\"u}del}, {Lahuis}, {Briggs}, {Carr},
  {Glassgold}, {Henning}, {Najita}, {van Boekel}, \& {van Dishoeck}}]{gudel10}
{G{\"u}del}, M., {Lahuis}, F., {Briggs}, K.~R., {Carr}, J., {Glassgold}, A.~E.,
  {Henning}, T., {Najita}, J.~R., {van Boekel}, R., \& {van Dishoeck}, E.~F.
  2010, \aap, 519, A113

\bibitem[{{G{\"u}del} \& {Telleschi}(2007)}]{gudel07b}
{G{\"u}del}, M. \& {Telleschi}, A. 2007, \aap, 474, L25

\bibitem[{{Gullbring} {et~al.}(2000){Gullbring}, {Calvet}, {Muzerolle}, \&
  {Hartmann}}]{2000ApJ...544..927G}
{Gullbring}, E., {Calvet}, N., {Muzerolle}, J., \& {Hartmann}, L. 2000, \apj,
  544, 927

\bibitem[{{Gullbring} {et~al.}(1998){Gullbring}, {Hartmann}, {Brice{\~n}o}, \&
  {Calvet}}]{1998ApJ...492..323G}
{Gullbring}, E., {Hartmann}, L., {Brice{\~n}o}, C., \& {Calvet}, N. 1998, \apj,
  492, 323

\bibitem[{{G{\"u}nther} \& {Schmitt}(2007)}]{gunther07}
{G{\"u}nther}, H.~M. \& {Schmitt}, J.~H.~M.~M. 2007, \memsai, 78, 359

\bibitem[{{Gustin} {et~al.}(2006){Gustin}, {Cowley}, {G{\'e}rard}, {Gladstone},
  {Grodent}, \& {Clarke}}]{gustin06}
{Gustin}, J., {Cowley}, S.~W.~H., {G{\'e}rard}, J., {Gladstone}, G.~R.,
  {Grodent}, D., \& {Clarke}, J.~T. 2006, Journal of Geophysical Research
  (Space Physics), 111, 9220

\bibitem[{{Gustin} {et~al.}(2004){Gustin}, {Feldman}, {G{\'e}rard}, {Grodent},
  {Vidal-Madjar}, {Ben Jaffel}, {Desert}, {Moos}, {Sahnow}, {Weaver}, {Wolven},
  {Ajello}, {Waite}, {Roueff}, \& {Abgrall}}]{gustin04}
{Gustin}, J., {Feldman}, P.~D., {G{\'e}rard}, J., {Grodent}, D.,
  {Vidal-Madjar}, A., {Ben Jaffel}, L., {Desert}, J., {Moos}, H.~W., {Sahnow},
  D.~J., {Weaver}, H.~A., {Wolven}, B.~C., {Ajello}, J.~M., {Waite}, J.~H.,
  {Roueff}, E., \& {Abgrall}, H. 2004, Icarus, 171, 336

\bibitem[{{Gustin} {et~al.}(2010){Gustin}, {Stewart}, {G{\'e}rard}, \&
  {Esposito}}]{gustin10}
{Gustin}, J., {Stewart}, I., {G{\'e}rard}, J.-C., \& {Esposito}, L. 2010,
  Icarus, 210, 270

\bibitem[{{Hartigan} {et~al.}(1995){Hartigan}, {Edwards}, \&
  {Ghandour}}]{1995ApJ...452..736H}
{Hartigan}, P., {Edwards}, S., \& {Ghandour}, L. 1995, \apj, 452, 736

\bibitem[{{Hartigan} \& {Kenyon}(2003)}]{hartigan03}
{Hartigan}, P. \& {Kenyon}, S.~J. 2003, \apj, 583, 334

\bibitem[{{Hartmann} {et~al.}(1998){Hartmann}, {Calvet}, {Gullbring}, \&
  {D'Alessio}}]{1998ApJ...495..385H}
{Hartmann}, L., {Calvet}, N., {Gullbring}, E., \& {D'Alessio}, P. 1998, \apj,
  495, 385

\bibitem[{{Haworth} {et~al.}(2016){Haworth}, {Ilee}, {Forgan}, {Facchini},
  {Price}, {Boneberg}, {Booth}, {Clarke}, {Gonzalez}, {Hutchison}, {Kamp},
  {Laibe}, {Lyra}, {Meru}, {Mohanty}, {Pani{\'c}}, {Rice}, {Suzuki}, {Teague},
  {Walsh}, {Woitke}, \& {Community authors}}]{panchrom_model_willacy}
{Haworth}, T.~J., {Ilee}, J.~D., {Forgan}, D.~H., {Facchini}, S., {Price},
  D.~J., {Boneberg}, D.~M., {Booth}, R.~A., {Clarke}, C.~J., {Gonzalez}, J.-F.,
  {Hutchison}, M.~A., {Kamp}, I., {Laibe}, G., {Lyra}, W., {Meru}, F.,
  {Mohanty}, S., {Pani{\'c}}, O., {Rice}, K., {Suzuki}, T., {Teague}, R.,
  {Walsh}, C., {Woitke}, P., \& {Community authors}. 2016, PASA, 33, e053

\bibitem[{{Herbst} {et~al.}(1994){Herbst}, {Herbst}, {Grossman}, \&
  {Weinstein}}]{herbst94}
{Herbst}, W., {Herbst}, D.~K., {Grossman}, E.~J., \& {Weinstein}, D. 1994, \aj,
  108, 1906

\bibitem[{{Herczeg} \& {Hillenbrand}(2008)}]{2008ApJ...681..594H}
{Herczeg}, G.~J. \& {Hillenbrand}, L.~A. 2008, \apj, 681, 594

\bibitem[{{Herczeg} \& {Hillenbrand}(2014)}]{herczeg14}
---. 2014, \apj, 786, 97

\bibitem[{{Herczeg} {et~al.}(2002){Herczeg}, {Linsky}, {Valenti},
  {Johns-Krull}, \& {Wood}}]{herczeg02}
{Herczeg}, G.~J., {Linsky}, J.~L., {Valenti}, J.~A., {Johns-Krull}, C.~M., \&
  {Wood}, B.~E. 2002, \apj, 572, 310

\bibitem[{{Herczeg} {et~al.}(2005){Herczeg}, {Walter}, {Linsky}, {Gahm},
  {Ardila}, {Brown}, {Johns-Krull}, {Simon}, \& {Valenti}}]{herczeg05}
{Herczeg}, G.~J., {Walter}, F.~M., {Linsky}, J.~L., {Gahm}, G.~F., {Ardila},
  D.~R., {Brown}, A., {Johns-Krull}, C.~M., {Simon}, M., \& {Valenti}, J.~A.
  2005, \aj, 129, 2777

\bibitem[{{Herczeg} {et~al.}(2004){Herczeg}, {Wood}, {Linsky}, {Valenti}, \&
  {Johns-Krull}}]{herczeg04}
{Herczeg}, G.~J., {Wood}, B.~E., {Linsky}, J.~L., {Valenti}, J.~A., \&
  {Johns-Krull}, C.~M. 2004, \apj, 607, 369

\bibitem[{{Hoadley} {et~al.}(2015){Hoadley}, {France}, {Alexander}, {McJunkin},
  \& {Schneider}}]{hoadley15}
{Hoadley}, K., {France}, K., {Alexander}, R.~D., {McJunkin}, M., \&
  {Schneider}, P.~C. 2015, \apj, 812, 41

\bibitem[{{Hughes} {et~al.}(1994){Hughes}, {Hartigan}, {Krautter}, \&
  {Kelemen}}]{1994AJ....108.1071H}
{Hughes}, J., {Hartigan}, P., {Krautter}, J., \& {Kelemen}, J. 1994, \aj, 108,
  1071

\bibitem[{{Ingleby} {et~al.}(2011{\natexlab{a}}){Ingleby}, {Calvet}, {Bergin},
  {Herczeg}, {Brown}, {Alexander}, {Edwards}, {Espaillat}, {France}, {Gregory},
  {Hillenbrand}, {Roueff}, {Valenti}, {Walter}, {Johns-Krull}, {Brown},
  {Linsky}, {McClure}, {Ardila}, {Abgrall}, {Bethell}, {Hussain}, \&
  {Yang}}]{ingleby11b}
{Ingleby}, L., {Calvet}, N., {Bergin}, E., {Herczeg}, G., {Brown}, A.,
  {Alexander}, R., {Edwards}, S., {Espaillat}, C., {France}, K., {Gregory},
  S.~G., {Hillenbrand}, L., {Roueff}, E., {Valenti}, J., {Walter}, F.,
  {Johns-Krull}, C., {Brown}, J., {Linsky}, J., {McClure}, M., {Ardila}, D.,
  {Abgrall}, H., {Bethell}, T., {Hussain}, G., \& {Yang}, H.
  2011{\natexlab{a}}, ArXiv e-prints

\bibitem[{{Ingleby} {et~al.}(2009){Ingleby}, {Calvet}, {Bergin}, {Yerasi},
  {Espaillat}, {Herczeg}, {Roueff}, {Abgrall}, {Hern{\'a}ndez}, {Brice{\~n}o},
  {Pascucci}, {Miller}, {Fogel}, {Hartmann}, {Meyer}, {Carpenter}, {Crockett},
  \& {McClure}}]{ingleby09}
{Ingleby}, L., {Calvet}, N., {Bergin}, E., {Yerasi}, A., {Espaillat}, C.,
  {Herczeg}, G., {Roueff}, E., {Abgrall}, H., {Hern{\'a}ndez}, J.,
  {Brice{\~n}o}, C., {Pascucci}, I., {Miller}, J., {Fogel}, J., {Hartmann}, L.,
  {Meyer}, M., {Carpenter}, J., {Crockett}, N., \& {McClure}, M. 2009, \apjl,
  703, L137

\bibitem[{{Ingleby} {et~al.}(2013){Ingleby}, {Calvet}, {Herczeg}, {Blaty},
  {Walter}, {Ardila}, {Alexander}, {Edwards}, {Espaillat}, {Gregory},
  {Hillenbrand}, \& {Brown}}]{ingleby13}
{Ingleby}, L., {Calvet}, N., {Herczeg}, G., {Blaty}, A., {Walter}, F.,
  {Ardila}, D., {Alexander}, R., {Edwards}, S., {Espaillat}, C., {Gregory},
  S.~G., {Hillenbrand}, L., \& {Brown}, A. 2013, ArXiv e-prints

\bibitem[{{Ingleby} {et~al.}(2011{\natexlab{b}}){Ingleby}, {Calvet},
  {Hern{\'a}ndez}, {Brice{\~n}o}, {Espaillat}, {Miller}, {Bergin}, \&
  {Hartmann}}]{ingleby11}
{Ingleby}, L., {Calvet}, N., {Hern{\'a}ndez}, J., {Brice{\~n}o}, C.,
  {Espaillat}, C., {Miller}, J., {Bergin}, E., \& {Hartmann}, L.
  2011{\natexlab{b}}, \aj, 141, 127

\bibitem[{{Ingleby} {et~al.}(2014){Ingleby}, {Calvet}, {Hern{\'a}ndez},
  {Hartmann}, {Briceno}, {Miller}, {Espaillat}, \& {McClure}}]{ingleby14}
{Ingleby}, L., {Calvet}, N., {Hern{\'a}ndez}, J., {Hartmann}, L., {Briceno},
  C., {Miller}, J., {Espaillat}, C., \& {McClure}, M. 2014, \apj, 790, 47

\bibitem[{{Johns-Krull} \& {Valenti}(2001)}]{2001ApJ...561.1060J}
{Johns-Krull}, C.~M. \& {Valenti}, J.~A. 2001, \apj, 561, 1060

\bibitem[{{Johns-Krull} {et~al.}(2000){Johns-Krull}, {Valenti}, \&
  {Linsky}}]{2000ApJ...539..815J}
{Johns-Krull}, C.~M., {Valenti}, J.~A., \& {Linsky}, J.~L. 2000, \apj, 539, 815

\bibitem[{{Jura}(1975)}]{jura75}
{Jura}, M. 1975, \apj, 197, 575

\bibitem[{{Komasa} {et~al.}(2011){Komasa}, {Piszczatowski}, {Lach}, \&
  et~al.}]{komasa11}
{Komasa}, J., {Piszczatowski}, K., {Lach}, G., \& et~al. 2011, Chem. Theory
  Comput., 7, 3105

\bibitem[{{Kraus} \& {Hillenbrand}(2009)}]{2009ApJ...704..531K}
{Kraus}, A.~L. \& {Hillenbrand}, L.~A. 2009, \apj, 704, 531

\bibitem[{{Lamzin}(2006)}]{lamzin06}
{Lamzin}, S.~A. 2006, Astronomy Letters, 32, 176

\bibitem[{{Lawson} {et~al.}(2001){Lawson}, {Crause}, {Mamajek}, \&
  {Feigelson}}]{2001MNRAS.321...57L}
{Lawson}, W.~A., {Crause}, L.~A., {Mamajek}, E.~E., \& {Feigelson}, E.~D. 2001,
  \mnras, 321, 57

\bibitem[{{Lawson} {et~al.}(2004){Lawson}, {Lyo}, \&
  {Muzerolle}}]{2004MNRAS.351L..39L}
{Lawson}, W.~A., {Lyo}, A.-R., \& {Muzerolle}, J. 2004, \mnras, 351, L39

\bibitem[{{Le Bourlot} {et~al.}(1995){Le Bourlot}, {Pineau des Forets},
  {Roueff}, {Dalgarno}, \& {Gredel}}]{bourlot95}
{Le Bourlot}, J., {Pineau des Forets}, G., {Roueff}, E., {Dalgarno}, A., \&
  {Gredel}, R. 1995, \apj, 449, 178

\bibitem[{{Liu} \& {Dalgarno}(1996)}]{liu96}
{Liu}, W. \& {Dalgarno}, A. 1996, \apj, 462, 502

\bibitem[{{Liu} {et~al.}(2007){Liu}, {Shemansky}, {Hallett}, \&
  {Weaver}}]{liu07}
{Liu}, X., {Shemansky}, D.~E., {Hallett}, J.~T., \& {Weaver}, H.~A. 2007,
  \apjs, 169, 458

\bibitem[{{Loinard} {et~al.}(2007){Loinard}, {Torres}, {Mioduszewski},
  {Rodr{\'{\i}}guez}, {Gonz{\'a}lez-L{\'o}pezlira}, {Lachaume}, {V{\'a}zquez},
  \& {Gonz{\'a}lez}}]{2007ApJ...671..546L}
{Loinard}, L., {Torres}, R.~M., {Mioduszewski}, A.~J., {Rodr{\'{\i}}guez},
  L.~F., {Gonz{\'a}lez-L{\'o}pezlira}, R.~A., {Lachaume}, R., {V{\'a}zquez},
  V., \& {Gonz{\'a}lez}, E. 2007, \apj, 671, 546

\bibitem[{{L{\'o}pez-Santiago} {et~al.}(2010){L{\'o}pez-Santiago}, {Albacete
  Colombo}, \& {L{\'o}pez-Garc{\'{\i}}a}}]{lopez10}
{L{\'o}pez-Santiago}, J., {Albacete Colombo}, J.~F., \&
  {L{\'o}pez-Garc{\'{\i}}a}, M.~A. 2010, \aap, 524, A97

\bibitem[{{Luhman}(2004)}]{2004ApJ...602..816L}
{Luhman}, K.~L. 2004, \apj, 602, 816

\bibitem[{{Luhman} \& {Steeghs}(2004)}]{2004ApJ...609..917L}
{Luhman}, K.~L. \& {Steeghs}, D. 2004, \apj, 609, 917

\bibitem[{{Mamajek} {et~al.}(1999){Mamajek}, {Lawson}, \&
  {Feigelson}}]{1999PASA...16..257M}
{Mamajek}, E.~E., {Lawson}, W.~A., \& {Feigelson}, E.~D. 1999, PASA, 16, 257

\bibitem[{{Mandell} {et~al.}(2012){Mandell}, {Bast}, {van Dishoeck}, {Blake},
  {Salyk}, {Mumma}, \& {Villanueva}}]{mandell12}
{Mandell}, A.~M., {Bast}, J., {van Dishoeck}, E.~F., {Blake}, G.~A., {Salyk},
  C., {Mumma}, M.~J., \& {Villanueva}, G. 2012, \apj, 747, 92

\bibitem[{{Manset} {et~al.}(2009){Manset}, {Bastien}, {M{\'e}nard}, {Bertout},
  {Le van Suu}, \& {Boivin}}]{manset09}
{Manset}, N., {Bastien}, P., {M{\'e}nard}, F., {Bertout}, C., {Le van Suu}, A.,
  \& {Boivin}, L. 2009, \aap, 499, 137

\bibitem[{{McJunkin} {et~al.}(2016){McJunkin}, {France}, {Schindhelm},
  {Herczeg}, {Schneider}, \& {Brown}}]{mcj16}
{McJunkin}, M., {France}, K., {Schindhelm}, E., {Herczeg}, G., {Schneider},
  P.~C., \& {Brown}, A. 2016, \apj, 828, 69

\bibitem[{{McJunkin} {et~al.}(2014){McJunkin}, {France}, {Schneider},
  {Herczeg}, {Brown}, {Hillenbrand}, {Schindhelm}, \& {Edwards}}]{mcj14}
{McJunkin}, M., {France}, K., {Schneider}, P.~C., {Herczeg}, G.~J., {Brown},
  A., {Hillenbrand}, L., {Schindhelm}, E., \& {Edwards}, S. 2014, \apj, 780,
  150

\bibitem[{{Muzerolle} {et~al.}(2003){Muzerolle}, {Calvet}, {Hartmann}, \&
  {D'Alessio}}]{2003ApJ...597L.149M}
{Muzerolle}, J., {Calvet}, N., {Hartmann}, L., \& {D'Alessio}, P. 2003, \apjl,
  597, L149

\bibitem[{{Najita} {et~al.}(2008){Najita}, {Crockett}, \&
  {Carr}}]{2008ApJ...687.1168N}
{Najita}, J.~R., {Crockett}, N., \& {Carr}, J.~S. 2008, \apj, 687, 1168

\bibitem[{{Nomura} {et~al.}(2007){Nomura}, {Aikawa}, {Tsujimoto}, {Nakagawa},
  \& {Millar}}]{nomura07}
{Nomura}, H., {Aikawa}, Y., {Tsujimoto}, M., {Nakagawa}, Y., \& {Millar}, T.~J.
  2007, \apj, 661, 334

\bibitem[{{Nomura} \& {Millar}(2005)}]{nomura05}
{Nomura}, H. \& {Millar}, T.~J. 2005, \aap, 438, 923

\bibitem[{{Osterman} {et~al.}(2011){Osterman}, {Green}, {Froning},
  {B{\'e}land}, {Burgh}, {France}, {Penton}, {Delker}, {Ebbets}, {Sahnow},
  {Bacinski}, {Kimble}, {Andrews}, {Wilkinson}, {McPhate}, {Siegmund}, {Ake},
  {Aloisi}, {Biagetti}, {Diaz}, {Dixon}, {Friedman}, {Ghavamian}, {Goudfrooij},
  {Hartig}, {Keyes}, {Lennon}, {Massa}, {Niemi}, {Oliveira}, {Osten},
  {Proffitt}, {Smith}, \& {Soderblom}}]{osterman11}
{Osterman}, S., {Green}, J., {Froning}, C., {B{\'e}land}, S., {Burgh}, E.,
  {France}, K., {Penton}, S., {Delker}, T., {Ebbets}, D., {Sahnow}, D.,
  {Bacinski}, J., {Kimble}, R., {Andrews}, J., {Wilkinson}, E., {McPhate}, J.,
  {Siegmund}, O., {Ake}, T., {Aloisi}, A., {Biagetti}, C., {Diaz}, R., {Dixon},
  W., {Friedman}, S., {Ghavamian}, P., {Goudfrooij}, P., {Hartig}, G., {Keyes},
  C., {Lennon}, D., {Massa}, D., {Niemi}, S., {Oliveira}, C., {Osten}, R.,
  {Proffitt}, C., {Smith}, T., \& {Soderblom}, D. 2011, \apss, 306

\bibitem[{{Pascucci} {et~al.}(2013){Pascucci}, {Herczeg}, {Carr}, \&
  {Bruderer}}]{pascucci13}
{Pascucci}, I., {Herczeg}, G., {Carr}, J.~S., \& {Bruderer}, S. 2013, \apj,
  779, 178

\bibitem[{{Pontoppidan} {et~al.}(2011){Pontoppidan}, {Blake}, \&
  {Smette}}]{pontoppidan11}
{Pontoppidan}, K.~M., {Blake}, G.~A., \& {Smette}, A. 2011, \apj, 733, 84

\bibitem[{{Pontoppidan} {et~al.}(2008){Pontoppidan}, {Blake}, {van Dishoeck},
  {Smette}, {Ireland}, \& {Brown}}]{2008ApJ...684.1323P}
{Pontoppidan}, K.~M., {Blake}, G.~A., {van Dishoeck}, E.~F., {Smette}, A.,
  {Ireland}, M.~J., \& {Brown}, J. 2008, \apj, 684, 1323

\bibitem[{{Pontoppidan} {et~al.}(2010){Pontoppidan}, {Salyk}, {Blake},
  {Meijerink}, {Carr}, \& {Najita}}]{pontoppidan10b}
{Pontoppidan}, K.~M., {Salyk}, C., {Blake}, G.~A., {Meijerink}, R., {Carr},
  J.~S., \& {Najita}, J. 2010, \apj, 720, 887

\bibitem[{{Quast} {et~al.}(2000){Quast}, {Torres}, {de La Reza}, {da Silva}, \&
  {Mayor}}]{2000IAUS..200P..28Q}
{Quast}, G.~R., {Torres}, C.~A.~O., {de La Reza}, R., {da Silva}, L., \&
  {Mayor}, M. 2000, in IAU Symposium, Vol. 200, IAU Symposium, 28

\bibitem[{{Ramsay Howat} \& {Greaves}(2007)}]{2007MNRAS.379.1658R}
{Ramsay Howat}, S.~K. \& {Greaves}, J.~S. 2007, \mnras, 379, 1658

\bibitem[{{Raymond} {et~al.}(1997){Raymond}, {Blair}, \& {Long}}]{raymond97}
{Raymond}, J.~C., {Blair}, W.~P., \& {Long}, K.~S. 1997, \apj, 489, 314

\bibitem[{{Ricci} {et~al.}(2010){Ricci}, {Testi}, {Natta}, {Neri}, {Cabrit}, \&
  {Herczeg}}]{2010A&A...512A..15R}
{Ricci}, L., {Testi}, L., {Natta}, A., {Neri}, R., {Cabrit}, S., \& {Herczeg},
  G.~J. 2010, \aap, 512, A15

\bibitem[{{Rodriguez} {et~al.}(2010){Rodriguez}, {Kastner}, {Wilner}, \&
  {Qi}}]{2010ApJ...720.1684R}
{Rodriguez}, D.~R., {Kastner}, J.~H., {Wilner}, D., \& {Qi}, C. 2010, \apj,
  720, 1684

\bibitem[{{Salyk} {et~al.}(2009){Salyk}, {Blake}, {Boogert}, \&
  {Brown}}]{salyk09}
{Salyk}, C., {Blake}, G.~A., {Boogert}, A.~C.~A., \& {Brown}, J.~M. 2009, \apj,
  699, 330

\bibitem[{{Salyk} {et~al.}(2008){Salyk}, {Pontoppidan}, {Blake}, {Lahuis}, {van
  Dishoeck}, \& {Evans}}]{salyk08}
{Salyk}, C., {Pontoppidan}, K.~M., {Blake}, G.~A., {Lahuis}, F., {van
  Dishoeck}, E.~F., \& {Evans}, II, N.~J. 2008, \apjl, 676, L49

\bibitem[{{Salyk} {et~al.}(2011{\natexlab{a}}){Salyk}, {Pontoppidan}, {Blake},
  {Najita}, \& {Carr}}]{salyk11b}
{Salyk}, C., {Pontoppidan}, K.~M., {Blake}, G.~A., {Najita}, J.~R., \& {Carr},
  J.~S. 2011{\natexlab{a}}, \apj, 731, 130

\bibitem[{{Salyk} {et~al.}(2011{\natexlab{b}}){Salyk}, {Pontoppidan}, {Blake},
  {Najita}, \& {Carr}}]{salyk11a}
---. 2011{\natexlab{b}}, \apj, 731, 130

\bibitem[{{Schindhelm} {et~al.}(2012{\natexlab{a}}){Schindhelm}, {France},
  {Herczeg}, \& {Bergin}}]{schindhelm12a}
{Schindhelm}, E., {France}, K., {Herczeg}, G., \& {Bergin}, E.
  2012{\natexlab{a}}, \apjl

\bibitem[{{Schindhelm} {et~al.}(2012{\natexlab{b}}){Schindhelm}, {France},
  {Herczeg}, {Bergin}, {Yang}, {Brown}, {Brown}, {Linsky}, \&
  {Valenti}}]{schindhelm12b}
{Schindhelm}, E., {France}, K., {Herczeg}, G.~J., {Bergin}, E., {Yang}, H.,
  {Brown}, A., {Brown}, J.~M., {Linsky}, J.~L., \& {Valenti}, J.
  2012{\natexlab{b}}, \apjl, 756, L23

\bibitem[{{Schneider} {et~al.}(2015{\natexlab{a}}){Schneider}, {France},
  {G{\"u}nther}, {Herczeg}, {Robrade}, {Bouvier}, {McJunkin}, \&
  {Schmitt}}]{schneider15}
{Schneider}, P.~C., {France}, K., {G{\"u}nther}, H.~M., {Herczeg}, G.,
  {Robrade}, J., {Bouvier}, J., {McJunkin}, M., \& {Schmitt}, J.~H.~M.~M.
  2015{\natexlab{a}}, \aap, 584, A51

\bibitem[{{Schneider} {et~al.}(2015{\natexlab{b}}){Schneider}, {G{\"u}nther},
  {Robrade}, {Facchini}, {Hodapp}, {Manara}, {Perdelwitz}, {Schmitt},
  {Skinner}, \& {Wolk}}]{schneider15b}
{Schneider}, P.~C., {G{\"u}nther}, H.~M., {Robrade}, J., {Facchini}, S.,
  {Hodapp}, K.~W., {Manara}, C.~F., {Perdelwitz}, V., {Schmitt}, J.~H.~M.~M.,
  {Skinner}, S., \& {Wolk}, S.~J. 2015{\natexlab{b}}, \aap, 584, L9

\bibitem[{{Schwenke}(1988)}]{schwenke88}
{Schwenke}, D.~W. 1988, Theoretica Chimica Acta, 74, 381

\bibitem[{{Sicilia-Aguilar} {et~al.}(2016){Sicilia-Aguilar}, {Banzatti},
  {Carmona}, {Stolker}, {Kama}, {Mendigut{\'{\i}}a}, {Garufi}, {Flaherty}, {van
  der Marel}, \& {Greaves}}]{aguilar16}
{Sicilia-Aguilar}, A., {Banzatti}, A., {Carmona}, A., {Stolker}, T., {Kama},
  M., {Mendigut{\'{\i}}a}, I., {Garufi}, A., {Flaherty}, K., {van der Marel},
  N., \& {Greaves}, J. 2016, PASA, 33, e059

\bibitem[{{Simon} {et~al.}(2000){Simon}, {Dutrey}, \&
  {Guilloteau}}]{2000ApJ...545.1034S}
{Simon}, M., {Dutrey}, A., \& {Guilloteau}, S. 2000, \apj, 545, 1034

\bibitem[{{Simon} {et~al.}(2016){Simon}, {Pascucci}, {Edwards}, {Feng},
  {Gorti}, {Hollenbach}, {Rigliaco}, \& {Keane}}]{simon16}
{Simon}, M.~N., {Pascucci}, I., {Edwards}, S., {Feng}, W., {Gorti}, U.,
  {Hollenbach}, D., {Rigliaco}, E., \& {Keane}, J.~T. 2016, \apj, 831, 169

\bibitem[{{Skinner} \& {G{\"u}del}(2013)}]{skinner13}
{Skinner}, S.~L. \& {G{\"u}del}, M. 2013, \apj, 765, 3

\bibitem[{{Slanger} \& {Black}(1982)}]{slanger82}
{Slanger}, T.~G. \& {Black}, G. 1982, \jcp, 77, 2432

\bibitem[{{Stempels} {et~al.}(2007){Stempels}, {Gahm}, \&
  {Petrov}}]{2007A&A...461..253S}
{Stempels}, H.~C., {Gahm}, G.~F., \& {Petrov}, P.~P. 2007, \aap, 461, 253

\bibitem[{{Stephens} \& {Dalgarno}(1972)}]{stephens:72}
{Stephens}, T.~L. \& {Dalgarno}, A. 1972, \jqsrt, 12, 569

\bibitem[{{Tielens} \& {Hollenbach}(1985)}]{tielens85}
{Tielens}, A.~G.~G.~M. \& {Hollenbach}, D. 1985, \apj, 291, 747

\bibitem[{{van Boekel} {et~al.}(2005){van Boekel}, {Min}, {Waters}, {de Koter},
  {Dominik}, {van den Ancker}, \& {Bouwman}}]{2005A&A...437..189V}
{van Boekel}, R., {Min}, M., {Waters}, L.~B.~F.~M., {de Koter}, A., {Dominik},
  C., {van den Ancker}, M.~E., \& {Bouwman}, J. 2005, \aap, 437, 189

\bibitem[{{van den Ancker} {et~al.}(1998){van den Ancker}, {de Winter}, \&
  {Tjin A Djie}}]{1998A&A...330..145V}
{van den Ancker}, M.~E., {de Winter}, D., \& {Tjin A Djie}, H.~R.~E. 1998,
  \aap, 330, 145

\bibitem[{{van Harrevelt} \& {van Hemert}(2008)}]{harrevelt08}
{van Harrevelt}, R. \& {van Hemert}, M. 2008, Jounal of Phys. Chem. A, 112,
  3002

\bibitem[{{van Leeuwen}(2007)}]{2007A&A...474..653V}
{van Leeuwen}, F. 2007, \aap, 474, 653

\bibitem[{{Venuti} {et~al.}(2015){Venuti}, {Bouvier}, {Irwin}, {Stauffer},
  {Hillenbrand}, {Rebull}, {Cody}, {Alencar}, {Micela}, {Flaccomio}, \&
  {Peres}}]{venuti15}
{Venuti}, L., {Bouvier}, J., {Irwin}, J., {Stauffer}, J.~R., {Hillenbrand},
  L.~A., {Rebull}, L.~M., {Cody}, A.~M., {Alencar}, S.~H.~P., {Micela}, G.,
  {Flaccomio}, E., \& {Peres}, G. 2015, \aap, 581, A66

\bibitem[{{Walter} \& {Kuhi}(1981)}]{walter81}
{Walter}, F.~M. \& {Kuhi}, L.~V. 1981, \apj, 250, 254

\bibitem[{{Webb} {et~al.}(1999){Webb}, {Zuckerman}, {Platais}, {Patience},
  {White}, {Schwartz}, \& {McCarthy}}]{1999ApJ...512L..63W}
{Webb}, R.~A., {Zuckerman}, B., {Platais}, I., {Patience}, J., {White}, R.~J.,
  {Schwartz}, M.~J., \& {McCarthy}, C. 1999, \apjl, 512, L63

\bibitem[{{White} \& {Ghez}(2001)}]{2001ApJ...556..265W}
{White}, R.~J. \& {Ghez}, A.~M. 2001, \apj, 556, 265

\bibitem[{{Witt} {et~al.}(1989){Witt}, {Stecher}, {Boroson}, \&
  {Bohlin}}]{witt89}
{Witt}, A.~N., {Stecher}, T.~P., {Boroson}, T.~A., \& {Bohlin}, R.~C. 1989,
  \apjl, 336, L21

\bibitem[{{Woitke} {et~al.}(2016){Woitke}, {Min}, {Pinte}, {Thi}, {Kamp},
  {Rab}, {Anthonioz}, {Antonellini}, {Baldovin-Saavedra}, {Carmona}, {Dominik},
  {Dionatos}, {Greaves}, {G{\"u}del}, {Ilee}, {Liebhart}, {M{\'e}nard},
  {Rigon}, {Waters}, {Aresu}, {Meijerink}, \& {Spaans}}]{woitke16}
{Woitke}, P., {Min}, M., {Pinte}, C., {Thi}, W.-F., {Kamp}, I., {Rab}, C.,
  {Anthonioz}, F., {Antonellini}, S., {Baldovin-Saavedra}, C., {Carmona}, A.,
  {Dominik}, C., {Dionatos}, O., {Greaves}, J., {G{\"u}del}, M., {Ilee}, J.~D.,
  {Liebhart}, A., {M{\'e}nard}, F., {Rigon}, L., {Waters}, L.~B.~F.~M.,
  {Aresu}, G., {Meijerink}, R., \& {Spaans}, M. 2016, \aap, 586, A103

\bibitem[{{Wolniewicz}(1995)}]{wolniewicz95}
{Wolniewicz}, L. 1995, \jcp, 103, 1792

\bibitem[{{Yi} {et~al.}(2007){Yi}, {Park}, \& {Lee}}]{yi07}
{Yi}, W., {Park}, J., \& {Lee}, J. 2007, Chemical Physics Letters, 439, 46

\bibitem[{{Yung} {et~al.}(1982){Yung}, {Gladstone}, {Chang}, {Ajello}, \&
  {Srivastava}}]{yung82}
{Yung}, Y.~L., {Gladstone}, G.~R., {Chang}, K.~M., {Ajello}, J.~M., \&
  {Srivastava}, S.~K. 1982, \apjl, 254, L65

\bibitem[{{Zhang} {et~al.}(2013){Zhang}, {Pontoppidan}, {Salyk}, \&
  {Blake}}]{zhang13}
{Zhang}, K., {Pontoppidan}, K.~M., {Salyk}, C., \& {Blake}, G.~A. 2013, \apj,
  766, 82

\bibitem[{{Zhang} {et~al.}(2005){Zhang}, {Rheinecker}, \& {Bowman}}]{zhang05}
{Zhang}, X., {Rheinecker}, J.~L., \& {Bowman}, J.~M. 2005, \jcp, 122, 114313

\end{thebibliography}



\begin{figure}
\figurenum{A.1a}
\begin{center}
\epsfig{figure=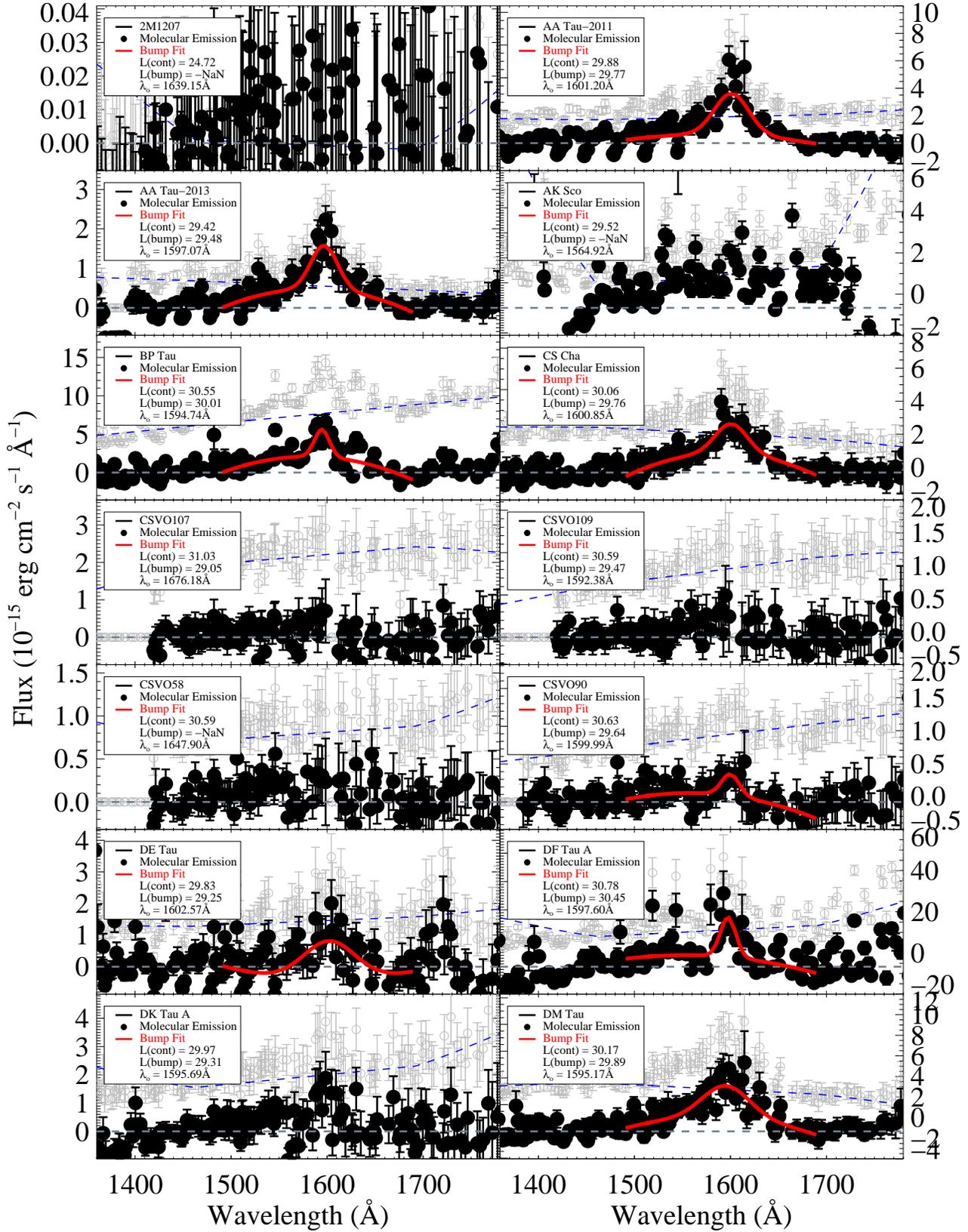,width=6in,angle=00}
\vspace{+0.45in}
\caption{ 1360~--1780~\AA\ spectral montage of all 41 observations (of 37 individual targets) analyzed in this work.     The reddening-corrected extracted continuum spectrum is shown in gray, the second order polynomial fit to the underlying FUV continuum spectrum is shown as the dashed blue line.  The continuum-subtracted Bump spectrum is displayed as the solid black circles and a Gaussian plus second-order polynomial fit to the Bump is shown in red, when applicable.  24/41 observations find Bump fits with FWHM~$>$~8~\AA, 1580~$\leq$~$\lambda_{o}$(CG)~$\leq$~1620~\AA, and a positive Bump flux.  However, there are examples of stars without clear bump emission that are fitted (e.g., CSVO109, DK Tau, DR Tau, and TW Hya-STIS), and stars that demonstrate Bumps that are missed by this FWHM cut (e.g., DN Tau, DQ Tau, and LkCa15).  Taken together, we estimate that 19/37 total sources show clear Bump emission.    
\label{cosovly}}
\end{center}
\end{figure}

\begin{figure}
\figurenum{A.1b}
\begin{center}
\epsfig{figure=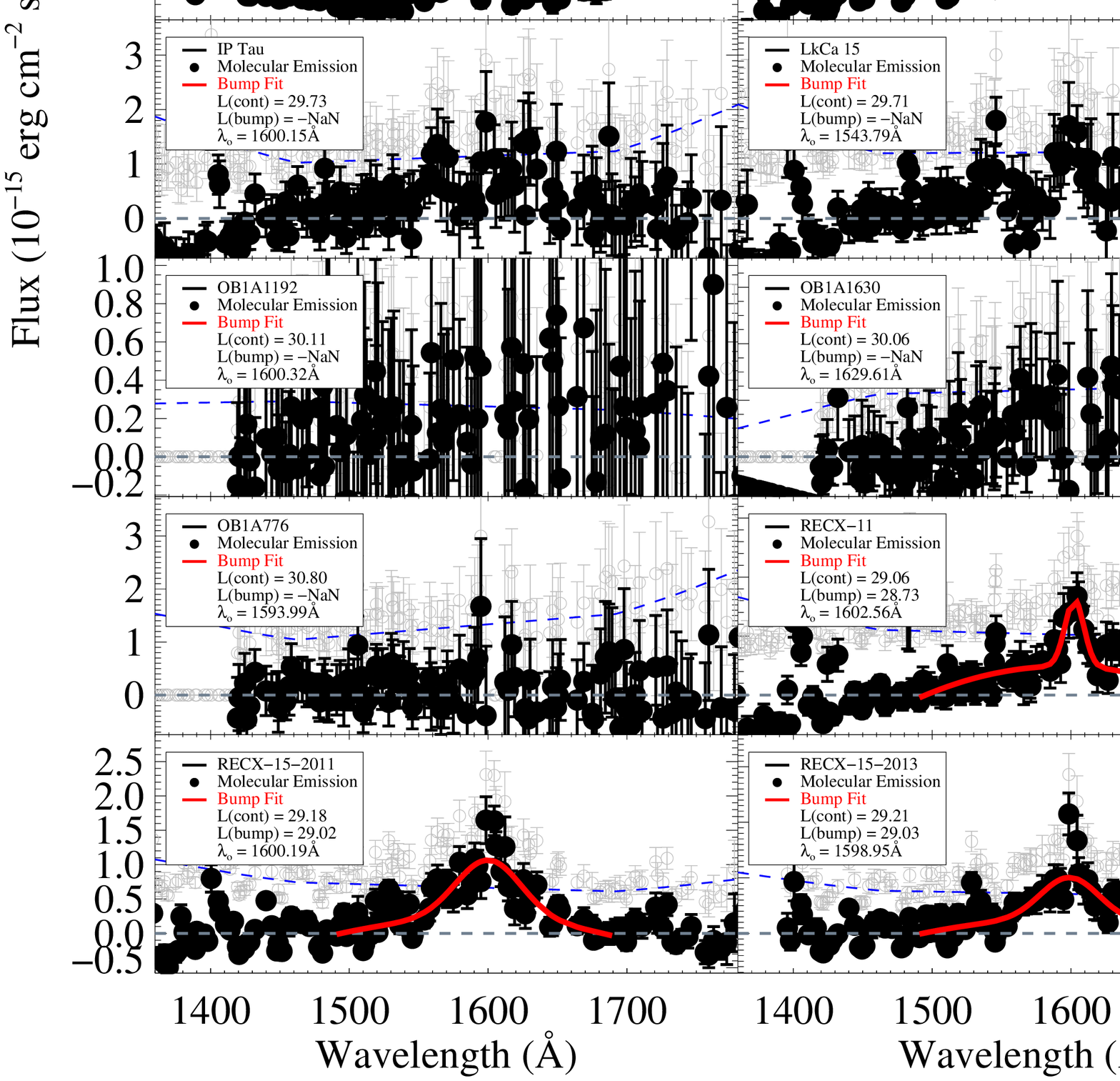,width=6in,angle=00}
\vspace{+0.45in}
\caption{1600~\AA\ Bump montage, continued.       
\label{cosovly}}
\end{center}
\end{figure}

\begin{figure}
\figurenum{A.1c}
\begin{center}
\epsfig{figure=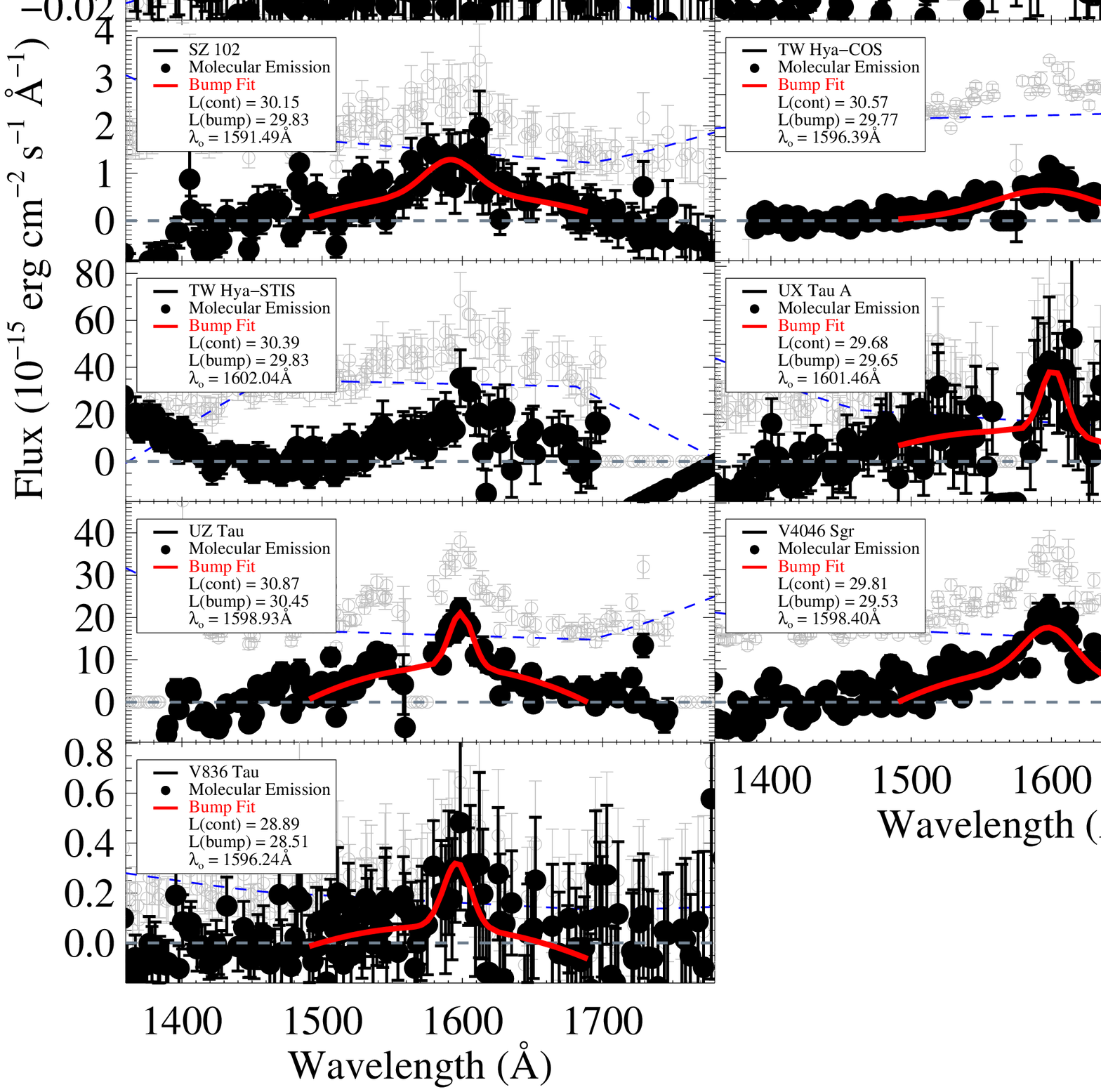,width=6in,angle=00}
\vspace{+0.45in}
\caption{1600~\AA\ Bump montage, continued.       
\label{cosovly}}
\end{center}
\end{figure}

\clearpage

\begin{deluxetable}{lccccccc}
\tabletypesize{\footnotesize}
\tablecaption{ $HST$ Target List \label{lya_targets}}
\tablewidth{0pt}
\tablehead{ \colhead{Target}   &    \colhead{d} &  \colhead{A$_V, lit$}    &   \colhead{A$_V, N(HI)$}    &   \colhead{M$_*$}  &  \colhead{$\dot M$}      &   \colhead{$L_{X}$\tablenotemark{a}}   &   \colhead{Ref.\tablenotemark{a}}  \\ 
   & (pc)& (mag)  &  (mag)    &     (M$_{\odot}$)    &       (10$^{-8}$ M$_{\odot}$ yr$^{-1}$)        &   (10$^{30}$ erg s$^{-1}$)   &    } 
\startdata
2M1207  & 52 &  0.00   &   0.00   &   0.04   &   0.00   &   0.00   &          75   \\ 
AA Tau-2011  & 140   &  0.50   &   0.34   &   0.80   &   0.33   &   1.00   &          2,4,7,12,16,58,59,83?   \\ 
AA Tau-2013  & 140   &  0.50   &   0.34   &   0.80   &   0.33   &   0.22   &          2,4,7,12,16,58,59,71   \\ 
AK Sco  & 103   &  0.50   &   0.08   &   1.35   &   0.09   &   0.10   &          18,20,34,62,74   \\ 
BP Tau  & 140   &  0.50   &   0.17   &   0.73   &   2.88   &   1.40   &          7,12,38,58,59,83   \\ 
CS Cha  & 160   &  0.80   &   0.16   &   1.05   &   1.20   &   2.57   &          21,35,60,73   \\ 
CSVO107  & 450   &  0.32   &   0.32   &   0.80   &   1.09   &   -1.00   &          21,22,84   \\ 
CSVO109  & 450   &  0.00   &   0.00   &   0.60   &   2.52   &   -1.00   &          21,22,84   \\ 
CSVO58  & 450   &  0.12   &   0.12   &   0.80   &   0.45   &   -1.00   &          21,22,84   \\ 
CSVO90  & 450   &  0.00   &   0.00   &   0.50   &   1.77   &   -1.00   &          21,22,84   \\ 
DE Tau  & 140   &  0.60   &   0.31   &   0.59   &   2.64   &   0.90   &         7,10,12,58,59,78   \\ 
DF Tau A  & 140   &  0.60   &   0.54   &   0.19   &   17.70   &   0.67   &         7,10,58,59,78   \\ 
DK Tau A  & 140   &  0.80   &   0.46   &   0.71   &   3.79   &   0.84   &         7,10,12,58,59,83   \\ 
DM Tau  & 140   &  0.00   &   0.48   &   0.50   &   0.29   &   2.00   &         16,29,32,58,59,83   \\ 
DN Tau  & 140   &  1.90   &   0.18   &   0.60   &   0.35   &   1.10   &        7,16,32,39,58,59,83   \\ 
DQ Tau  & 140   &  0.70   &   0.70   &   0.60   &   2.50   &   1.00   &         79   \\ 
DR Tau  & 140   &  3.20   &   0.48   &   0.80   &   3.16   &   0.40   &         2,8,16,58,59,73   \\ 
GM Aur  & 140   &  0.10   &   0.51   &   1.20   &   0.96   &   0.69   &         7,16,32,58,59,73  \\ 
HD 135344B  & 140   &  0.30   &   0.12   &   1.60   &   0.54   &   0.28   &         19,31,42,64,73   \\ 
HN Tau A  & 140   &  0.50   &   0.36   &   0.85   &   0.13   &   0.32   &         6,7,12,58,59,83   \\ 
IP Tau  & 140   &  0.20   &   0.72   &   0.68   &   0.08   &   1.26   &         7,12,58,59,73   \\ 
LkCa 15  & 140   &  0.60   &   0.31   &   0.85   &   0.13   &   2.50   &         12,29,32,58,59,80   \\ 
OB1A1192  & 450   &  0.85   &   0.85   &   0.77   &   0.17   &   0.83   &         21,22,84   \\ 
OB1A1630  & 450   &  0.00   &   0.00   &   0.40   &   0.08   &   0.04   &         21,22,84   \\ 
OB1A776  & 450   &  0.60   &   0.60   &   0.80   &   0.74   &   0.13   &         21,22,84   \\ 
RECX-11  & 97   &  0.00   &   0.03   &   0.80   &   0.03   &   3.20   &         13,24,47,61,77   \\ 
RECX-15-2011  & 97   &  0.00   &   0.02   &   0.40   &   0.10   &   0.06   &         13,14,15,61,77   \\ 
RECX-15-2013  & 97   &  0.00   &   0.02   &   0.40   &   0.10   &   0.06   &         13,14,15,61,77   \\ 
RU Lup  & 120   &  0.07   &   0.07   &   0.80   &   3.00   &   1.40   &         25,30,41,62,81   \\ 
RW Aur A-2011  & 140   &  1.60   &   0.11   &   1.40   &   3.16   &   0.39   &         5,9,11,12,17,58,59,73   \\ 
RW Aur A-2013  & 140   &  1.60   &   0.11   &   1.40   &   3.16   &   1.60   &         5,9,11,12,17,58,59,72   \\ 
RY Lup  & 120   &  0.40   &   0.10   &   1.71   &   1.00   &   9.80   &         23, 82   \\ 
SCHJ0439  & 140   &  0.00   &   0.00   &   0.08   &   0.01   &   -1.00   &         5   \\ 
SU Aur  & 140   &  0.90   &   0.29   &   2.30   &   0.45   &   11.64   &         1,3,8,11,12,58,59,70   \\ 
SZ 102  & 200   &  1.13   &   0.24   &   0.75   &   0.08   &   0.18   &         26,37,43,48,83   \\ 
TW Hya-COS  & 56   &  0.00   &   0.00   &   0.60   &   0.02   &   2.00   &         27,30,42,62,76   \\ 
TW Hya-STIS  & 56   &  0.00   &   0.00   &   0.60   &   0.02   &   2.00   &         27,30,42,62,76   \\ 
UX Tau A  & 140   &  0.20   &   0.51   &   1.30   &   1.00   &   2.00   &         12,32,58,59,73   \\ 
UZ Tau  & 140   &  1.50   &   1.50   &   0.60   &   3.00   &   0.74   &         70   \\ 
V4046 Sgr  & 83   &  0.00   &   0.00   &   0.86   &   1.30   &   2.00   &        28,33,44,76   \\ 
V836 Tau  & 140   &  1.70   &   0.38   &   0.75   &   0.01   &   1.70   &         12,30,46,58,59,83   
\enddata
\tablenotetext{a}{\rule{0mm}{5mm}
(1) \citet{akeson02}; (2) \citet{2007ApJ...659..705A}; (3) \citet{1988ApJ...330..350B}; (4) \citet{1999A&A...349..619B}; (5) \citet{pascucci13}; (6) \citet{france11b}; (7) \citet{1998ApJ...492..323G}; (8) \citet{2000ApJ...544..927G}; (9) \citet{1995ApJ...452..736H}; (10) \citet{2001ApJ...561.1060J}; (11) \citet{2000ApJ...539..815J}; (12) \citet{2009ApJ...704..531K}; (13) \citet{2004MNRAS.351L..39L}; (14) \citet{2004ApJ...609..917L}; (15) \citet{2007MNRAS.379.1658R}; (16) \citet{2010A&A...512A..15R}; (17) \citet{2001ApJ...556..265W}; (18) \citet{1998A&A...330..145V}; (19) \citet{2005A&A...437..189V}; (20) \citet{2003A&A...409.1037A}; (21) \cite{calvet05}; (22) \citet{ingleby14}; (23) \citet{manset09}; (24) \citet{2001MNRAS.321...57L}; (25) \citet{herczeg05}; (26) \citet{2010A&A...511A..10C}; (27) \citet{1999ApJ...512L..63W}; (28) \citet{2000IAUS..200P..28Q}; (29) \citet{1998ApJ...495..385H}; (30) \citet{2008ApJ...681..594H}; (31) \citet{2006A&A...459..837G}; (32) \citet{2011ApJ...732...42A}; (33) \citet{france11a}; (34) \citet{2009ApJ...698L.108G}; (35) \citet{2007ApJ...664L.111E};  (37) \citet{2003A&A...406.1001C}; (38) \citet{2000ApJ...545.1034S}; (39) \citet{2003ApJ...597L.149M}; (40) \citet{2010ApJ...717..441E}; (41) \citet{2007A&A...461..253S}; (42) \citet{2008ApJ...684.1323P}; (43) \citet{2004ApJ...604..758C}; (44) \citet{2010ApJ...720.1684R}; (45) \citet{2004ApJ...608..809G}; (46) \citet{2008ApJ...687.1168N}; (47) \citet{ingleby11}; (48) \citet{1994AJ....108.1071H}; (58) \citet{1999A&A...352..574B}; (59) \citet{2007ApJ...671..546L}; (60) \citet{2004ApJ...602..816L}; (61) \citet{1999PASA...16..257M}; (62) \citet{2007A&A...474..653V}; (64) \citet{2009ApJ...699.1822G}; 
(70)~\citet{gudel07}; (71)~\citet{schneider15}; (72)~\citet{schneider15b}; (73)~\citet{mcj14}; 
(74)~\citet{castro13}; (75)~\citet{gizis04}; (76)~\citet{gunther07}; (77)~\citet{lopez10}; (78)~\citet{walter81};
(79)~\citet{getman11}; (80)~\citet{skinner13}; (81)~\citet{gudel07b}; (82)~\citet{antonellini16};
(83)~\citet{gudel10}; (84)~\citet{ingleby11b}; 
}
\end{deluxetable}

\begin{deluxetable}{lccccccc}
\tabletypesize{\tiny}
\tablecaption{ 1600~\AA\ Bump Parameters and FUV Luminosities\tablenotemark{a}\label{lya_lines}}
\tablewidth{0pt}
\tablehead{ \colhead{Target}   &   \colhead{$L$(Bump)}    &   \colhead{$\lambda_{o}$\tablenotemark{b}}    &   \colhead{$FWHM$\tablenotemark{b}}  &  \colhead{$L$(Cont)}      &   \colhead{$L$(C IV)\tablenotemark{b}}  &   \colhead{$L$(H$_{2}$)}   &   \colhead{$L$(Ly$\alpha$)\tablenotemark{c}}  \\ 
   & (10$^{29}$ erg s$^{-1}$)  &  (\AA)    &     (\AA)    &       (10$^{29}$ erg s$^{-1}$)        &   (10$^{29}$ erg s$^{-1}$)   &    (10$^{29}$ erg s$^{-1}$)   & (10$^{29}$ erg s$^{-1}$)   
} 
\startdata
2M1207  &  0.00~$\pm$~0.00   &   1639.15~$\pm$~2.75   &   -8.60~$\pm$~8.03   &   0.00   &   0.01   &   0.01~$\pm$~0.00   &   0.21~    \\ 
AA Tau-2011\tablenotemark{d}  &  5.93~$\pm$~0.91   &   1601.20~$\pm$~1.62   &   42.17~$\pm$~4.82   &   7.52   &   4.34   &   24.15~$\pm$~1.60   &   385.32~$\pm$~69.01   \\ 
AA Tau-2013\tablenotemark{d}  &  3.03~$\pm$~0.62   &   1597.07~$\pm$~2.07   &   32.91~$\pm$~4.99   &   2.61   &   0.44   &   18.67~$\pm$~0.89   &   247.42~    \\ 
AK Sco  &  6.68~$\pm$~1.15   &   1564.92~$\pm$~3.61   &   -72.44~$\pm$~9.56   &   3.32   &   6.05   &   1.74~$\pm$~0.24   &   19.34~    \\ 
BP Tau\tablenotemark{d}  &  10.20~$\pm$~2.28   &   1594.74~$\pm$~1.47   &   18.29~$\pm$~3.28   &   35.80   &   18.60   &   18.18~$\pm$~1.98   &   159.10~$\pm$~28.20   \\ 
CS Cha\tablenotemark{d}  &  5.76~$\pm$~1.70   &   1600.85~$\pm$~3.07   &   38.95~$\pm$~8.55   &   11.50   &   4.47   &   12.81~$\pm$~0.87   &   131.32~    \\ 
CSVO107  &  1.11~$\pm$~11.80   &   1676.18~$\pm$~0.49   &   -0.80~$\pm$~2.00   &   189   &   24.50   &   50.99~$\pm$~2.41   &   742.57~    \\ 
CSVO109  &  2.96~$\pm$~3.38   &   1592.38~$\pm$~11.12   &    104.28~$\pm$~63.07   &   38.60   &   23.60   &   18.39~$\pm$~1.07   &   267.81~    \\ 
CSVO58  &  2.05~$\pm$~4.66   &   1647.90~$\pm$~1.08   &   -3.06~$\pm$~2.52   &   38.90   &   11.10   &   12.40~$\pm$~0.82   &   180.57~    \\ 
CSVO90\tablenotemark{d}  &  4.42~$\pm$~3.09   &   1599.99~$\pm$~4.55   &   21.81~$\pm$~12.24   &   42.70   &   16.30   &   26.83~$\pm$~2.04   &   390.72~    \\ 
DE Tau\tablenotemark{d}  &  1.77~$\pm$~0.66   &   1602.57~$\pm$~4.55   &   63.62~$\pm$~17.81   &   6.78   &   3.96   &   5.99~$\pm$~0.89   &   91.32~$\pm$~26.80   \\ 
DF Tau A\tablenotemark{d}  &  28.10~$\pm$~3.93   &   1597.60~$\pm$~0.69   &   19.18~$\pm$~1.91   &   60.10   &   20.80   &   73.74~$\pm$~3.54   &   1041.73~    \\ 
DK Tau A  &  2.03~$\pm$~1.34   &   1595.69~$\pm$~3.38   &   14.24~$\pm$~7.00   &   9.25   &   2.14   &   4.72~$\pm$~0.27   &   58.53~    \\ 
DM Tau\tablenotemark{d}  &  7.80~$\pm$~2.33   &   1595.17~$\pm$~3.29   &   57.42~$\pm$~13.03   &   14.80   &   6.16   &   62.04~$\pm$~4.48   &   933.74~$\pm$~102.58   \\ 
DN Tau\tablenotemark{d}  &  1.41~$\pm$~1.76   &   1629.61~$\pm$~0.47   &   -1.73~$\pm$~1.87   &   4.91   &   4.69   &   0.05~$\pm$~0.01   &   0.40~    \\ 
DQ Tau\tablenotemark{d}  &  2.15~$\pm$~0.56   &   1548.37~$\pm$~0.72   &   -4.21~$\pm$~0.57   &   3.47   &   2.59   &   5.45~$\pm$~1.50   &   79.36~    \\ 
DR Tau  &  1.18~$\pm$~0.18   &   1597.79~$\pm$~0.57   &   11.24~$\pm$~1.29   &   44.50   &   6.09   &   17.84~$\pm$~2.01   &  31.73~    \\ 
GM Aur\tablenotemark{d}  &  15.60~$\pm$~5.47   &   1596.07~$\pm$~2.12   &   19.83~$\pm$~5.36   &   38.10   &   20.50   &   93.67~$\pm$~9.11   &   1880.25~$\pm$~462.67   \\ 
HD 135344B  &  1.87~$\pm$~0.54   &   1550.10~$\pm$~1.29   &   -15.46~$\pm$~2.74   &   20.60   &   17.60   &   7.59~$\pm$~0.65   &   97.88~    \\ 
HN Tau A\tablenotemark{d}  &  4.59~$\pm$~1.35   &   1594.86~$\pm$~3.39   &   42.24~$\pm$~9.57   &   10.20   &   1.89   &   10.71~$\pm$~0.62   &   158.68~$\pm$~31.15   \\ 
IP Tau  &  2.47~$\pm$~3.87   &   1600.15~$\pm$~2.92   &   -5.76~$\pm$~5.63   &   5.35   &   4.20   &   35.20~$\pm$~1.76   &   686.45~    \\ 
LkCa 15\tablenotemark{d}  &  1.93~$\pm$~1.25   &   1543.79~$\pm$~3.04   &   -10.25~$\pm$~5.84   &   5.12   &   4.61   &   7.27~$\pm$~0.39   &   101.98~$\pm$~16.79   \\ 
OB1A1192  &  4.17~$\pm$~2.47   &   1600.32~$\pm$~1.33   &   -5.66~$\pm$~4.37   &   12.80   &   9.53   &   17.64~$\pm$~1.84   &   256.89~    \\ 
OB1A1630  &  1.91~$\pm$~36.70   &   1629.61~$\pm$~3.87   &   -1.69~$\pm$~21.38   &   11.50   &   6.76   &   15.73~$\pm$~1.41   &   229.07~    \\ 
OB1A776  &  2.48~$\pm$~7.59   &   1593.99~$\pm$~2.68   &   -3.09~$\pm$~5.08   &   63.60   &   14.50   &   22.43~$\pm$~1.50   &   326.64~    \\ 
RECX-11\tablenotemark{d}  &  0.54~$\pm$~0.18   &   1602.56~$\pm$~1.52   &   16.09~$\pm$~3.53   &   1.16   &   1.88   &   2.11~$\pm$~0.11   &   66.28~$\pm$~4.43   \\ 
RECX-15-2011\tablenotemark{d}  &  1.06~$\pm$~0.19   &   1600.19~$\pm$~2.20   &   55.34~$\pm$~7.52   &   1.51   &   0.38   &   5.80~$\pm$~0.32   &   142.24~$\pm$~19.39   \\ 
RECX-15-2013\tablenotemark{d}  &  1.08~$\pm$~0.25   &   1598.95~$\pm$~2.72   &   48.63~$\pm$~8.40   &   1.61   &   0.43   &   5.14~$\pm$~0.59   &   75.96~    \\ 
RU Lup  &  5.54~$\pm$~1.59   &   1540.45~$\pm$~0.61   &   -8.01~$\pm$~1.82   &   49.40   &   7.68   &   27.20~$\pm$~15.20   &   653.30~$\pm$~148.30   \\ 
RW Aur A-2011  &  10.10~$\pm$~3.51   &   1592.80~$\pm$~0.39   &   -5.50~$\pm$~1.43   &   15.00   &   1.78   &   0.77~$\pm$~0.21   &   6.56~    \\ 
RW Aur A-2013  &  3.29~$\pm$~0.30   &   1536.29~$\pm$~0.32   &   -11.00~$\pm$~0.80   &   66.60   &   5.95   &   1.04~$\pm$~0.57   &   8.79~    \\ 
RY Lup\tablenotemark{d}  &  1.55~$\pm$~0.80   &   1601.34~$\pm$~2.65   &   20.52~$\pm$~7.20   &   1.61   &   1.22   &   3.05~$\pm$~0.36   &   36.47~    \\ 
SCHJ0439  &  0.00~$\pm$~0.00   &   1593.07~$\pm$~0.84   &   -1.83~$\pm$~1.32   &   0.01   &   0.02   &   0.10~$\pm$~0.01   &   1.46~    \\ 
SU Aur\tablenotemark{d}  &  1.70~$\pm$~1.05   &   1604.61~$\pm$~3.68   &   17.78~$\pm$~8.13   &   5.72   &   4.06   &   2.53~$\pm$~0.28   &   41.60~$\pm$~9.81   \\ 
SZ 102\tablenotemark{d}  &  6.72~$\pm$~3.53   &   1591.49~$\pm$~5.61   &   42.99~$\pm$~18.09   &   14.10   &   7.36   &   3.55~$\pm$~0.25   &   35.01~    \\ 
TW Hya-COS\tablenotemark{d}  &  5.88~$\pm$~1.71   &   1596.39~$\pm$~1.78   &   87.29~$\pm$~13.22   &   37.60   &   8.91   &   12.44~$\pm$~1.57   &   181.16~    \\ 
TW Hya-STIS  &  6.81~$\pm$~4.03   &   1602.04~$\pm$~1.60   &   10.09~$\pm$~3.79   &   24.70   &   17.20   &   16.80~$\pm$~2.00   &   199.60~$\pm$~34.30   \\ 
UX Tau A\tablenotemark{d}  &  4.44~$\pm$~3.17   &   1601.46~$\pm$~4.01   &   21.29~$\pm$~11.28   &   4.78   &   5.00   &   43.90~$\pm$~4.39   &   617.25~$\pm$~51.89   \\ 
UZ Tau\tablenotemark{d}  &  28.30~$\pm$~5.09   &   1598.93~$\pm$~0.94   &   19.51~$\pm$~2.68   &   74.20   &   26.90   &   174.82~$\pm$~0.30   &   2545.92~    \\ 
V4046 Sgr\tablenotemark{d}  &  3.36~$\pm$~0.42   &   1598.40~$\pm$~1.34   &   42.24~$\pm$~3.85   &   6.50   &   4.55   &   19.80~$\pm$~0.90   &   383.30~$\pm$~42.10   \\ 
V836 Tau\tablenotemark{d}  &  0.33~$\pm$~0.27   &   1596.24~$\pm$~5.55   &   22.86~$\pm$~14.78   &   0.77   &   0.76   &   0.12~$\pm$~0.01   &   1.18~$\pm$~0.00
\enddata
\tablenotetext{a}{N(HI)-based reddening corrections used except where noted in Section 3.2.} 
\tablenotetext{b}{The Bump spectra are parameterized by a Central Gaussian and a second-order polynomial, fitted over 1520~--~1670~\AA\ bandpass.  The central wavelength ($\lambda_{o}$) and full-width at half-maximum ($FWHM$) refer to the Central Gaussian component.    } 
\tablenotetext{c}{Intrinsic Ly$\alpha$ luminosity values with error bars were taken from the literature~\citep{schindhelm12b, france14a}.  Ly$\alpha$ values without error bars were extrapolated from direct H$_{2}$ fluorescence measurements based on the $L$(Ly$\alpha$) vs. $L$(H$_{2}$) relationship (see Section 3.1.2).  
}
\tablenotetext{d}{The 24 spectra with visually-defined Bumps  (stars with FWHM $>$~0, except CSVO109, DK Tau, DR Tau, and TW Hya-STIS, plus DN Tau, DQ Tau, and LkCa15; Section 3.1.1).  These spectra are the solid black diamonds in Figures 4, 5, 7, and 8. } 
\end{deluxetable}

\begin{deluxetable}{lcccccc}
\tabletypesize{\small}
\tablecaption{Non-thermal H$_{2}$ Coincidences with Ly$\alpha$~\label{lya_lines}}
\tablewidth{0pt}
\tablehead{
\colhead{[$v$,$J$]} & \colhead{$\lambda$} &   \colhead{$A_{vJ \rightarrow v'J'}$}   & 
\colhead{Electronic Band\tablenotemark{a}} & \colhead{($v^{'}$~--~$v$)} & \colhead{$\Delta$$J$} & \colhead{$P_{diss}$\tablenotemark{b}}  \\ 
   &    (\AA)    &   (10$^{7}$ s$^{-1}$)    &     &        &      &      (\% )   }
\startdata
$[0,20]$   &         1217.72   &    5.43   &    Ly   &   (5~--~0)   &   P(20)   &   47.6    \\
$[0,21]$   &         1218.84   &    6.12   &    Ly   &   (6~--~0)   &   P(21)   &   51.8    \\
$[0,24]$   &         1214.94   &    7.77   &    Ly   &   (9~--~0)   &   R(24)   &   61.6    \\
$[0,24]$   &         1214.16   &    5.72   &    Ly   &   (10~--~0)   &   P(24)   &   60.9    \\
$[2,19]$   &         1212.92   &    4.08   &    Ly   &   (10~--~2)   &   R(19)   &   59.6    \\
$[2,20]$   &         1214.17   &    4.32   &    Ly   &   (11~--~2)   &   R(20)   &   61.6    \\
$[3,16]$   &         1213.52   &    4.85   &    Ly   &   (11~--~3)   &   P(16)   &   48.4    \\
$[3,16]$   &         1214.57   &    4.74   &    Wp   &   (0~--~3)   &   P(16)   &   5.5    \\
$[3,17]$   &         1213.33   &    2.94   &    Ly   &   (12~--~3)   &   P(17)   &   57.9    \\
$[3,18]$   &         1213.84   &    3.12   &    Ly   &   (13~--~3)   &   P(18)   &   61.4    \\
$[3,18]$   &         1216.70   &    17.85   &    Wm   &   (0~--~3)   &   Q(18)   &   0.0    \\
$[4,13]$   &         1213.32   &    2.59   &    Ly   &   (11~--~4)   &   R(13)   &   55.5    \\
$[4,14]$   &         1214.57   &    13.77   &    Wp   &   (1~--~4)   &   P(14)   &   1.1    \\
$[4,16]$   &         1216.93   &    32.07   &    Wm   &   (1~--~4)   &   Q(16)   &   0.0    \\
$[5,8]$   &         1219.45   &    3.94   &    Ly   &   (11~--~5)   &   P(8)   &   52.3    \\
$[5,8]$   &         1214.62   &    10.16   &    Wp   &   (1~--~5)   &   R(8)   &   13.4    \\
$[5,9]$   &         1213.39   &    4.06   &    Ly   &   (12~--~5)   &   P(9)   &   51.9    \\
$[5,9]$   &         1217.00   &    8.02   &    Wp   &   (1~--~5)   &   R(9)   &   1.7    \\
$[5,11]$   &         1217.37   &    3.68   &    Ly   &   (13~--~5)   &   P(11)   &   55.7    \\
$[5,12]$   &         1215.93   &    16.27   &    Wp   &   (2~--~5)   &   P(12)   &   6.7    \\
$[5,13]$   &         1213.07   &    32.75   &    Wm   &   (2~--~5)   &   Q(13)   &   0.0    \\
\tableline
\enddata
\tablenotetext{a}{ `Ly' refers to the $B$$^{1}\Sigma^{+}_{u}$~--~$X$$^{1}\Sigma^{+}_{g}$ electronic transition system, `Wp' and `Wm' refer to the $C$$^{1}\Pi_{u}$~--~$X$$^{1}\Sigma^{+}_{g}$ electronic transition systems, where the `p' and `m' distinguish the $C^{+}$ and  $C^{-}$ states, respectively.  } 
\tablenotetext{b}{ $P_{diss}$ is the molecular dissociation probability following absorption into the [$v^{'}$,$J^{'}$]
level of the upper electronic state, $P_{diss}$ = $\frac{A_{n'v'J' \rightarrow continuum}}{\Sigma A_{n'v'J'}}$.  }
\end{deluxetable}

\clearpage

\begin{deluxetable}{ccccccl}
\tabletypesize{\small}
\tablecaption{Brightest Predicted Non-thermal H$_{2}$ Fluorescence of Ly$\alpha$ Following H$_{2}$O Photodissociation~\label{lya_lines}}
\tablewidth{0pt}
\tablehead{
\colhead{$\lambda$} & \colhead{$A_{v'J' \rightarrow v''J''}$, (10$^{7}$ s$^{-1}$) } &   \colhead{Electronic Band\tablenotemark{a} }   & 
\colhead{($v^{'}$~--~$v$)} & \colhead{$\Delta$$J$} & \colhead{$B_{mn}$\tablenotemark{b}}  &
\colhead{Detection~~~~~~~~~~~~~~~~~~~~~~~~~~~~~~~~~~~~~~}     }
\startdata
      1217.72   &    5.43   &    Ly   &   (5~--~0)   &   P(20)   &   0.05  &     Ly$\alpha$-pump  \\
      1176.07   &    5.08   &    Ly   &   (5~--~0)   &   R(18)   &   0.04  &     \ion{C}{3}\tablenotemark{c} \\
      1265.82   &    6.79   &    Ly   &   (5~--~1)   &   P(20)   &   0.06  &     Thermal\tablenotemark{c}  \\
      1223.42   &    10.30   &    Ly   &   (5~--~1)   &   R(18)   &   0.09  &     Thermal\tablenotemark{c}  \\
      1361.61   &    5.21   &    Ly   &   (5~--~3)   &   P(20)   &   0.05  &     Y  \\
      1318.66   &    2.98   &    Ly   &   (5~--~3)   &   R(18)   &   0.03  &     CO\tablenotemark{c}  \\
      1365.53   &    4.49   &    Ly   &   (5~--~4)   &   R(18)   &   0.04  &     Y  \\
      1452.19   &    4.14   &    Ly   &   (5~--~5)   &   P(20)   &   0.04  &     Y  \\
      1453.80   &    4.23   &    Ly   &   (5~--~6)   &   R(18)   &   0.04  &     Y/N  \\
      1528.18   &    4.05   &    Ly   &   (5~--~7)   &   P(20)   &   0.04  &     Y/N  \\
      1260.14   &    6.20   &    Ly   &   (9~--~0)   &   P(26)   &   0.07  &     N  \\
      1214.94   &    7.77   &    Ly   &   (9~--~0)   &   R(24)   &   0.09  &     Ly$\alpha$-pump  \\
      1256.96   &    2.72   &    Ly   &   (9~--~1)   &   R(24)   &   0.03  &     Thermal\tablenotemark{c}  \\
      1342.23   &    3.54   &    Ly   &   (9~--~2)   &   P(26)   &   0.04  &     Thermal\tablenotemark{c}  \\
      1298.00   &    2.59   &    Ly   &   (9~--~2)   &   R(24)   &   0.03  &     N  \\
      1373.95   &    2.82   &    Ly   &   (9~--~4)   &   R(24)   &   0.03  &     Y  \\
      1214.16   &    5.72   &    Ly   &   (10~--~0)   &   P(24)   &   0.07  &     Ly$\alpha$-pump  \\
      1171.89   &    8.47   &    Ly   &   (10~--~0)   &   R(22)   &   0.10  &     Thermal\tablenotemark{c}  \\
      1297.10   &    3.22   &    Ly   &   (10~--~2)   &   P(24)   &   0.04  &     Y/N  \\
      1255.02   &    3.73   &    Ly   &   (10~--~2)   &   R(22)   &   0.04  &     N  \\
      1333.54   &    3.27   &    Ly   &   (10~--~4)   &   R(22)   &   0.04  &     \ion{C}{2}\tablenotemark{c}  \\
      1166.90   &    5.52   &    Ly   &   (10~--~0)   &   P(21)   &   0.06  &     N  \\
      1252.19   &    3.38   &    Ly   &   (10~--~2)   &   P(21)   &   0.04  &     N  \\
      1212.92   &    4.08   &    Ly   &   (10~--~2)   &   R(19)   &   0.05  &     Ly$\alpha$-pump  \\
      1295.68   &    3.50   &    Ly   &   (10~--~4)   &   R(19)   &   0.04  &     Y/N  \\
      1170.71   &    4.92   &    Ly   &   (11~--~0)   &   P(22)   &   0.06  &     Y/N  \\
      1214.17   &    4.32   &    Ly   &   (11~--~2)   &   R(20)   &   0.05  &     Ly$\alpha$-pump  \\
      1294.03   &    3.10   &    Ly   &   (11~--~4)   &   R(20)   &   0.04  &     Thermal\tablenotemark{c}  \\
      1213.52   &    4.85   &    Ly   &   (11~--~3)   &   P(16)   &   0.05  &     Ly$\alpha$-pump  \\
      1171.88   &    21.99   &    Wp   &   (0~--~2)   &   P(16)   &   0.20  &     Thermal\tablenotemark{c}  \\
      1214.57   &    4.74   &    Wp   &   (0~--~3)   &   P(16)   &   0.04  &     Ly$\alpha$-pump  \\
      1182.36   &    10.48   &    Wp   &   (0~--~3)   &   R(14)   &   0.10  &     Y  \\
      1224.62   &    4.72   &    Wp   &   (0~--~4)   &   R(14)   &   0.04  &     N  \\
      1213.33   &    2.94   &    Ly   &   (12~--~3)   &   P(17)   &   0.03  &     Ly$\alpha$-pump  \\
      1213.84   &    3.12   &    Ly   &   (13~--~3)   &   P(18)   &   0.04  &     Ly$\alpha$-pump  \\
      1259.07   &    2.73   &    Ly   &   (13~--~5)   &   R(16)   &   0.03  &     N  \\
      1176.08   &    37.68   &    Wm   &   (0~--~2)   &   Q(18)   &   0.34  &     \ion{C}{3} \\
      1216.70   &    17.85   &    Wm   &   (0~--~3)   &   Q(18)   &   0.16  &     Ly$\alpha$-pump  \\
      1256.48   &    3.76   &    Wm   &   (0~--~4)   &   Q(18)   &   0.03  &     N  \\
      1213.32   &    2.59   &    Ly   &   (11~--~4)   &   R(13)   &   0.03  &     Ly$\alpha$-pump  \\
      1285.78   &    2.51   &    Ly   &   (11~--~5)   &   P(15)   &   0.03  &     Y  \\
      1172.99   &    15.19   &    Wp   &   (1~--~3)   &   P(14)   &   0.14  &     Y  \\
      1214.57   &    13.77   &    Wp   &   (1~--~4)   &   P(14)   &   0.12  &    Ly$\alpha$-pump  \\
      1186.23   &    17.36   &    Wp   &   (1~--~4)   &   R(12)   &   0.16  &     Y  \\
      1255.51   &    5.65   &    Wp   &   (1~--~5)   &   P(14)   &   0.05  &     Blend  \\
      1227.32   &    8.07   &    Wp   &   (1~--~5)   &   R(12)   &   0.07  &     Thermal\tablenotemark{c}  \\
      1177.27   &    24.92   &    Wm   &   (1~--~3)   &   Q(16)   &   0.23  &    \ion{C}{3}  \\
      1216.93   &    32.07   &    Wm   &   (1~--~4)   &   Q(16)   &   0.29  &     Ly$\alpha$-pump  \\
      1255.58   &    11.72   &    Wm   &   (1~--~5)   &   Q(16)   &   0.11  &     Blend  \\
      1193.34   &    13.19   &    Wp   &   (1~--~4)   &   P(10)   &   0.12  &     Thermal\tablenotemark{c}  \\
      1171.32   &    7.04   &    Wp   &   (1~--~4)   &   R(8)   &   0.06  &      Y/N \\
      1214.62   &    10.16   &    Wp   &   (1~--~5)   &   R(8)   &   0.09  &     Ly$\alpha$-pump  \\
      1213.39   &    4.06   &    Ly   &   (12~--~5)   &   P(9)   &   0.04  &    Ly$\alpha$-pump  \\
      1277.23   &    2.93   &    Ly   &   (12~--~7)   &   R(7)   &   0.03  &     Y/N  \\
      1154.91   &    14.12   &    Wp   &   (1~--~3)   &   P(11)   &   0.12  &     Thermal\tablenotemark{c}  \\
      1197.99   &    12.82   &    Wp   &   (1~--~4)   &   P(11)   &   0.11  &     Thermal\tablenotemark{c}  \\
      1174.20   &    18.15   &    Wp   &   (1~--~4)   &   R(9)   &   0.16  &     \ion{C}{3}  \\
      1240.87   &    6.12   &    Wp   &   (1~--~5)   &   P(11)   &   0.05  &     Thermal\tablenotemark{c} \\
      1217.00   &    8.02   &    Wp   &   (1~--~5)   &   R(9)   &   0.07  &     Ly$\alpha$-pump  \\
      1217.37   &    3.68   &    Ly   &   (13~--~5)   &   P(11)   &   0.04  &     Ly$\alpha$-pump  \\
      1175.59   &    6.42   &    Wp   &   (2~--~4)   &   P(12)   &   0.06  &     \ion{C}{3}  \\
      1215.93   &    16.27   &    Wp   &   (2~--~5)   &   P(12)   &   0.15  &     Ly$\alpha$-pump  \\
      1191.65   &    8.31   &    Wp   &   (2~--~5)   &   R(10)   &   0.08  &     Y  \\
      1255.38   &    5.13   &    Wp   &   (2~--~6)   &   P(12)   &   0.05  &     Blend  \\
      1231.39   &    15.04   &    Wp   &   (2~--~6)   &   R(10)   &   0.14  &     Y  \\
      1269.95   &    3.14   &    Wp   &   (2~--~7)   &   R(10)   &   0.03  &     Y/N  \\
      1173.83   &    5.17   &    Wm   &   (2~--~4)   &   Q(13)   &   0.05  &     Y  \\
      1213.07   &    32.75   &    Wm   &   (2~--~5)   &   Q(13)   &   0.30  &     Ly$\alpha$-pump  \\
      1251.25   &    22.03   &    Wm   &   (2~--~6)   &   Q(13)   &   0.20  &     Y  \\
      1287.70   &    3.33   &    Wm   &   (2~--~7)   &   Q(13)   &   0.03  &     Thermal\tablenotemark{c} 
\enddata
\clearpage
\tablenotetext{a}{`Ly' refers to the $B$$^{1}\Sigma^{+}_{u}$~--~$X$$^{1}\Sigma^{+}_{g}$...see Table 3.  } 
\tablenotetext{b}{The branching ratio is the ratio of the line transition probability to the total transition probability out of state [$v^{'}$,$J^{'}$], $B_{mn}$ = $\frac{A_{n'v'J' \rightarrow v''J''}}{\Sigma A_{n'v'J'}}$} 
\tablenotetext{c}{Blends with other atomic or molecular emission lines prevent reliable detection.  \ion{C}{3} is the 1175~\AA\ multiplet, \ion{C}{2} is the 1335~\AA\ multiplet, and CO is the $A$~--~$X$~(14~--~3) fluorescence band.  `Thermal' indicates that emission lines produced by Ly$\alpha$ fluorescence from $T$(H$_{2}$)~$\sim$~2500 K. }
\end{deluxetable}

\clearpage

\end{document}